\documentclass[11pt]{amsart}

\usepackage{amssymb}
\usepackage{epsfig}
\usepackage{comment}
\usepackage{amsmath}
\usepackage{subcaption}
\usepackage[section]{placeins}
\usepackage{graphicx}
\usepackage{units}
\usepackage{color}

    \newcommand{\mbs}[1]{\boldsymbol{#1}}

    \newtheorem{thm}{Theorem}[section]

    \newtheorem{defn}[thm]{Definition}

     \def\bB{{\mbs{B}}} \def\bC{{\mbs{C}}}
      \def\bF{{\mbs{F}}}
      \def\bI{{\mbs{I}}}

    \def\bP{{\mbs{P}}}

    \def\bY{{\mbs{Y}}}

    \def\bd{{\mbs{d}}}  
      
      \def\bl{{\mbs{l}}}
      
      \def\br{{\mbs{r}}}
    \def\bs{{\mbs{s}}}  
      
    \def\by{{\mbs{y}}} 

\begin{document}

\title{The anomalous yield behavior of fused silica glass}

\author{W.~Schill${}^1$, S.~Heyden${}^{1}$, S.~Conti${}^2$ and M.~Ortiz${}^{1}$}

\address
{
    ${}^1$Division of Engineering and Applied Science,
    California Institute of Technology,
    1200 E.~California Blvd., Pasadena, CA 91125.
}

\email{ortiz@caltech.edu}

\address
{
    ${}^2$Institut f\"ur Angewandte Mathematik,
    Universit\"at Bonn,
    Endenicher Allee 60, 53115 Bonn, Germany.
}

\begin{abstract}
We develop a critical-state model of fused silica plasticity on the basis of data mined from molecular dynamics (MD) calculations. The MD data is suggestive of an irreversible densification transition in volumetric compression resulting in permanent, or plastic, densification upon unloading. The MD data also reveals an evolution towards a critical state of constant volume under pressure-shear deformation. The trend towards constant volume is from above, when the glass is overconsolidated, or from below, when it is underconsolidated. We show that these characteristic behaviors are well-captured by a critical state model of plasticity, where the densification law for glass takes the place of the classical consolidation law of granular media and the locus of constant-volume states defines the critical-state line. A salient feature of the critical-state line of fused silica, as identified from the MD data, that renders its yield behavior anomalous is that it is strongly non-convex, owing to the existence of two well-differentiated phases at low and high pressures. We argue that this strong non-convexity of yield explains the patterning that is observed in molecular dynamics calculations of amorphous solids deforming in shear. We employ an explicit and exact rank-$2$ envelope construction to upscale the microscopic critical-state model to the macroscale. Remarkably, owing to the equilibrium constraint the resulting effective macroscopic behavior is still characterized by a non-convex critical-state line. Despite this lack of convexity, the effective macroscopic model is stable against microstructure formation and defines well-posed boundary-value problems.
\end{abstract}

\maketitle

%\tableofcontents

\section{Introduction}

\begin{figure}[h]
	\begin{subfigure}{0.30\textwidth}
         \includegraphics[width=0.99\linewidth]{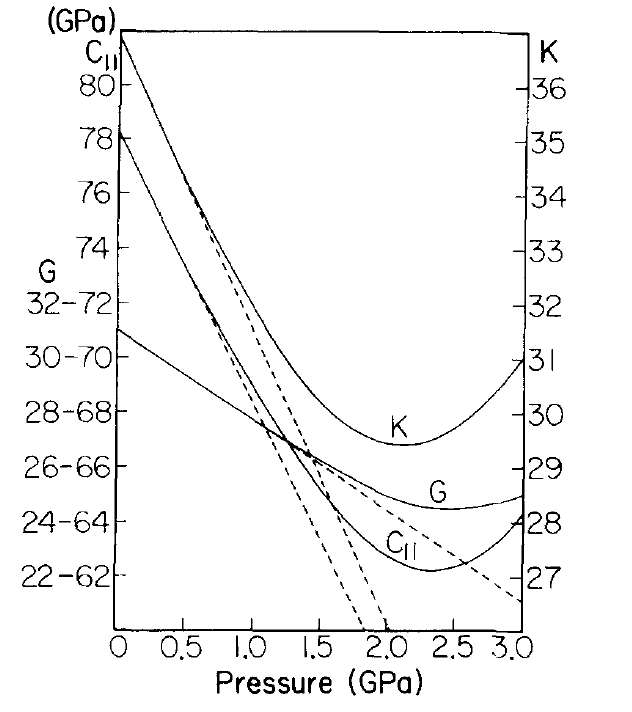}
	\end{subfigure}
	\begin{subfigure}{0.5\textwidth}
        \includegraphics[width=0.99\linewidth]{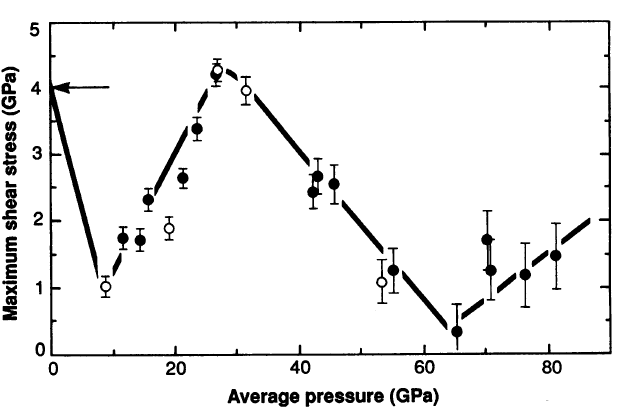}
	\end{subfigure}
	\caption{\small a) Elastic moduli {\sl vs}.~pressure as measured by Kondo {\sl et al.} (1981) \cite{doi:10.1063/1.329012}; b) Measurements of the yield strength of SiO${}_2$ glass at pressures as high as $81$ GPa at room temperature showing the variation of the strength of amorphous silica as it is compressed to denser structures with higher coordination \cite{MEADE1072}.} \label{8poaSL}
\end{figure}

The anomalous shear modulus behavior of silica glass has been a long-standing topic of investigation. For instance, Kondo {\sl et al.} \cite{doi:10.1063/1.329012} and references therein examined the non-monotonic dependence of the elastic moduli on pressure for fused quartz, cf.~Fig.~ \ref{8poaSL}a. Notably, between 0 and 2.5 GPa, the shear modulus and bulk modulus decreases. Likewise, the anomalous pressure dependence of the strength of amorphous silica has also received considerable attention. For instance, Meade and Jeanloz \cite{MEADE1072} made measurements of the yield strength at pressures up to $81$ GPa at room temperature and showed that the strength of amorphous silica decreases significantly as it is compressed to denser structures with higher coordination, Fig.~\ref{8poaSL}b. Clifton {\sl et al.} \cite{doi:10.1063/1.55558, doi:10.1063/1.322888, 1998AIPC..429..517S} and Simha and Gupta \cite{doi:10.1063/1.1763992} investigated the effect of pressure on failure waves in silica and soda-lime glass through angled flyer plate impact experiments and observed a loss of shear strength as the failure wave traversed the glass at pressures of 4-6 GPa.

These phenomena appear to be intimately linked to structural rearrangements occurring at the atomic level. Sato and Funamori \cite{PhysRevB.82.184102,Sato} performed structural measurements of SiO$ _{2} $ glass Si-O bond length and coordination number at pressures from 20 to 100 GPa using a diamond anvil cell and x-ray diffraction. They observed a transition from four-fold to six-fold coordinated structure that comes to completion at around 45 GPa. Wakabayashi {\sl et al.} \cite{PhysRevB.84.144103} studied the densification behavior again using a diamond anvil cell experimental setup and concluded that permanent densification occurs for pressures between 9 and 13 GPa. Vandembroucq \cite{0953-8984-20-48-485221} observed pressure-induced reorganizations of the amorphous network allowing a more efficient packing of tetrahedra that remain linked at their vertices only. Inamura {\sl et al.} \cite{2004PhRvL..93a5501I} studied transformations at pressures  of up to about 20 GPa and temperatures of up to about 700 C. Their results are indicative of the existence of a high pressure variant of silica glass. However, a sharp phase transformation was not observed, which is suggestive of a volumetric plastic hardening mechanism. Luo {\sl et al.} \cite{Luo2004} reported a novel dense silica polymorph retrieved from shock-wave and diamond-anvil cell experiments. The polymorph is composed of face-sharing polyhedra and it has a density similar to stishovite. Sterical constraints on the bond angles induce an intrinsic disorder in the Si positions and the resulting Si-coordination is transitional between four and sixfold.

Beyond the specific instance of fused silica, there exists an extensive literature on the microstructural mechanisms that mediate plastic deformation in amorphous solids. Demkowicz and Argon \cite{PhysRevB.72.245206} observed that in amorphous silicon plastic deformation is mediated by autocatalytic {\sl avalanches} of unit inelastic shearing events. They performed a bond-angle analysis in order to correlate changes in the average bond angle to discrete relaxation events. Langer \cite{PhysRevE.57.7192, PhysRevE.64.011504} formulated a theory of {\sl shear transformation zones} (STZ) to describe viscoplastic deformation in amorphous solids. Langer's theory accounts for the formation of deformation patterns such as shear banding in metallic glasses. An alternative theory of structural rearrangement in bulk metallic solids is based on {\sl free-volume} kinetics. Chen and Goldstein \cite{ChenGoldstein:1972} observed that the flow in metallic glasses is strongly inhomogeneous at high stresses and low temperatures, and attributed the patterning to local reductions in flow strength. Spaepen \cite{Spaepen:1977} later argued that these reductions are due to the formation of free volume, and that the attendant inhomogeneous flow is controlled by the competition between the stress-driven creation and diffusional annihilation of free volume \cite{PolkTurnbull:1972}. This hypothesis was later verified experimentally by Argon \cite{Argon:1979}.

\begin{figure}[h]
	\centering
    \includegraphics[width=0.6\textwidth]{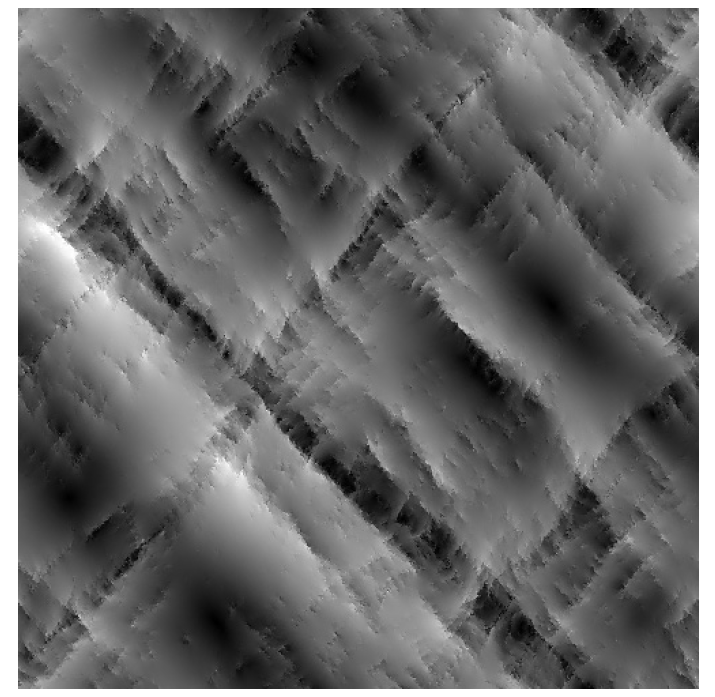}
    \caption{\small Molecular dynamics calculation of an idealized amorphous solid showing distinctive patterns in the deformation field (the darker color indicates larger non-affine displacements) \cite{0953-8984-20-24-244128}.} \label{Cho9pr}
\end{figure}

There have also been extensive molecular dynamics studies of the densification behavior and plastic deformations of amorphous silica. Pilla {\sl et al.} \cite{0953-8984-15-11-322}, Lacks \cite{PhysRevLett.80.5385}, Wu {\sl et al.} \cite{wu2012structure}, and Huang {\sl et al.}, \cite{PhysRevB.69.224203, PhysRevB.69.224204} computed pressure-density relationships over a broad range of pressures and temperatures. The attendant mechanisms of deformation entail transitions from four-fold to six-fold coordination. In particular, Wu {\sl et al.} \cite{wu2012structure} argued that the four-fold to six-fold transition is not direct but involves the formation of an intermediate five-fold coordinated structures at $\sim 12$ GPa and is only complete at $\sim 60$ GPa. Liang and co-workers \cite{PhysRevB.75.024205} noted anomalous behavior in the form of a minimum shear strength occurring at $\sim 10$ GPa and proposed a mechanism involving unquenchable 5-fold defects. Mantisi {\sl et al.} \cite{Mantisi2012} utilized an NVE ensemble along with monoclinic change in the simulation box orientation to study combined pressure-shear loading. They observed steps, or {\sl jerking}, in the shear stress {\sl vs}. shear strain response, which they attribute to either finite size effects or localized dissipative rearrangements. Several authors \cite{0953-8984-20-24-244128, PhysRevLett.103.065501} have performed molecular dynamics calculations on amorphous solids deforming under shear and found that the resulting deformation field forms distinctive patterns to accommodate permanent deformations, Fig.~\ref{Cho9pr}.

This past work strongly suggests that the plastic deformation of amorphous solids and, in particular, fused silica glass, is mediated by localized atomic-level instabilities that promote deformation patterning, Fig.~ \ref{Cho9pr}. Such fine-scale pattern formation is reminiscent of the microstructure attendant to the relaxation of non-convex energy functionals \cite{Dacorogna:1989:DMC:63481}. We argue that a critical state plasticity model \cite{Roscoe:1958, Schofield:1968} characterized by a {\sl strongly non-convex} critical-state line in pressure-shear space explains the observed patterning. In order to formulate the theory, we perform Molecular Dynamics (MD) calculations designed to mine data on the volume-pressure relation and the pressure-shear response of fused silica, Section~\ref{vlu4To}. In Section \ref{cont}, we formulate a critical state constitutive model that closely reproduces the phenomenology revealed by the MD data. The data suggest that the critical-state line in the pressure-shear plane is indeed strongly non-convex. The handling of non-convexity necessitates a fundamental extension of classical plasticity, which is based on the principle of maximum dissipation and is predicated on the assumption of convexity of the elastic domain. In Section \ref{Relax}, we consider the implications of this extension and utilize notions from the Direct Methods in the Calculus of Variations to characterize explicitly and exactly the effective, or {\sl relaxed}, behavior of fused silica at the macroscale. Remarkably, owing to the equilibrium constraint the effective macroscopic behavior of fused silica is still strongly non-convex, despite being stable with respect to microstructure formation. In particular, it defines well-posed boundary-value problems.

\section{Supporting Molecular Dynamics calculations} \label{vlu4To}

We use MD calculations for purposes of data mining, as well as to gain insight into the molecular basis of the inelasticity of glass.

{\bf NB} (Pressure sign convention): {\sl In keeping with the standard sign convention in experimental work and in MD, we take compressive pressure to be positive and tensile pressure to be negative.}

\subsection{Methodology}\label{section2}

\begin{figure}[h]
	\begin{subfigure}{0.49\textwidth}\caption{\small } \includegraphics[width=0.99\linewidth]{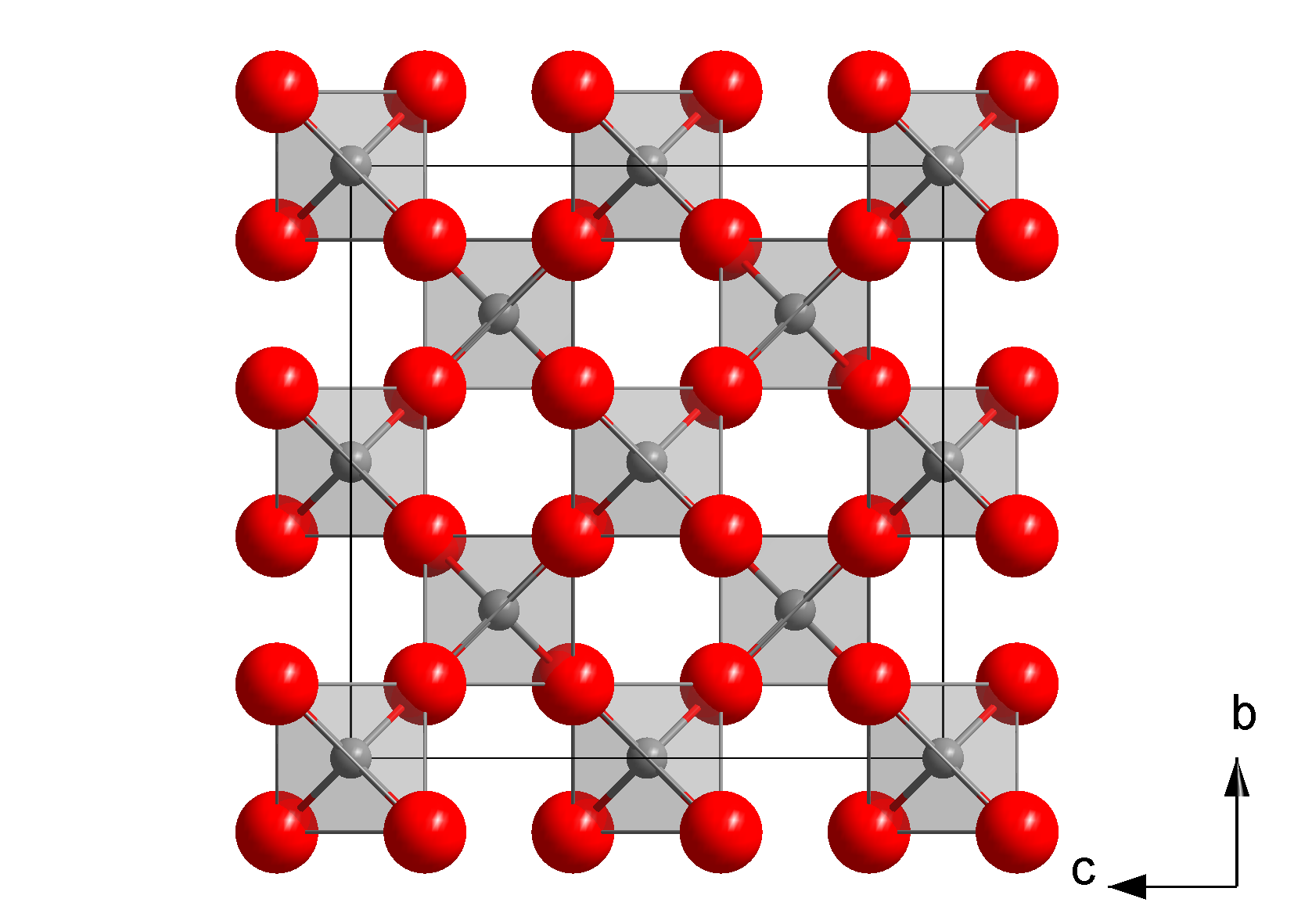}
	\end{subfigure}
	\begin{subfigure}{0.49\textwidth}\caption{\small } \includegraphics[width=0.99\linewidth]{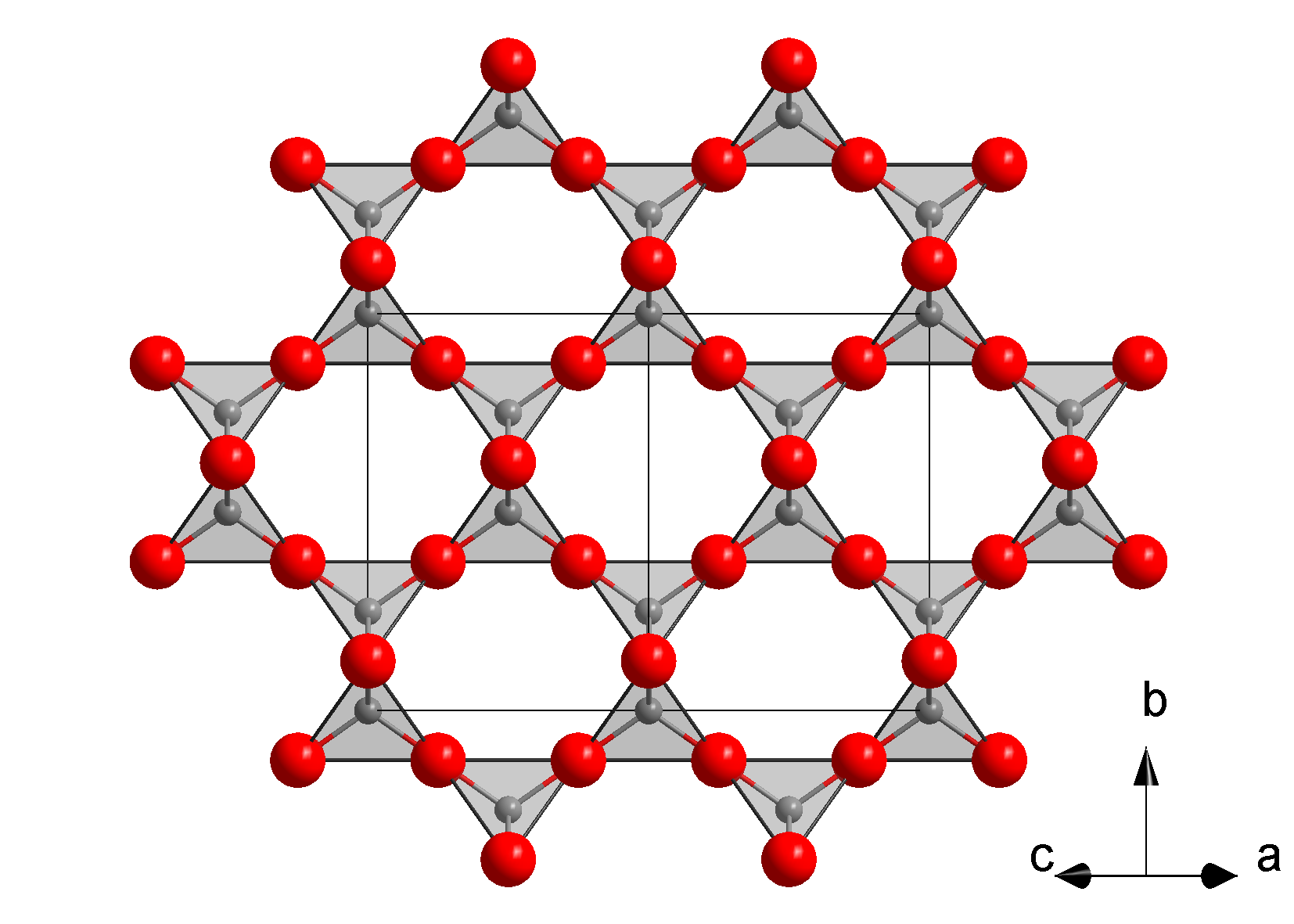}
	\end{subfigure}
	\caption{\small Two views of the crystal structure of $\beta$-cristobalite (By Solid State (Own work) [Public domain], via Wikimedia Commons). Si: red atoms; O: grey atoms.}
	\label{Cristobalite}
\end{figure}

\begin{figure}[h]
	\centering
	\includegraphics[width=0.95\textwidth]{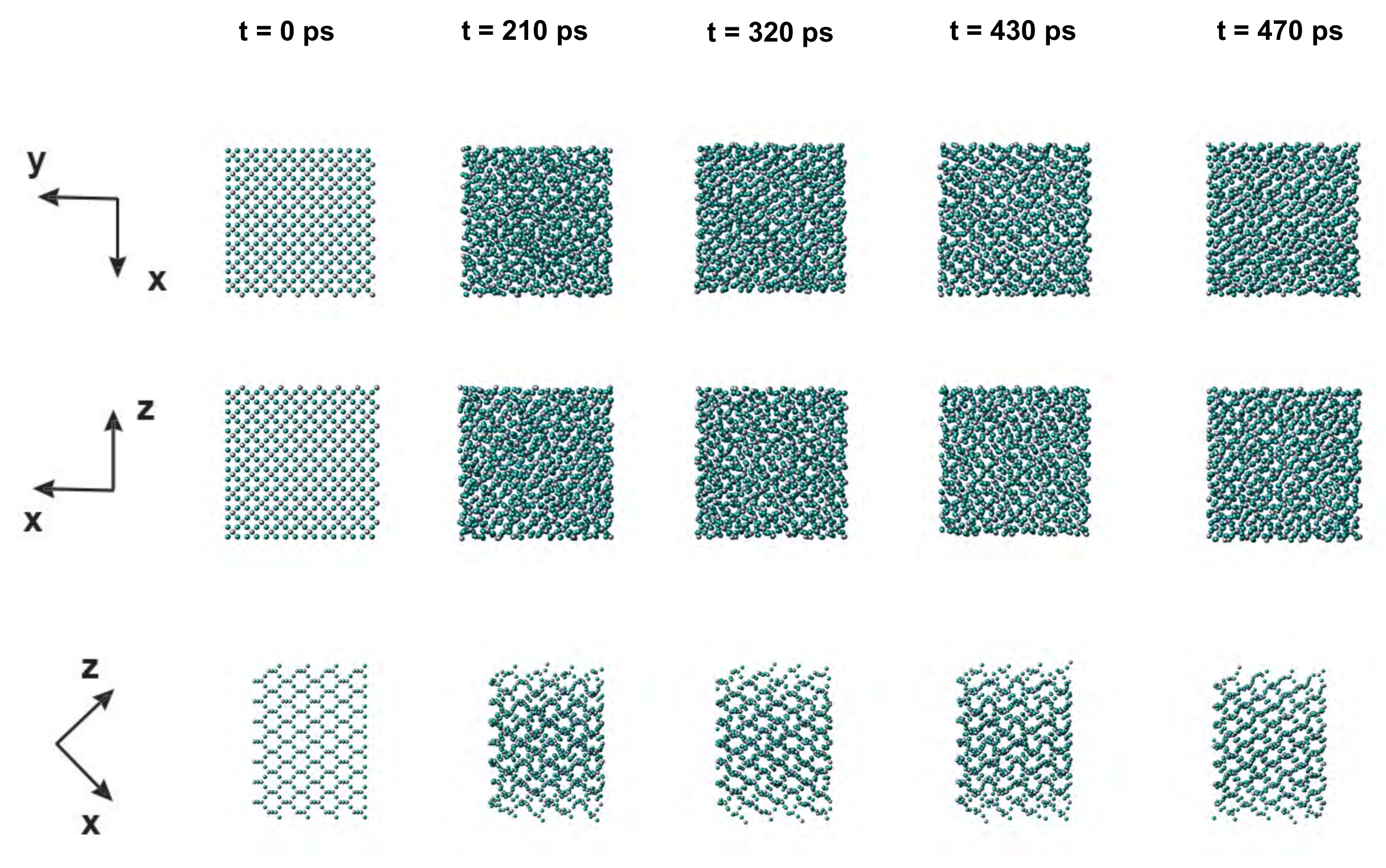}
	\caption{\small Rapid cooling of a $\beta$-cristobalite melt and generation of an amorphous structure. Sample is cooled from $\beta$-cristobalite structure at $T=5000$K to $T=300$K in $t=470$ ps.}
	\label{Quenching}
\end{figure}

All calculations are performed using Sandia National Laboratories (SNL) Large-scale Atomic/Molecular Massively Parallel Simulator (LAMMPS) \cite{PLIMPTON19951}. Calculations are carried out by explicit velocity-Verlet dynamics \cite{tuckerman1992reversible} with a time step of $0.5$ fs for a total of $10^6$ time steps up to maximum deformations of the order of 20$\%$, corresponding to strain rates of approximately $4 \times 10^{8}$ $1/\text{s}$. The representative volume element (RVE) contains $1,536$ atoms and is subjected to periodic boundary conditions. We utilize $4^3 $ primitive lattice cells of $ \beta $-cristobalite to construct RVEs $ 4\times7.16 = 28.64$ \AA~wide. We have verified that unit cells comprising $ 8^{3} $ lattice cells do not significantly alter the results of the calculations.

All calculations are performed at a temperature of $300$K. Long-range Coulombic interactions are evaluated by Ewald summation \cite{tuckerman2010statistical}. Short-range interactions are assumed to obey the modified BKS potential
\begin{equation*}
    E(r_{ij}) = A \exp(-r_{ij}/\rho) - C/r_{ij}^6 + D/r_{ij}^{12} ,
\end{equation*}
proposed by \cite{Malavasi2006285}, where $ r_{ij} $ represents the interatomic distance. This potential modifies the BKS potential proposed in \cite{PhysRevLett.64.1955} by the insertion of an additional repulsive short-range interaction term in order to increase calculation stability. The additional repulsive term additionally prevents the unphysical divergence of the potential at small interatomic distances. The parameters $ A $, $ C $, $ D $, and $ \rho $ used in calculations may be found in Table 4 of \cite{Malavasi2006285}.

In order to obtain an initial amorphous state of SiO$_{2}$, we utilize the melt quench procedure No.~2 of Malavasi \cite{Malavasi2006285}. This quench procedure is performed on an NVT ensemble (cf., e.~g., \cite{tuckerman2010statistical}) and consists of cooling a $\beta$-cristobalite melt, Fig.~\ref{Cristobalite}, from $5000$K to $300$K over $470$ fs with a time step of $2$ fs, Fig.~\ref{Quenching}.

\subsection{Volumetric behavior}

\begin{figure}[h]
	\centering
	\includegraphics[width=0.9\textwidth]{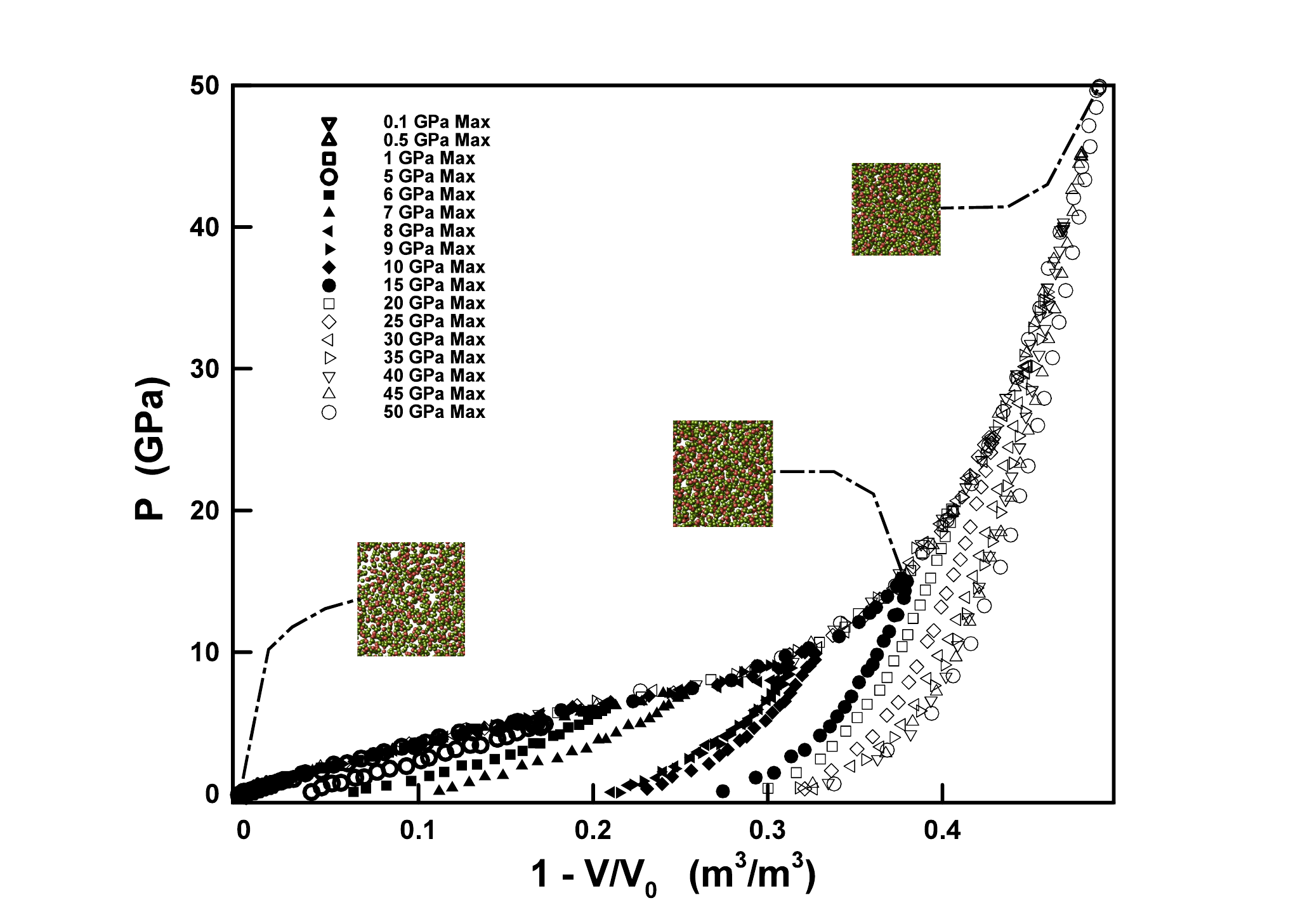}
    \caption{\small Pressure-compression response showing densification transition at $\sim 8$ GPa and unloading from several pressures showing permanent densification upon full unloading.}
	\label{pressurecompression1}
\end{figure}
\vspace{5 mm}

We begin by querying the behavior of amorphous silica under compressive volumetric loading and unloading.  Fig.~\ref{pressurecompression1} shows the computed dependence of pressure on volume, including unloading from a range of maximum pressures. At low maximum pressures, the material unloads ostensibly elastically and returns to its initial undeformed configuration upon unloading. By contrast, at pressures above $\sim 8$ GPa the material undergoes a distinctive permanent densification transition and the unloading curve exhibits permanent volumetric deformation.

Past studies \cite{0953-8984-15-11-322, PhysRevB.69.224203, PhysRevB.69.224204} have reported similar pressure-density relationships, but calculations to date have been limited to significantly smaller sample sizes and monotonic loading. We note that without unloading it is not possible to ascertain whether the material response is nonlinear elastic, and therefore governed by a simple equation of state, or elastic-plastic. The results collected in Fig.~\ref{pressurecompression1} clearly reveal that the latter is indeed the case and that the volumetric response of glass exhibits inelasticity in the form of loading-unloading irreversibility, path-dependency and hysteresis at sufficiently high pressures.

\begin{figure}[h]
\centering
    \includegraphics[width=0.85\linewidth]{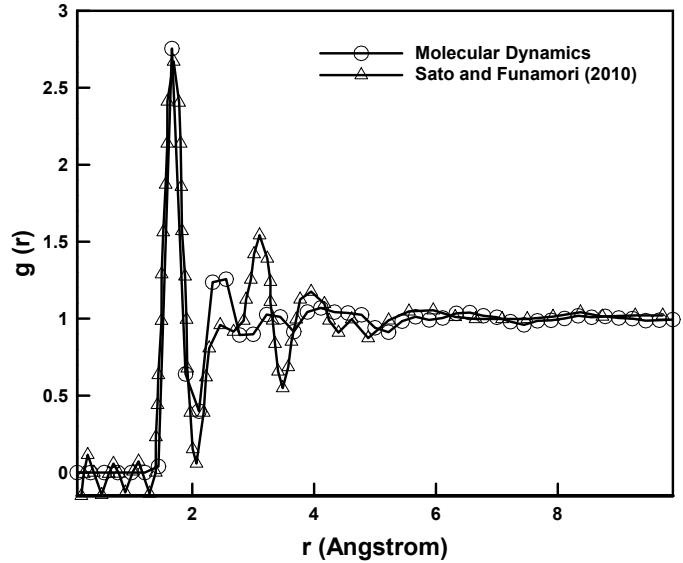}
    \caption{\small Computed and experimentally measured \cite{Sato} radial distribution function at pressure $p = 50$ GPa.}\label{radialDistributionFunction}
\end{figure}

{\sl Radial distribution functions} are commonly used as a validation and interpretation metric in MD simulations, e.~g., Jin {\sl et al.} \cite{jin1993structural, PhysRevB.50.118}.  Fig.~ \ref{radialDistributionFunction} shows the computed radial distribution function at $50$ GPa. By way of comparison, Fig.~ \ref{radialDistributionFunction} also shows corresponding experimental measurements performed by Sato and Funamori \cite{Sato}. As can be seen from the figure, the MD calculations accurately capture the location and amplitude of the first peak in the radial distribution, which determines the radius of the first shell of atoms, and, to a fair degree of approximation, the location and amplitude of the second peak. The tails of the computed and measured radial distributions differ in fine detail but exhibit a similar rate of decay.

\begin{figure}[h]
\begin{center}
	\begin{subfigure}{0.40\textwidth}\caption{\small }
        \includegraphics[width=0.99\linewidth]{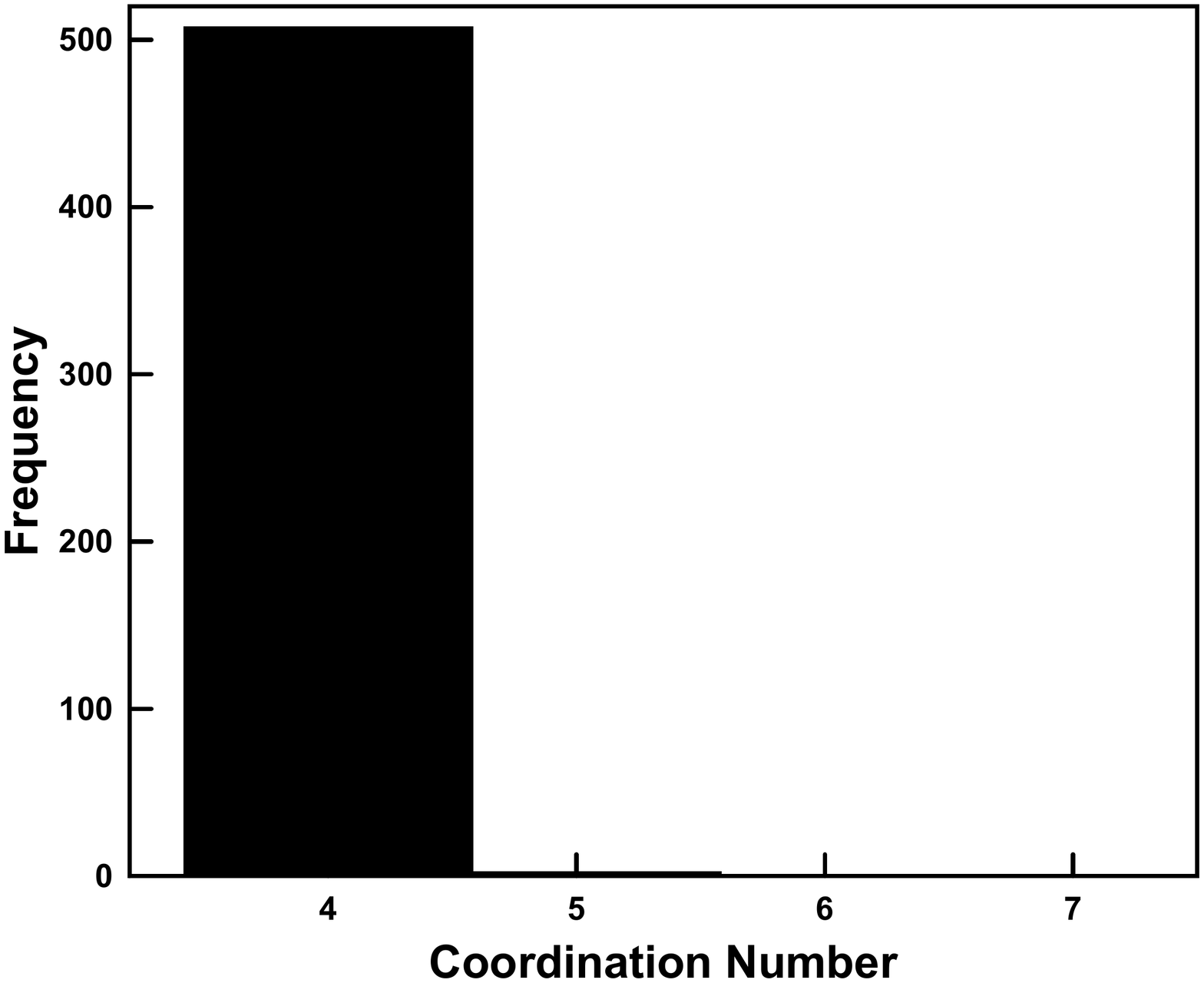}
	\end{subfigure}
	\begin{subfigure}{0.40\textwidth}\caption{\small }
        \includegraphics[width=0.99\linewidth]{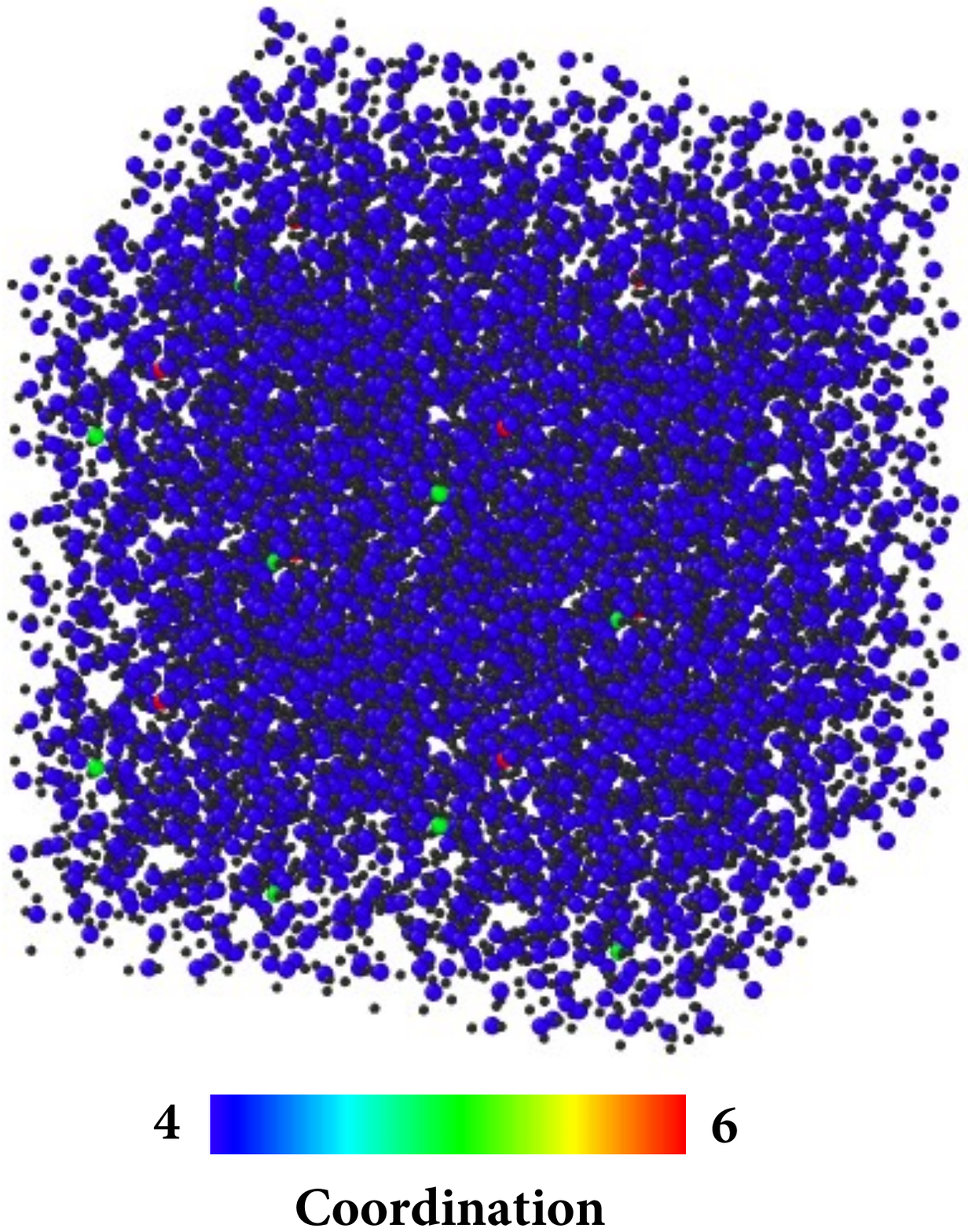}
	\end{subfigure}
	\begin{subfigure}{0.40\textwidth}\caption{\small }
        \includegraphics[width=0.99\linewidth]{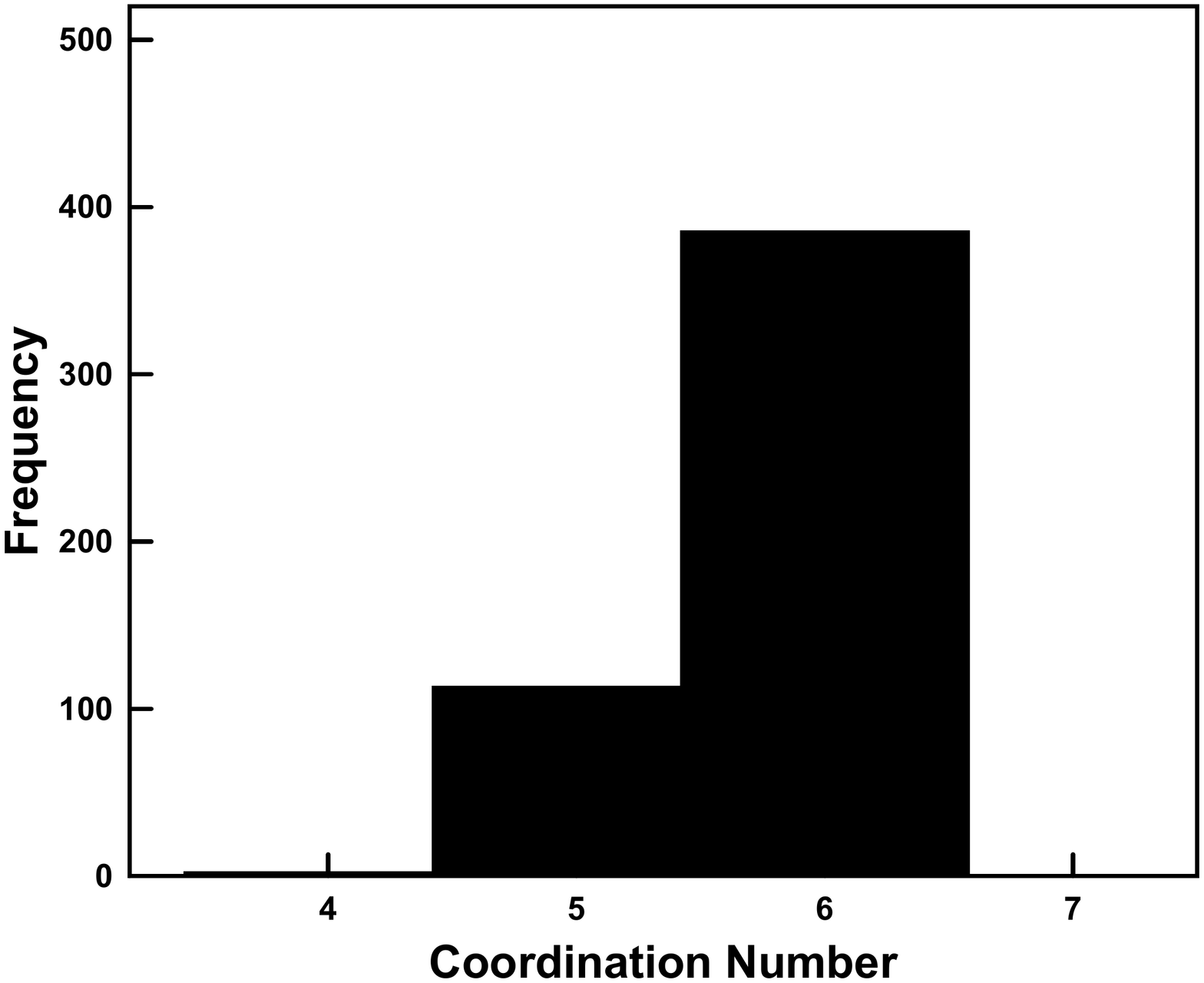}
	\end{subfigure}
	\begin{subfigure}{0.40\textwidth}\caption{\small }
        \includegraphics[width=0.99\linewidth]{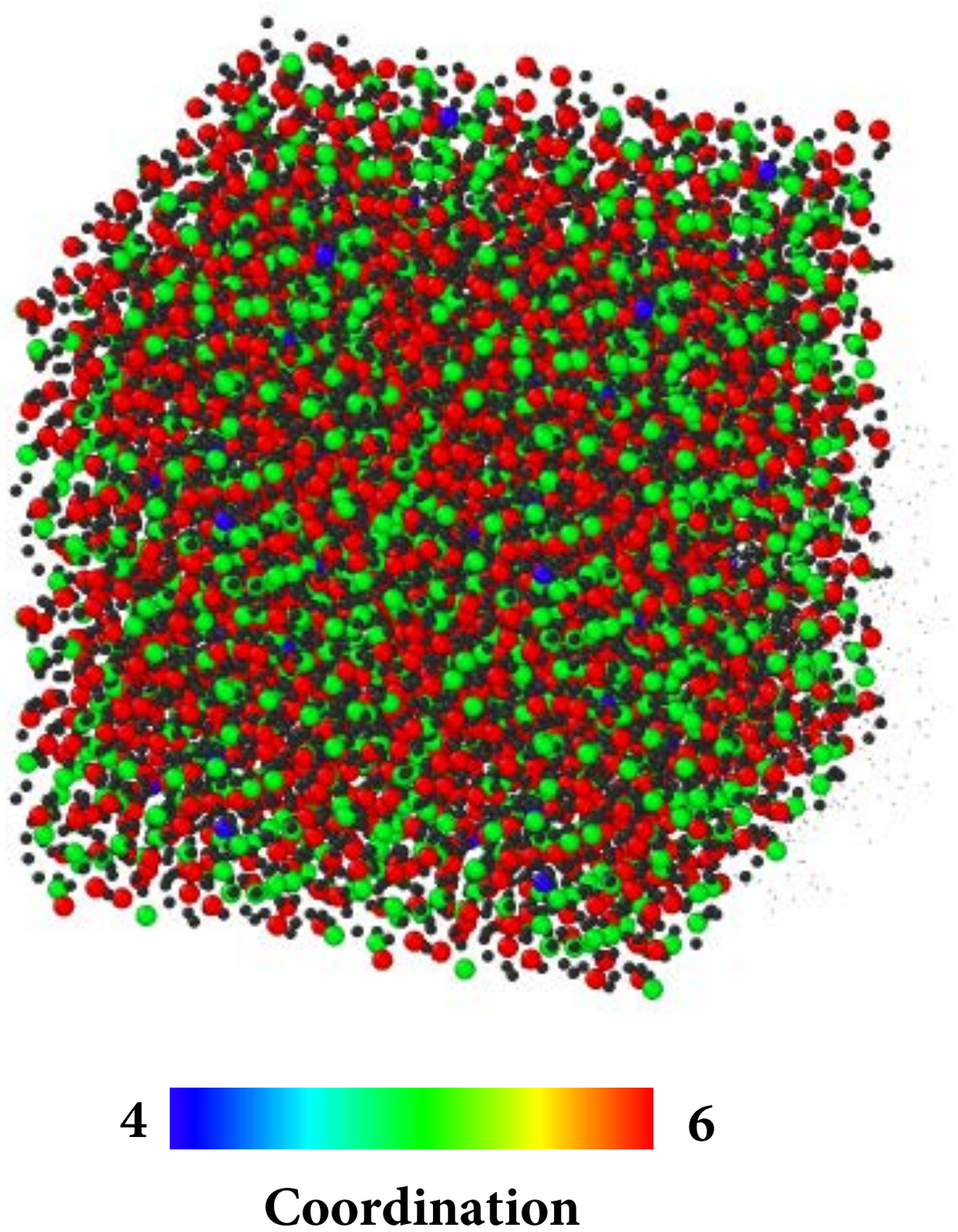}
	\end{subfigure}
	\begin{subfigure}{0.40\textwidth}\caption{\small }
        \includegraphics[width=0.99\linewidth]{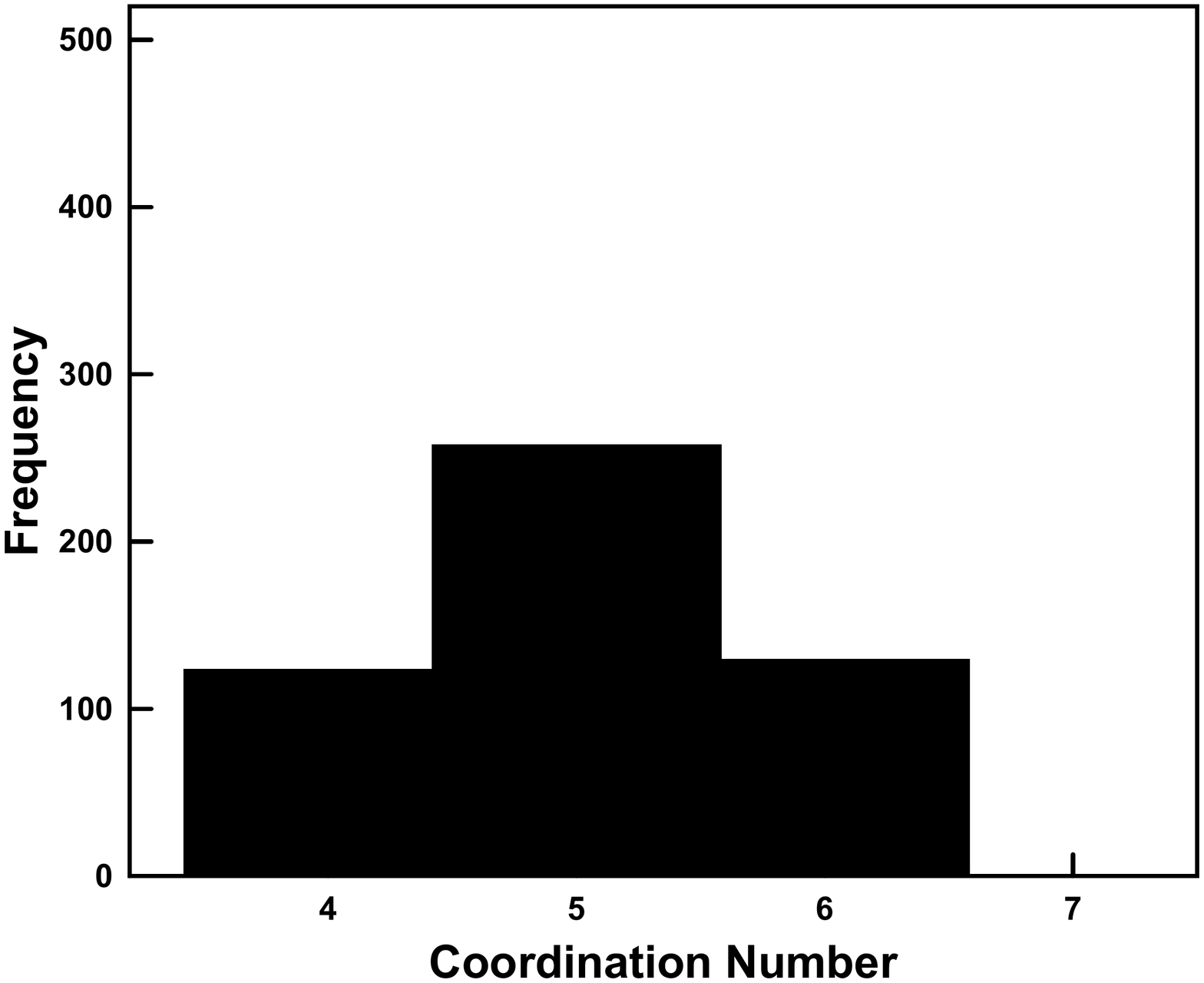}
	\end{subfigure}
	\begin{subfigure}{0.40\textwidth}\caption{\small }
        \includegraphics[width=0.99\linewidth]{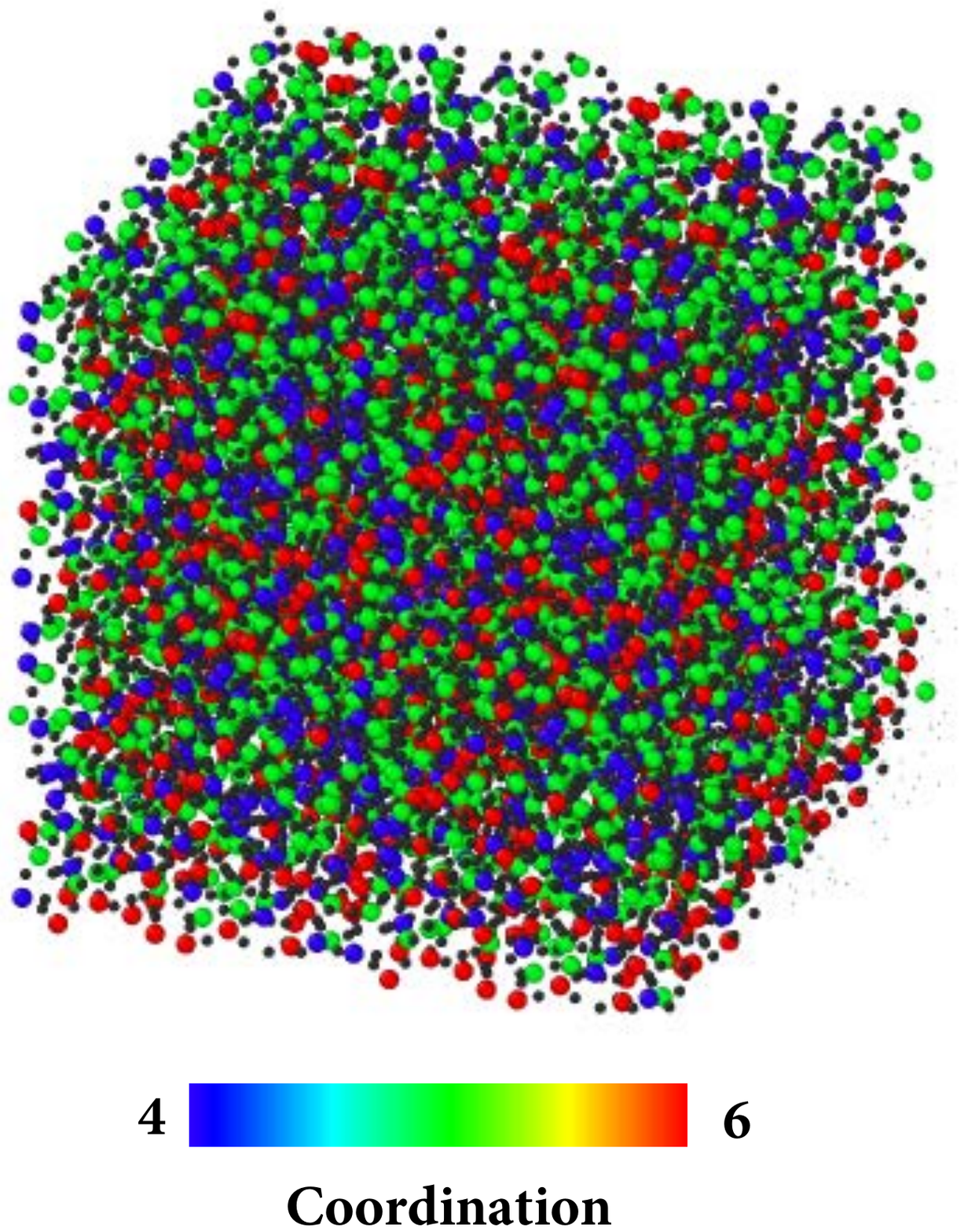}
	\end{subfigure}
    \caption{\small Evolution of the distribution of coordination numbers of the atoms in a sample during volumetric-compression loading and unloading up to a pressure of $50$ GPa. Si atom coordination numbers are illustrated by the color bar and the oxygen atoms are represented as black spheres. (a) and (b) Initial state; (c) and (d) Peak pressure. (e) and (f) Unloaded state.} \label{Coordination}
\end{center}
\end{figure}

In order to elucidate the atomic-level mechanisms underlying permanent volumetric deformation, we examine the evolution of the {\sl coordination number} (cf., e.~g., \cite{PhysRevB.50.118, van2005silica})
\begin{equation}\label{cn}
    CN = \int_0^{r_m} \rho g(r) \, 4\pi r^2 \, dr ,
\end{equation}
where $\rho$ is the particle density, or number of atoms per unit volume, $g(r)$ is the radial distribution function and $r_m$ is the location of the first minimum of $g(r)$. The coordination number measures the number of nearest-neighbors of an atom. A simple way to approximate equation \eqref{cn} given a set of atomic positions, is to perform a Voronoi tessellation of the atoms and then count the number of faces of individual Voronoi cells. In order to mitigate the effect of noise, a face is not counted if its area is below $1.3$ \AA${}^{2}$, if it has more than $10$ edges, or if one of its edges is shorter that $0.5$ \AA. Fig.~\ref{Coordination} shows the evolution of the distribution of coordination numbers in a sample during compressive volumetric loading and unloading up to a pressure of $50$ GPa. Initially, the entire sample consists of $4$-fold coordinated atoms, Figs.~\ref{Coordination}a and \ref{Coordination}b. At peak pressure, the coordination of most atoms changes from $4$-fold to $6$-fold, but a significant fraction of atoms exhibits an intermediate coordination. Remarkably, upon unloading, only a small fraction of atoms recovers a $4$-fold coordination, with the second largest fraction retaining $6$-fold coordination and the majority of the sample remaining in an intermediate $5$-fold coordination. These results evince the irreversible nature of the structural transitions attendant to permanent densification of glass, in agreement with experimental observations \cite{PhysRevB.82.184102, Sato, PhysRevB.84.144103, 0953-8984-20-48-485221, 2004PhRvL..93a5501I, Luo2004}. The prevalence of transitional structures with a preponderance of $5$-fold atoms upon unloading is also in agreement with the calculations of Wu {\sl et al.} \cite{wu2012structure} and the experimental observations of Luo {\sl et al.} \cite{Luo2004}.

\subsection{Pressure-shear coupling}
\label{FRoe1o}

Using the same initial amorphous configuration of atoms, we now subject the RVE to pressure followed by monotonic shear deformation. To impart the shear deformation, affine boundary conditions are applied to the boundary of the RVE while simultaneously controlling the pressure by means of a barostat. We generate shear stress-strain curves over a range of pressure and we average the curves over a sample of initial conditions.

\begin{figure}[h]
		\begin{subfigure}{0.49\textwidth}\caption{\small }
            \includegraphics[width=0.99\linewidth]{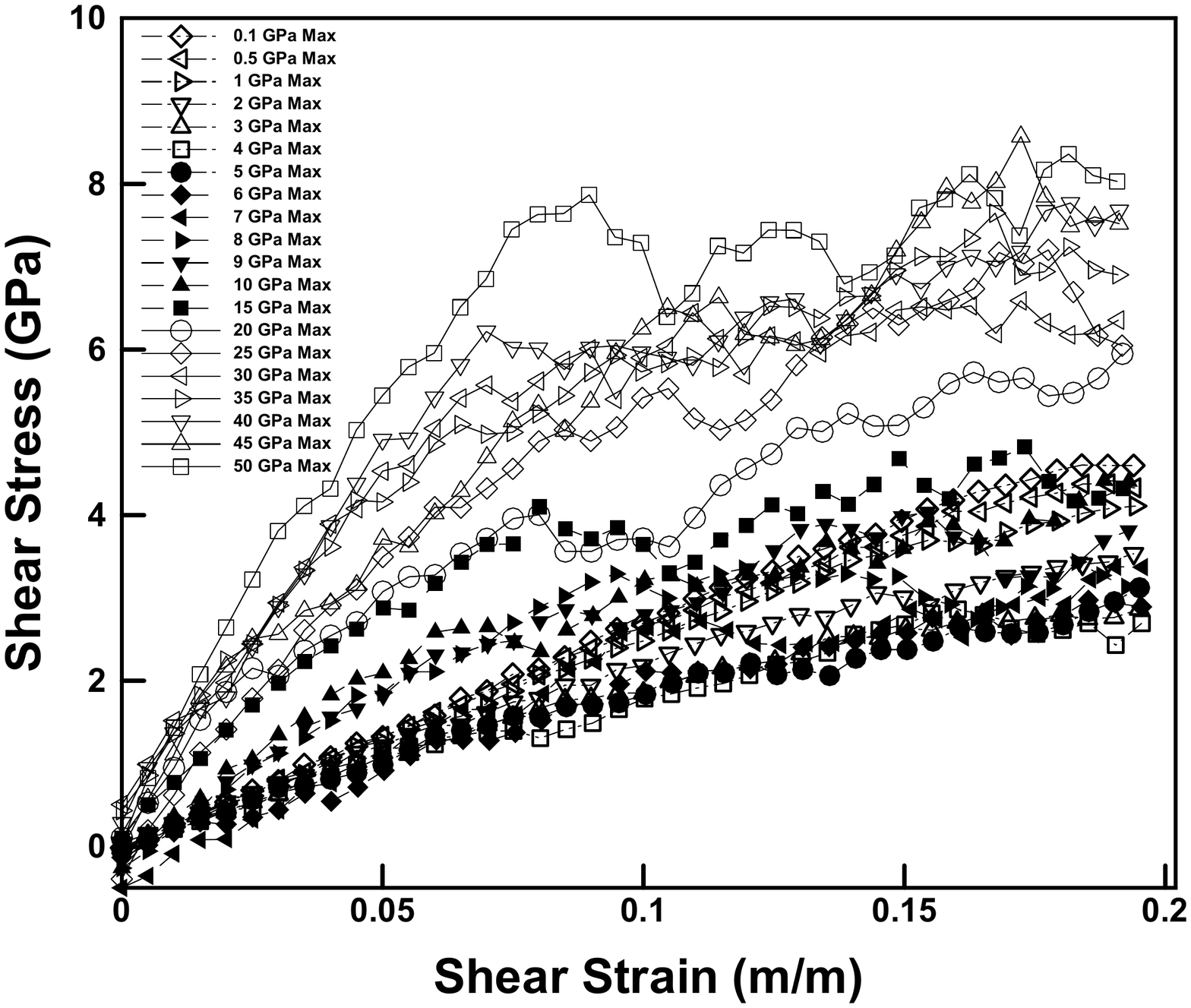}
		\end{subfigure}
		\begin{subfigure}{0.49\textwidth}\caption{\small }
            \includegraphics[width=0.99\linewidth]{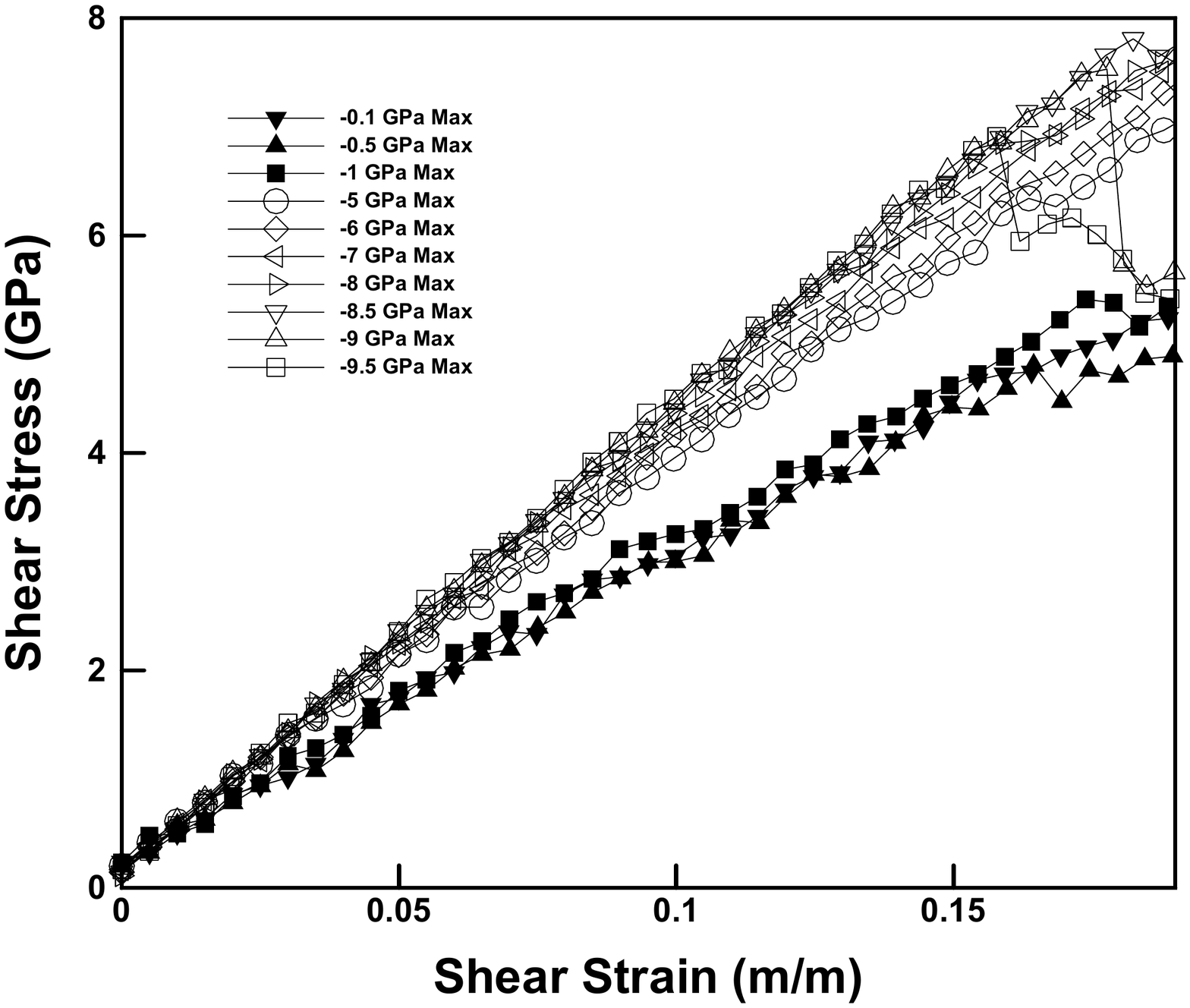}
		\end{subfigure}
	\centering
	\caption{\small a) Shear stress {\sl vs}.~shear strain under compressive pressure. b) Shear stress {\sl vs}.~shear strain under tensile pressure.}
	\label{shearBehaviorWithPressure}
\end{figure}

\begin{figure}[h]
		\begin{subfigure}{0.49\textwidth}\caption{\small }
            \includegraphics[width=0.99\linewidth]{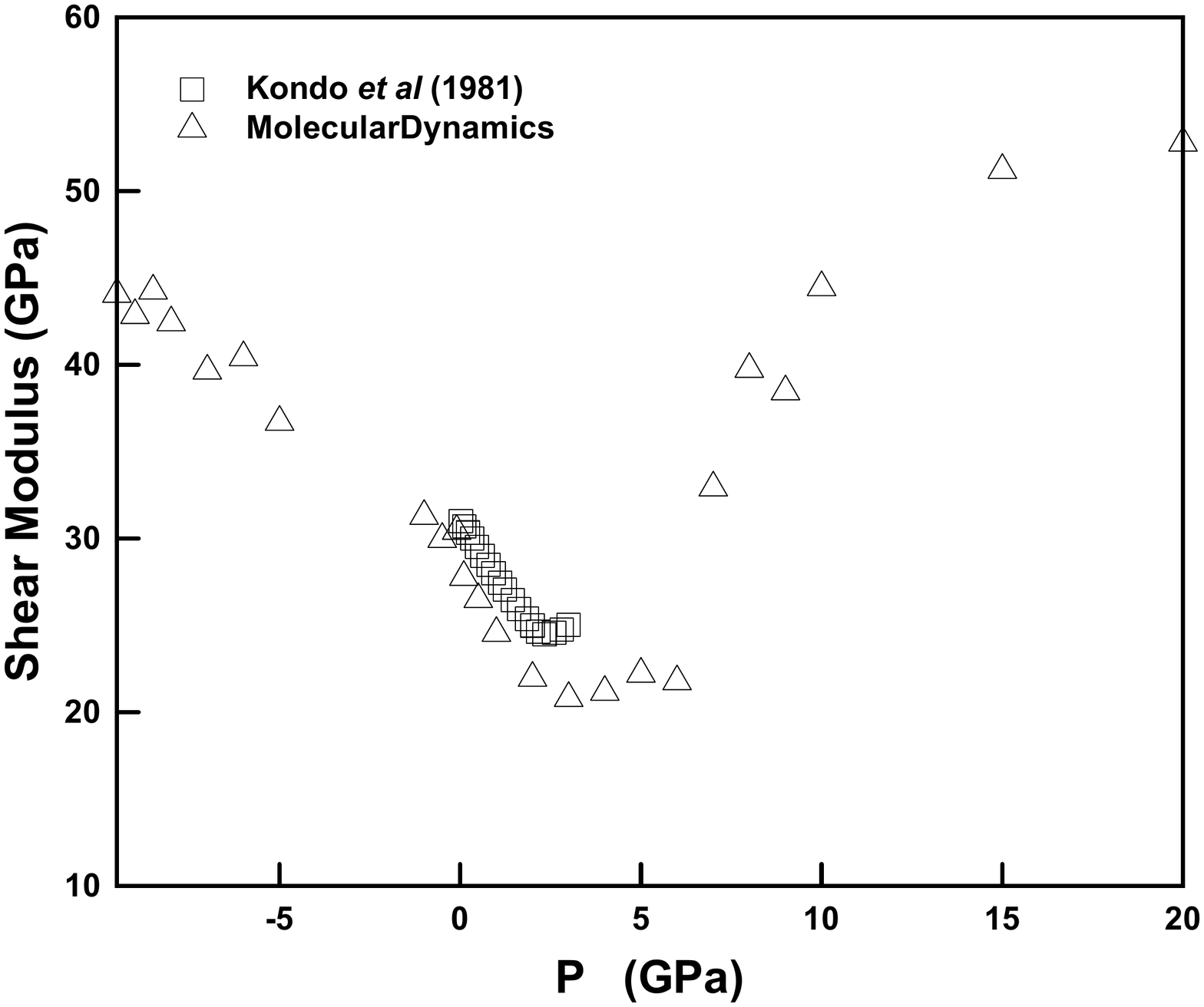}
		\end{subfigure}
		\begin{subfigure}{0.49\textwidth}\caption{\small }
            \includegraphics[width=0.99\linewidth]{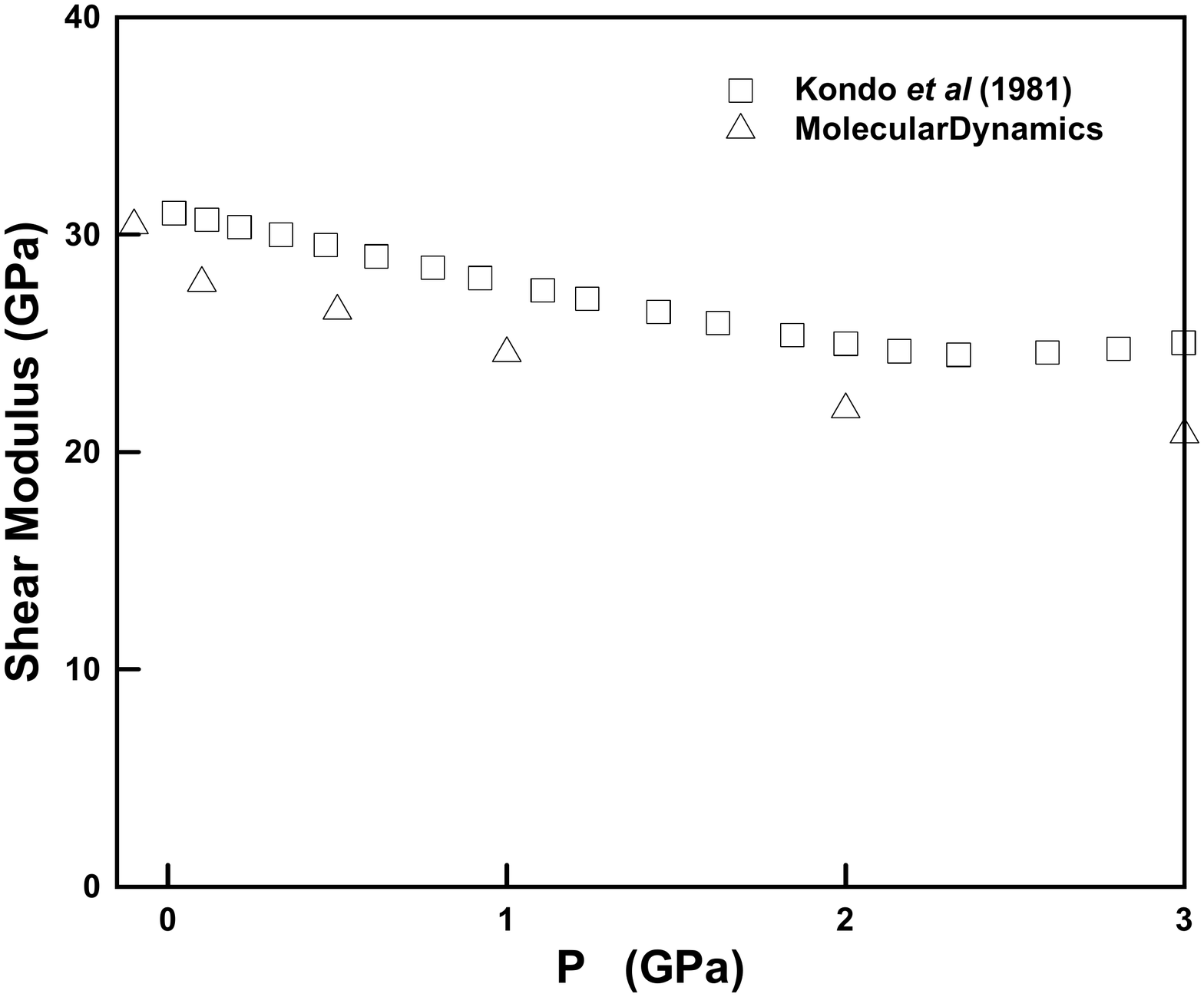}
		\end{subfigure}
	\centering
	\caption{\small Computed and experimentally measured \cite{doi:10.1063/1.329012} dependence of the shear modulus on pressure. a) Overall view showing initial anomalous dependence. b) Detail of the pressure range of 1-3 GPa.}
	\label{shearModulusVersusPressureKondo}
\end{figure}

The resulting average shear stress-strain curves are shown in ~Fig.~ \ref{shearBehaviorWithPressure}. The shear stress-strain curves exhibit an initial pressure-dependent elastic stage followed by yielding. The computed dependence of the shear modulus on pressure is shown in Fig.~\ref{shearModulusVersusPressureKondo}, which also includes measurements by Kondo {\sl et al.} \cite{doi:10.1063/1.329012} by way of comparison. As may be seen from the figure, the MD results capture the anomalous initial decrease of the shear modulus with pressure, cf.~\cite{doi:10.1063/1.55558}. Furthermore, the MD results closely match the experimental measurements, which provides a measure of model validation.

\begin{figure}[h]
	\centering
	\includegraphics[width=0.9 \textwidth]{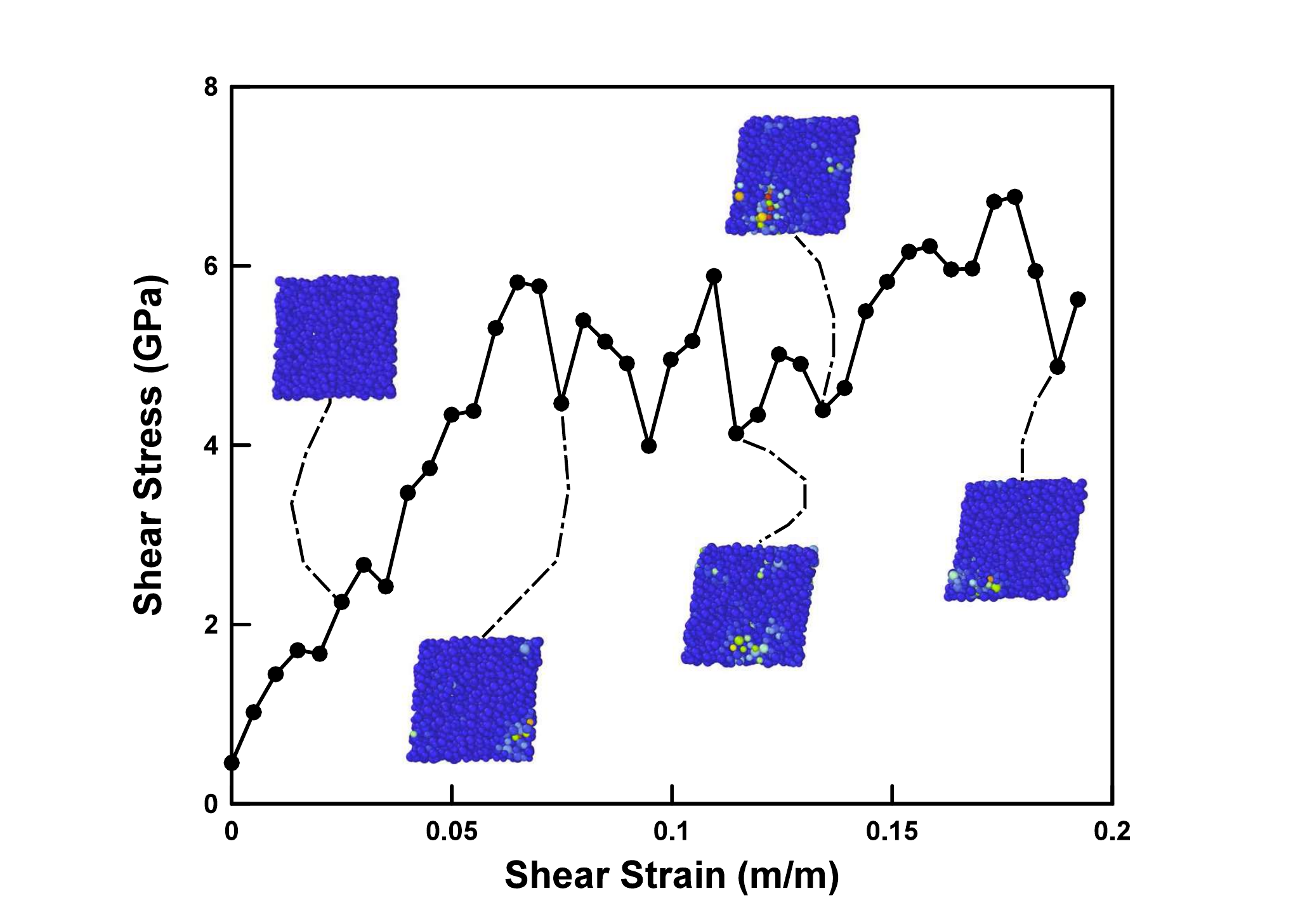}
    \caption{\small Shear stress {\sl vs.} shear strain curve and shear transitions at serrations. Blue indicates affine deformation whereas yellow and red indicate medium and large non-affine deformations, respectively. }
	\label{shearDeformations}
\end{figure}

A salient feature of the shear stress-strain curves is the serrated nature of the yield plateau, also known as {\sl jerky flow}, Fig.~\ref{shearBehaviorWithPressure}. These serrations have been associated with localized bursts of atomic movements, or {\sl avalanches} \cite{PhysRevB.72.245206}. In order to detect and quantify these avalanches, Falk and Langer \cite{PhysRevE.57.7192} proposed the parameter
\begin{equation}\label{Wie5ia}
    D(i)
    \equiv
    \min_{\mbs{\beta}\in\mathbb{R}^{3\times 3}}
    \left(
    \sum_{j}
    \Big|
    (\mbs{u}_{j} - \mbs{u}_{i})
    -
    \mbs{\beta}
    (\mbs{r}_{j} - \mbs{r}_{i})
    \Big|^{2}
    \right)^{1/2} ,
\end{equation}
which represents the deviation of the incremental displacements $\mbs{u}_j$ of the atoms in a neighborhood of a reference atom $i$ from an incremental affine deformation. Spikes in the distribution of $D(i)$ may therefore be identified with the occurrence of avalanches around atom $i$. Fig.~ \ref{shearDeformations} shows the distribution of $D(i)$ at points of a shear stress-strain curve when such avalanches occur. In this case, no averaging with respect to initial conditions is performed in order to preserve fluctuations. As may be seen from the figure, the occurrence of avalanches correlates closely with drops in the stress-strain curve, which identifies avalanches as the agents of plastic deformation and the mechanism underlying the observed jerky plastic flow.

\subsection{Volume evolution and critical state behavior}
\label{viUri7}

A fundamental characteristic of the pressure-shear response of glass, especially as regards the categorization of its plastic response, concerns the evolution of volume during shearing deformation. In order to ascertain this behavior, we deform samples volumetrically up to a maximum pressure $ p_{\text{max}} $, or {\sl preconsolidation pressure}, and subsequently unload to a lower pressure $ p \leq p_{\text{max}}$, or {\sl confining pressure}. The samples are then deformed in shear at constant confining pressure $ p $.

\begin{figure}[h]
	\begin{subfigure}{0.49\textwidth}\caption{\small }
		\includegraphics[width=0.99\linewidth]{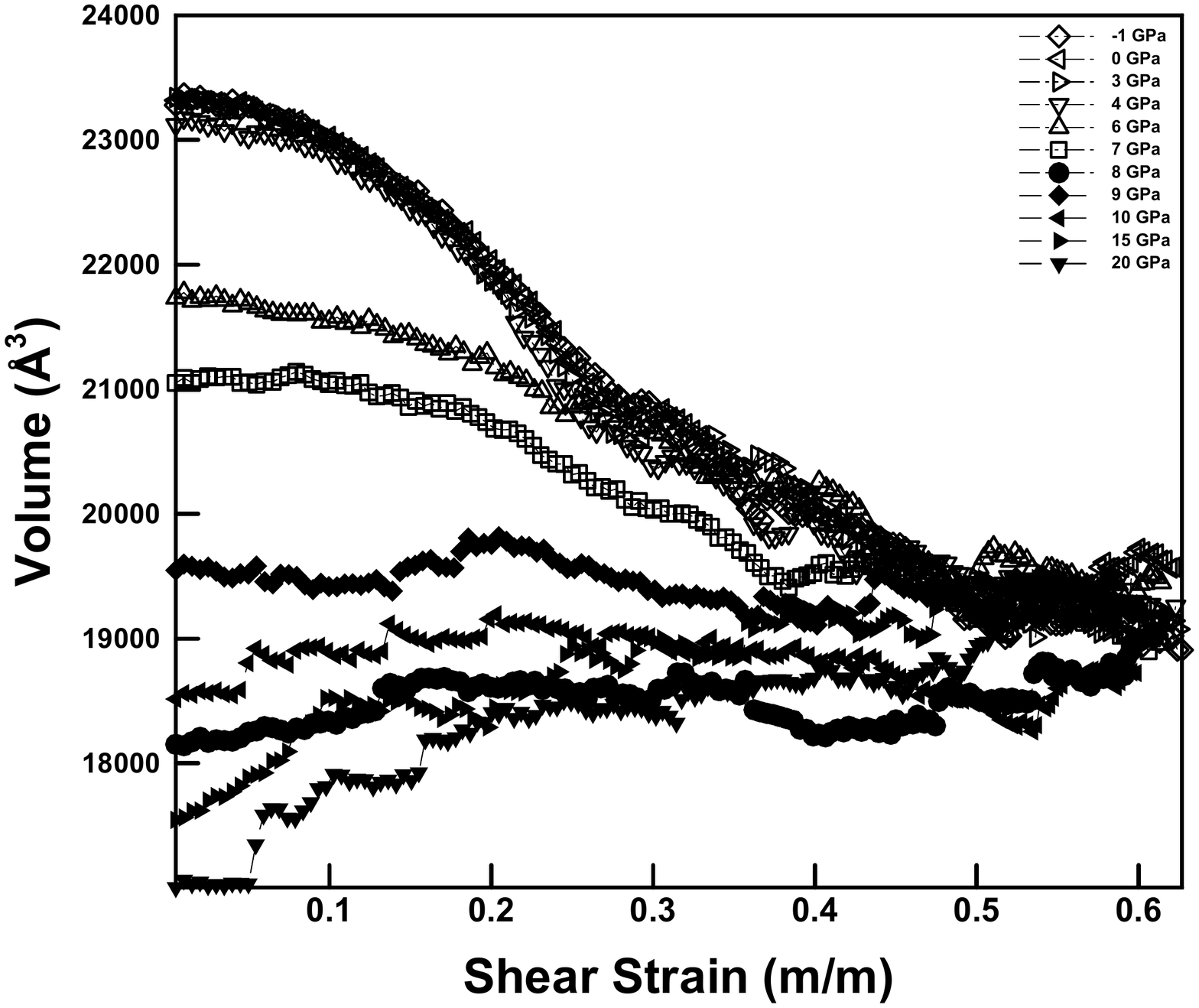}
	\end{subfigure}
	\begin{subfigure}{0.49\textwidth}\caption{\small }
		\includegraphics[width=0.99\linewidth]{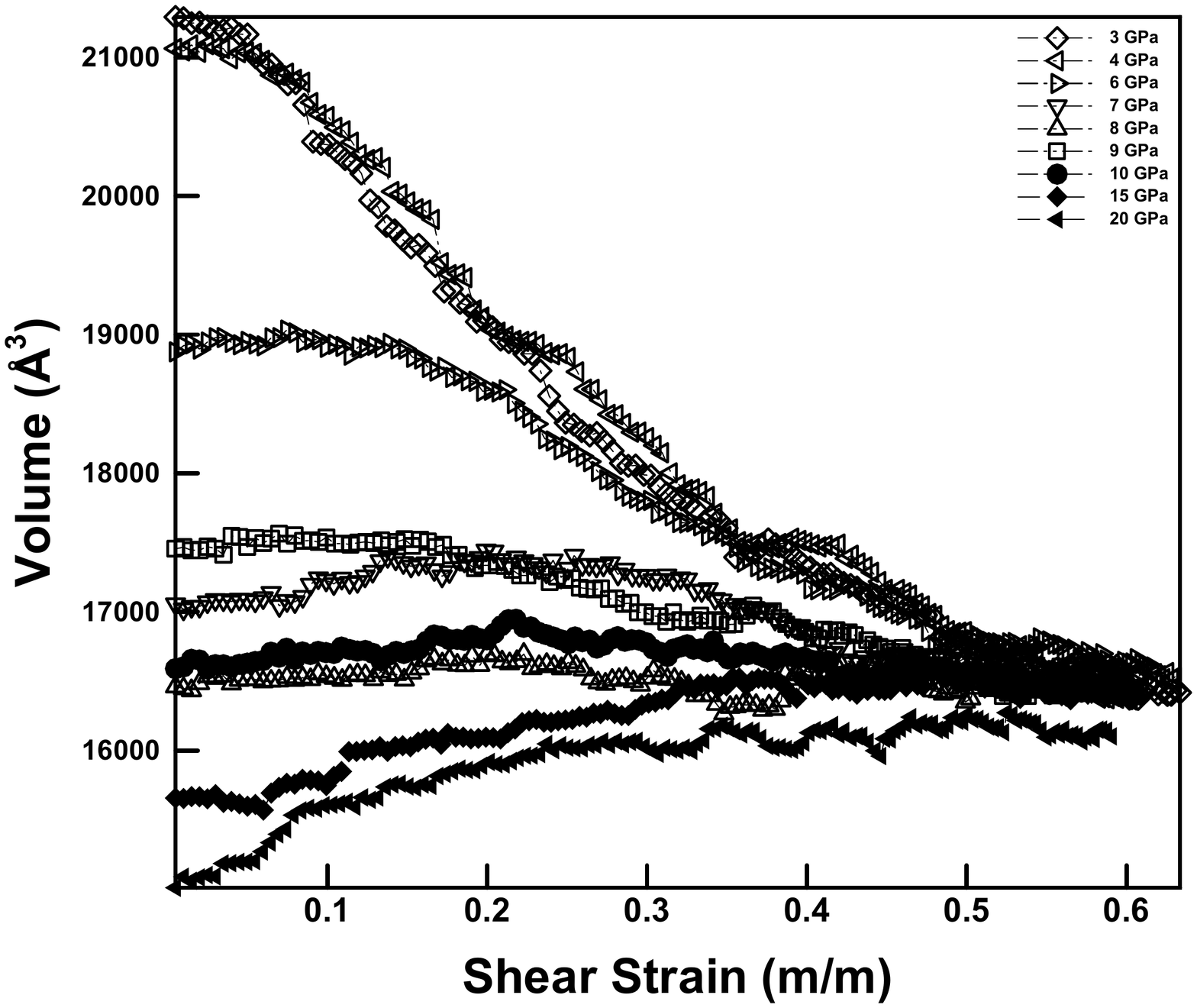}
	\end{subfigure}
	\begin{subfigure}{0.49\textwidth}\caption{\small }
		\includegraphics[width=0.99\linewidth]{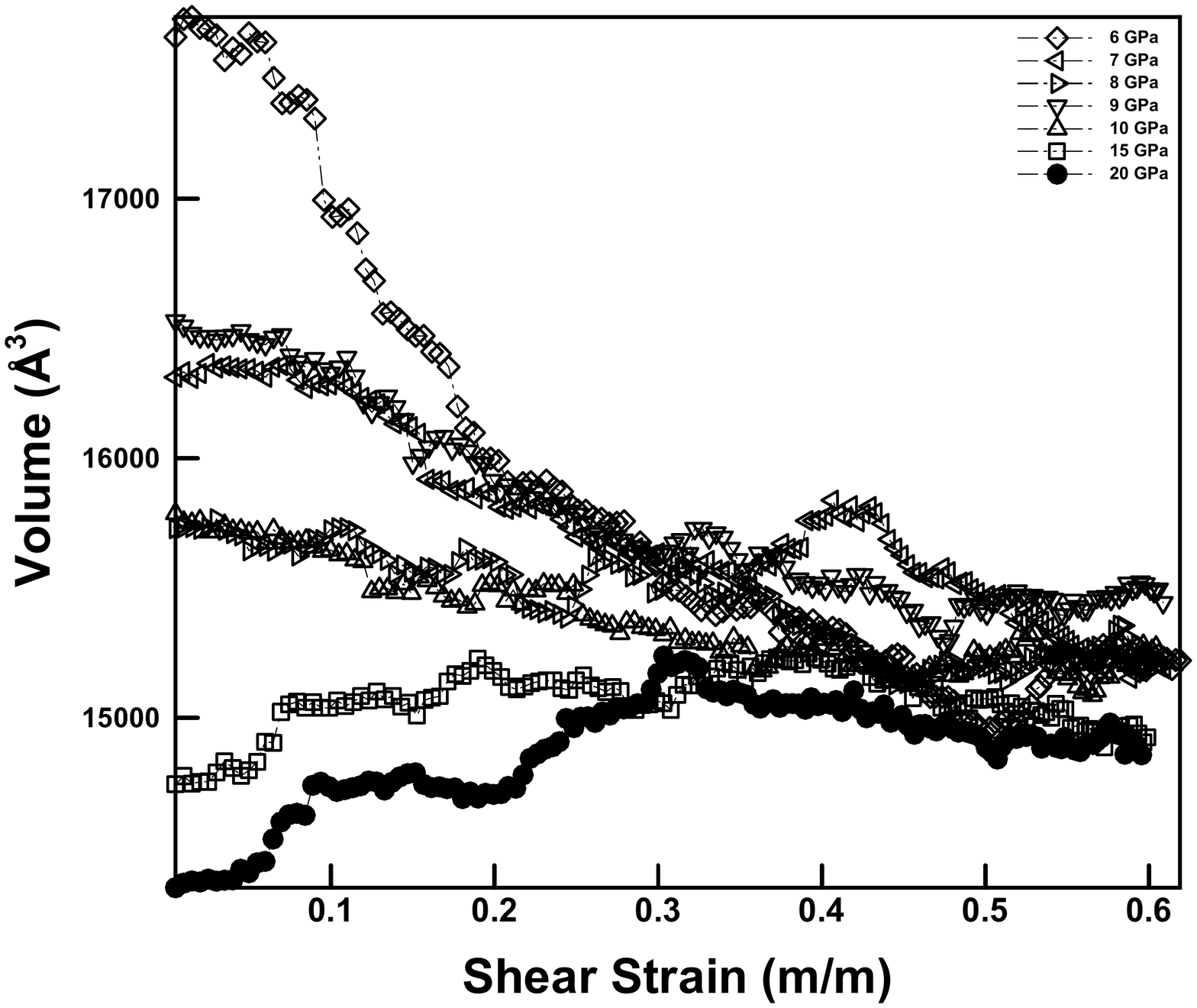}
	\end{subfigure}
	\begin{subfigure}{0.49\textwidth}\caption{\small }
		\includegraphics[width=0.99\linewidth]{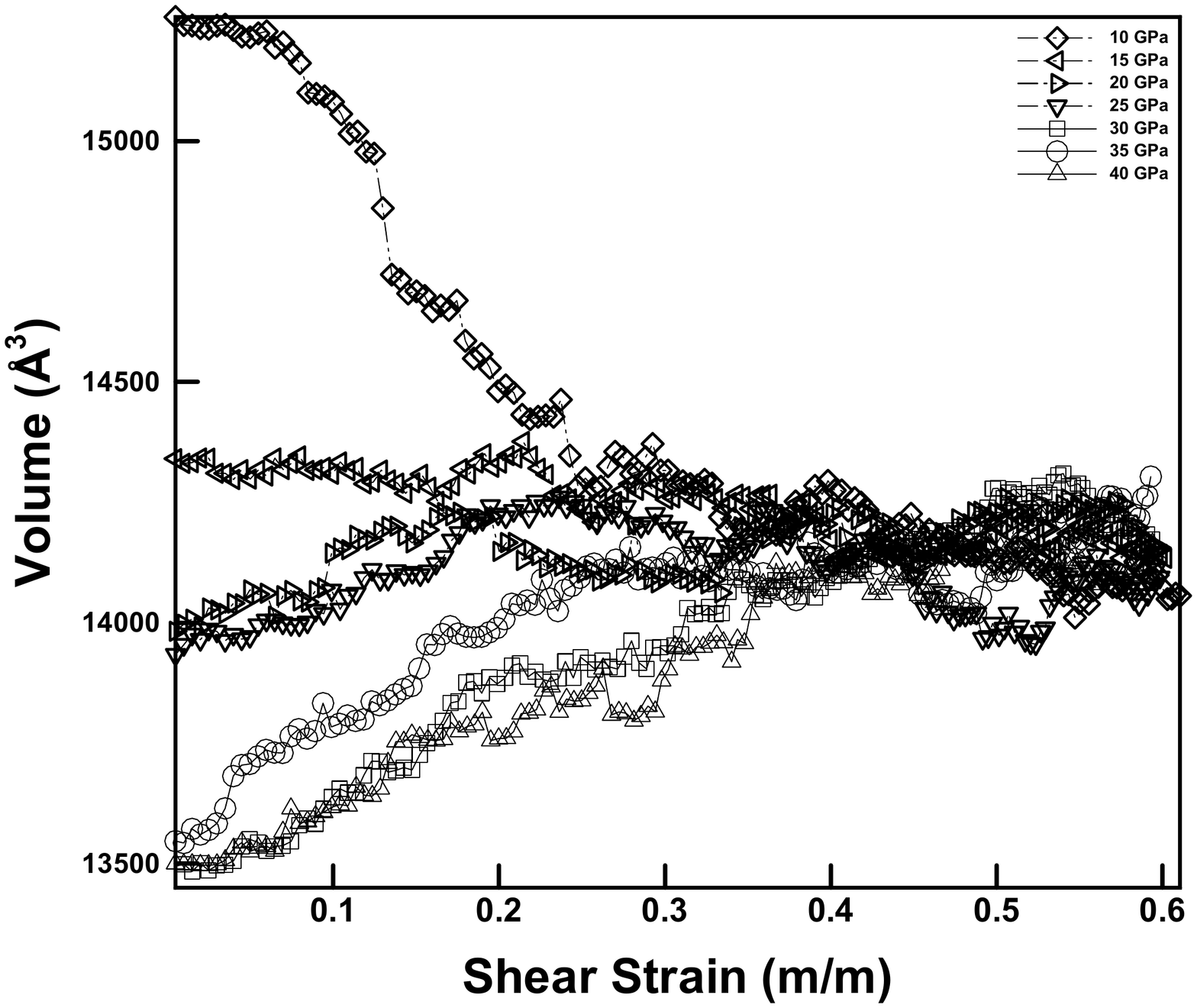}
	\end{subfigure}
	\centering
	\caption{\small Evolution of volume during pressure-shear response for different values of preconsolidation pressure $p_{\rm max}$ (shown inset in the figures) and confining pressure $p$. a) $p = -1$ GPa. b) $p = 3$ GPa. c) $p = 6$ GPa. d) $p = 9$ GPa.}
	\label{volumeandshear}
\end{figure}

\begin{figure}[h]
	\begin{subfigure}{0.49\textwidth}\caption{\small }
		\includegraphics[width=0.99\linewidth]{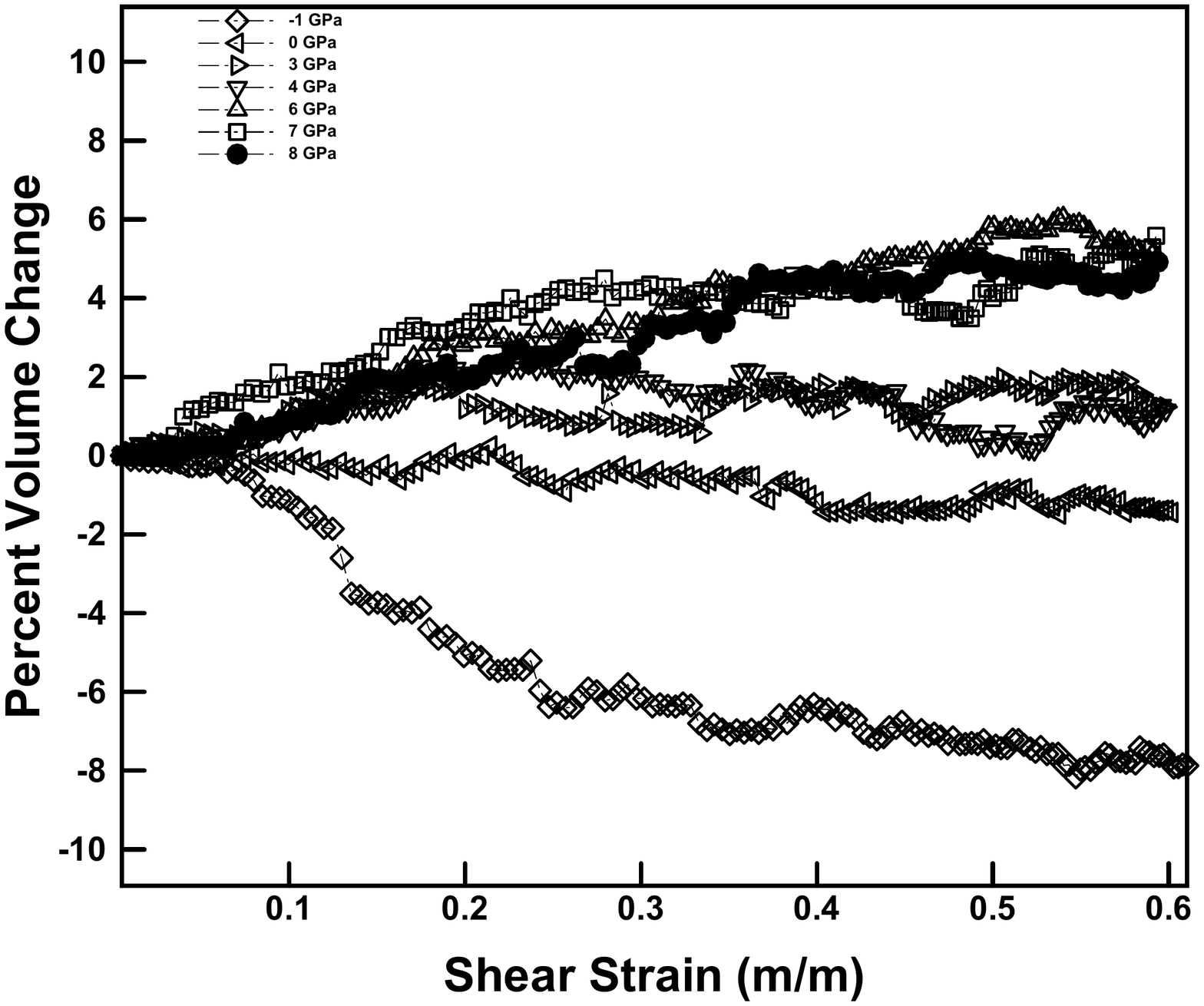}
	\end{subfigure}
	\begin{subfigure}{0.49\textwidth}\caption{\small }
		\includegraphics[width=0.99\linewidth]{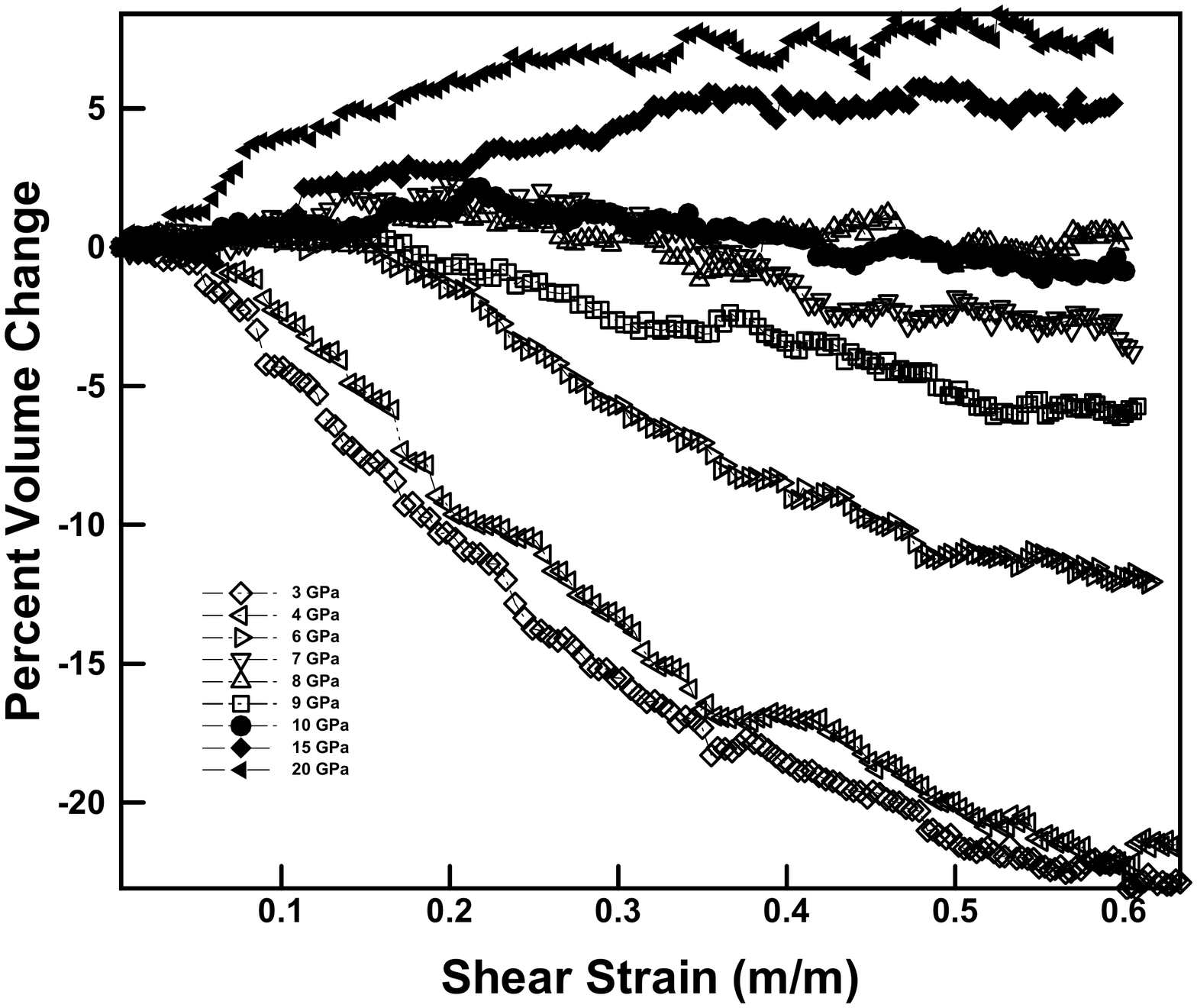}
	\end{subfigure}
	\begin{subfigure}{0.49\textwidth}\caption{\small }
			\includegraphics[width=0.99\linewidth]{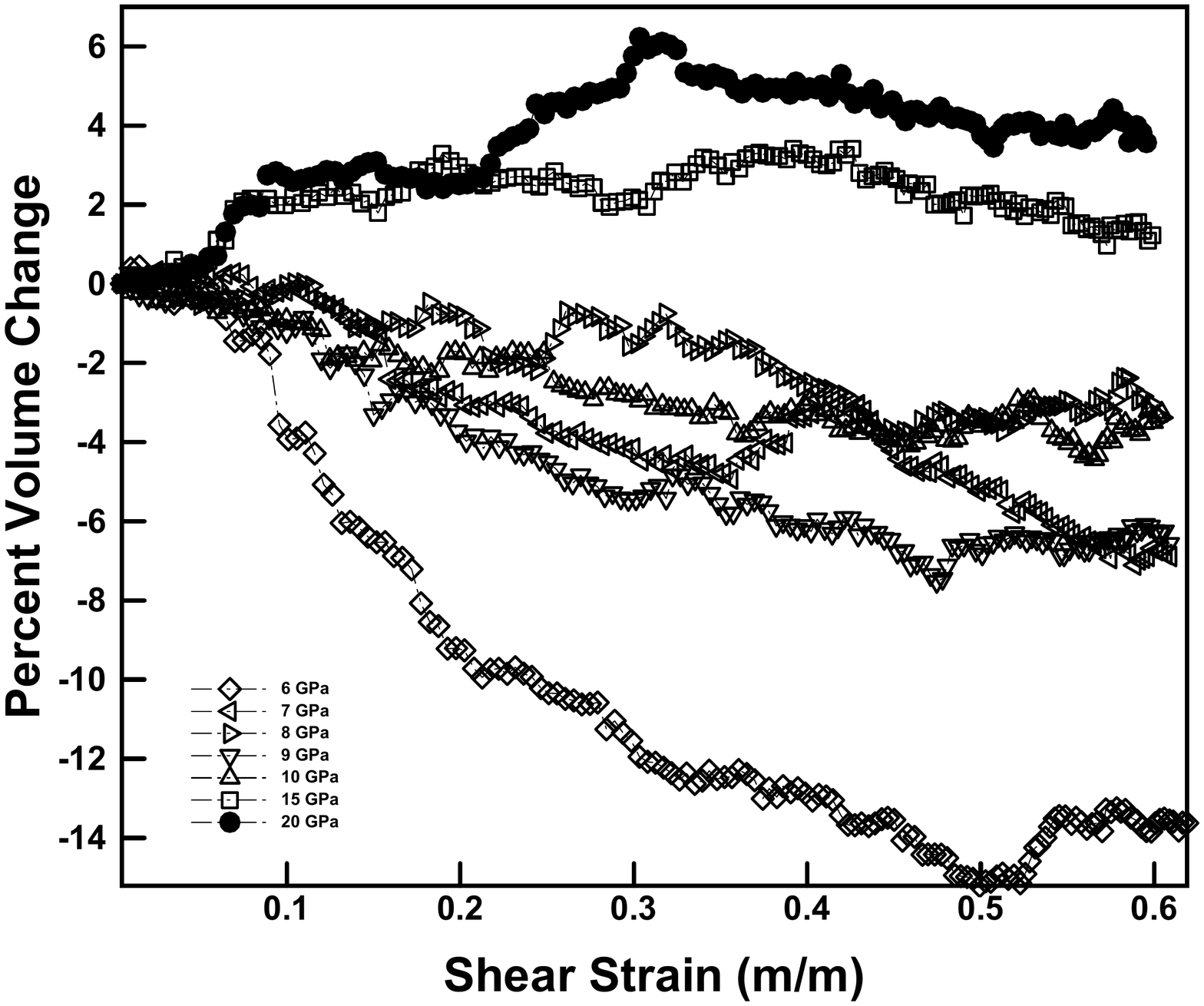}
	\end{subfigure}
	\begin{subfigure}{0.49\textwidth}\caption{\small }
			\includegraphics[width=0.99\linewidth]{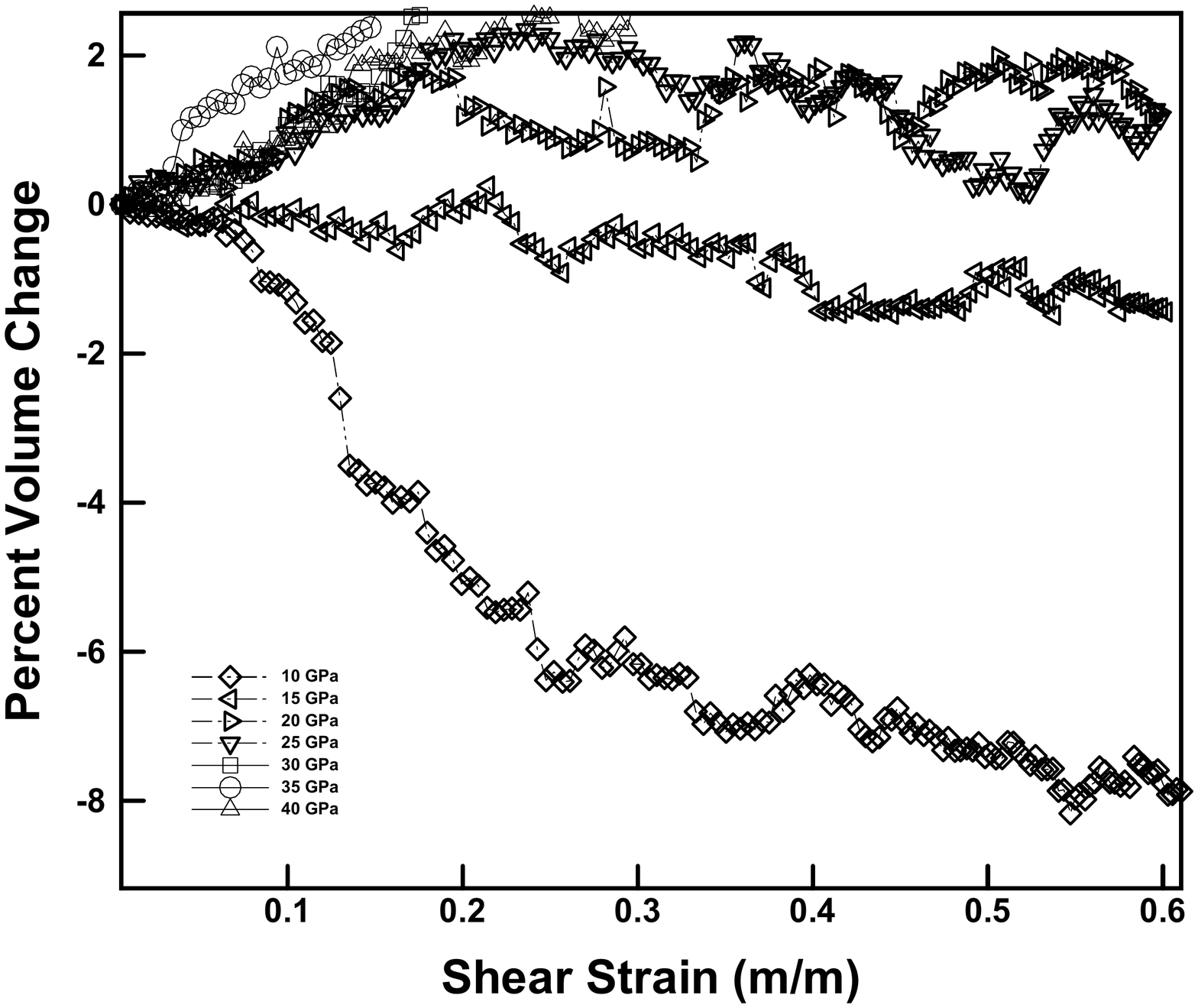}
	\end{subfigure}
	\centering
	\caption{\small  Evolution of volumetric strain during pressure-shear response for different values of preconsolidation pressure $p_{\rm max}$ (shown inset in the figures) and confining pressure $p$. a) $p = -1$ GPa. b) $p = 3$ GPa. c) $p = 6$ GPa. d) $p = 9$ GPa.}
	\label{pervolumeandshear}
\end{figure}

\begin{figure}[h]
	\begin{subfigure}{0.49\textwidth}\caption{\small }
		\includegraphics[width=0.99\linewidth]{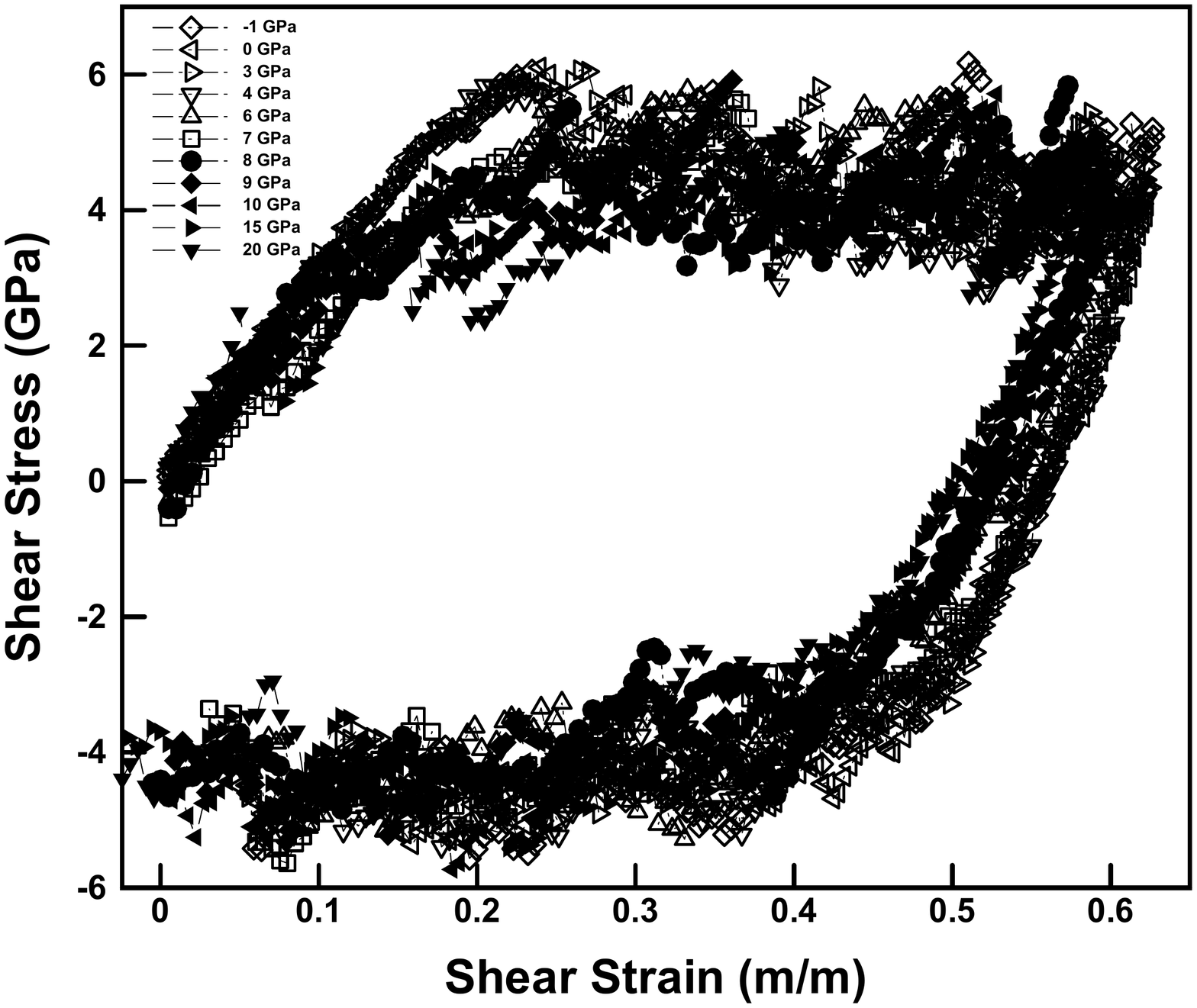}
	\end{subfigure}
	\begin{subfigure}{0.49\textwidth}\caption{\small }
		\includegraphics[width=0.99\linewidth]{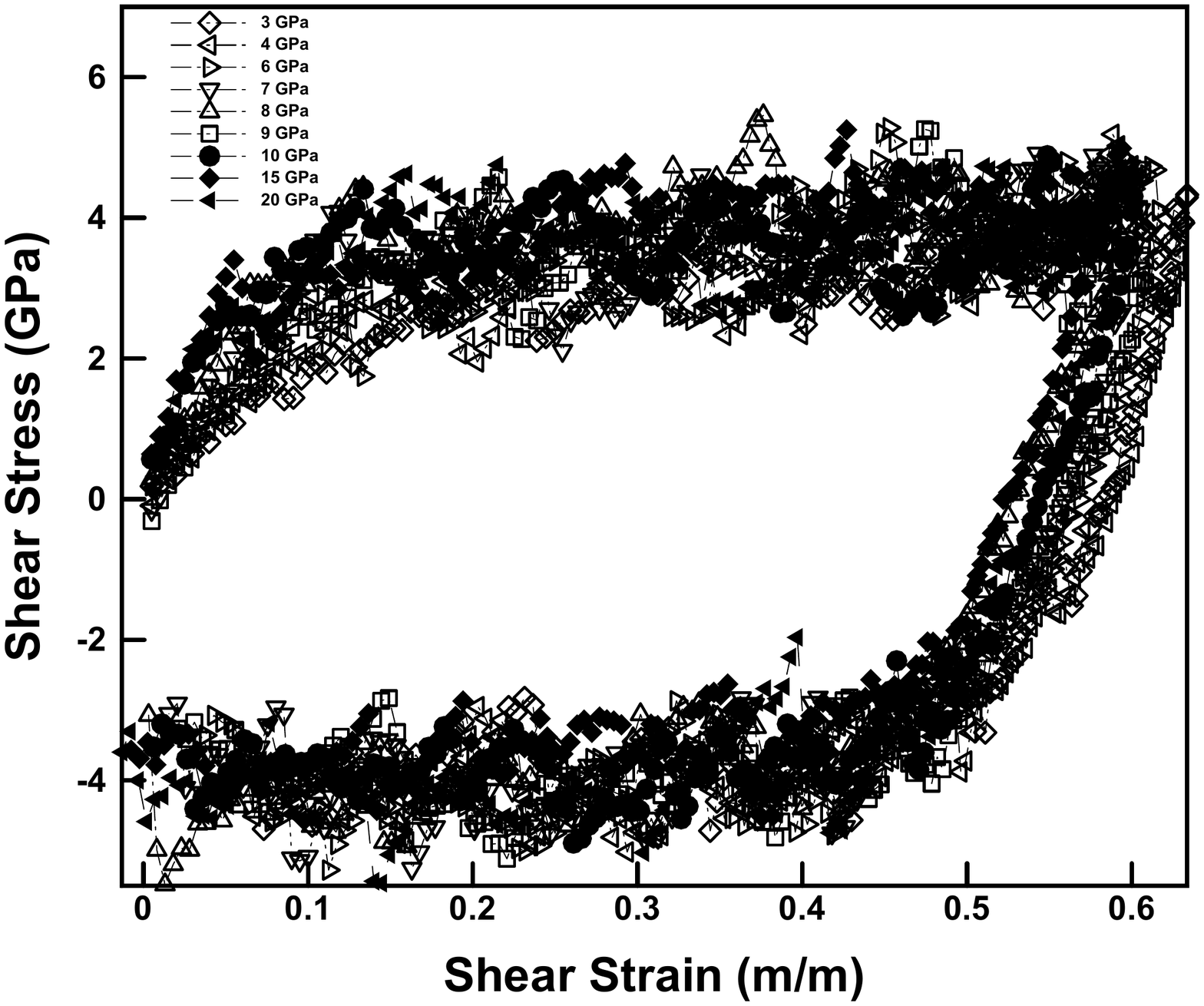}
	\end{subfigure}
	\begin{subfigure}{0.49\textwidth}\caption{\small }
		\includegraphics[width=0.99\linewidth]{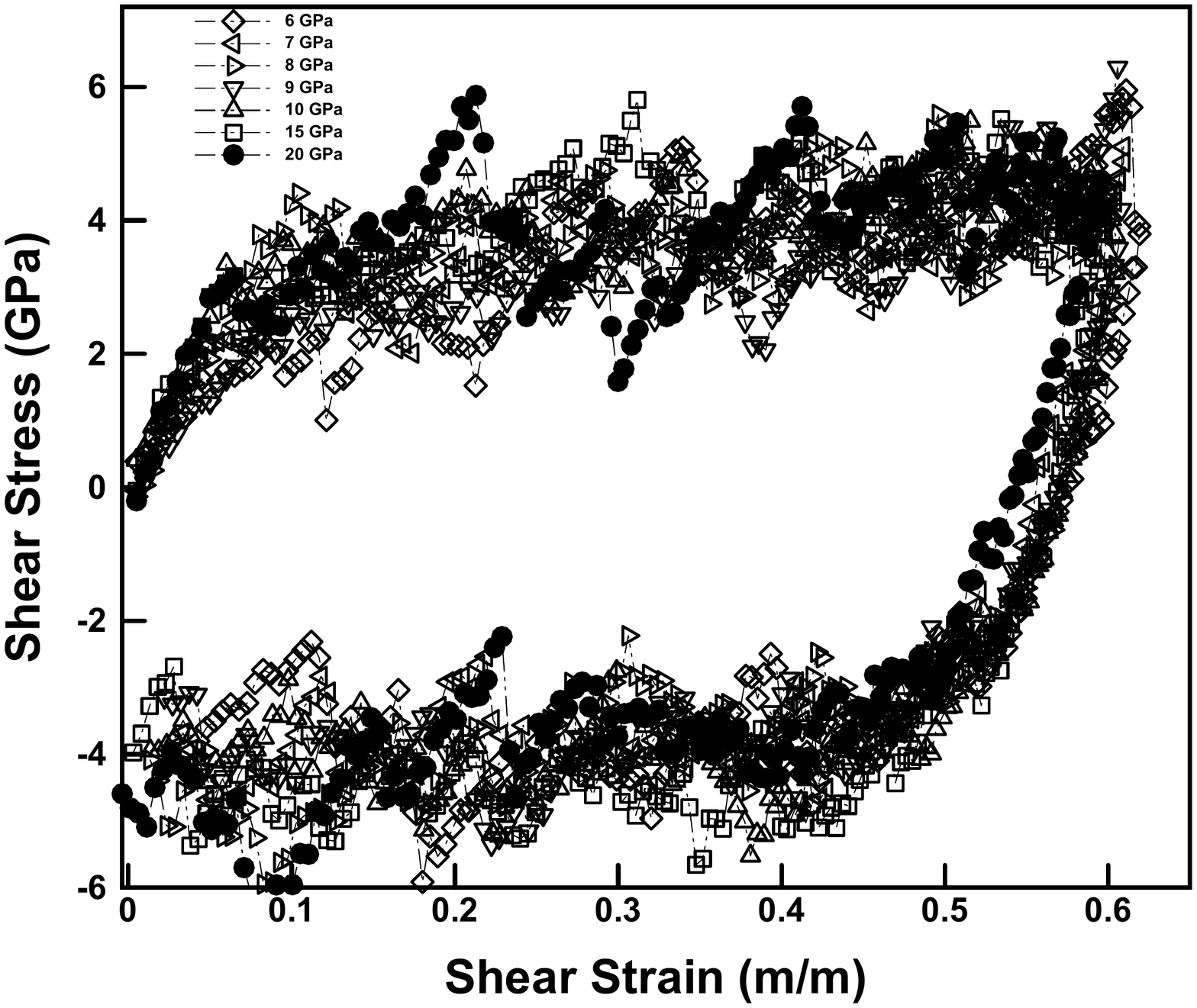}
	\end{subfigure}
	\begin{subfigure}{0.49\textwidth}\caption{\small }
		\includegraphics[width=0.99\linewidth]{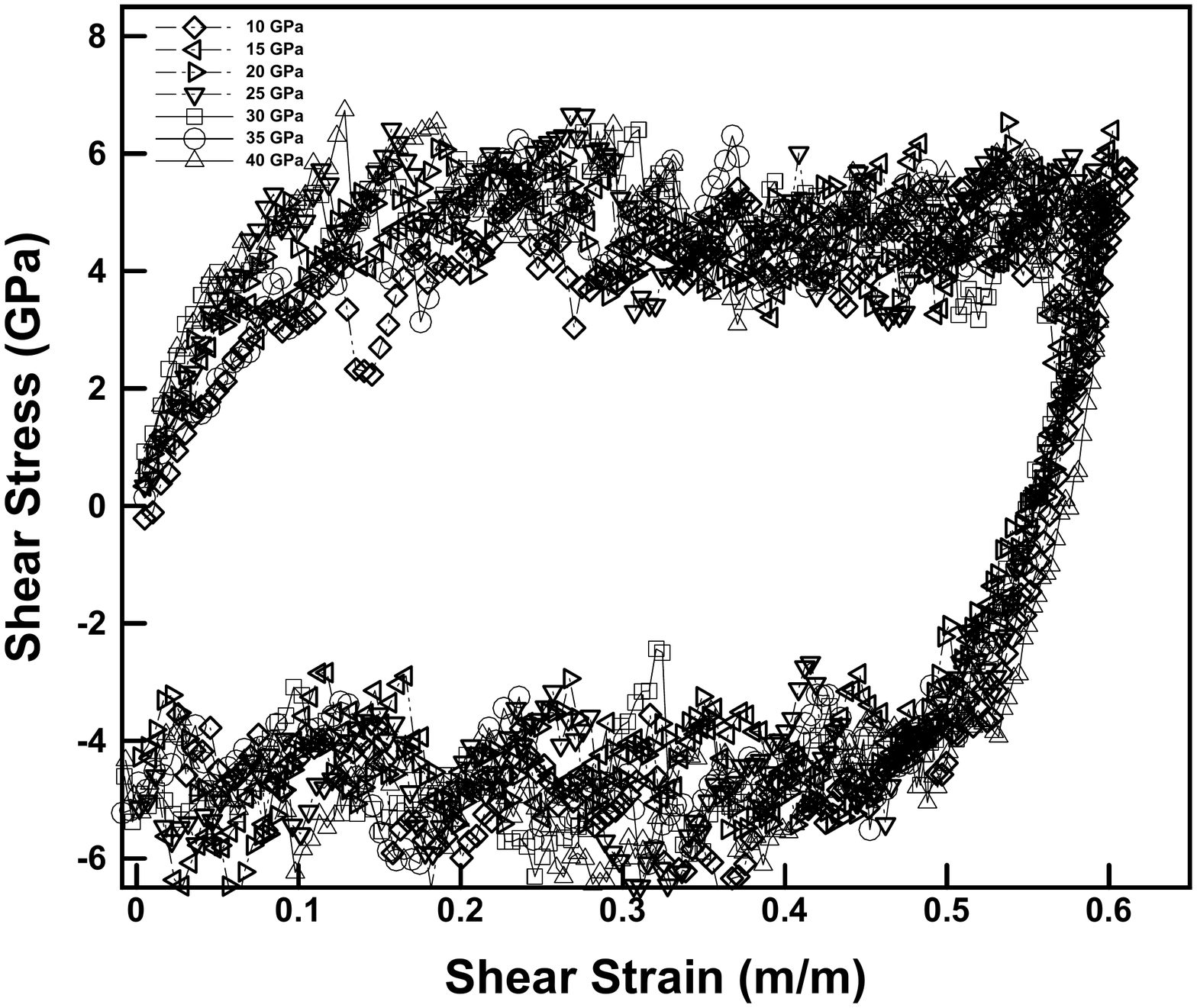}
	\end{subfigure}
	\centering
	\caption{\small  Shear stress {\sl vs}. shear strain for different values of preconsolidation pressure $p_{\rm max}$ (shown inset in the figures) and confining pressure $p$. a) $p = -1$ GPa. b) $p = 3$ GPa. c) $p = 6$ GPa. d) $p = 9$ GPa.}
	\label{pvolumeandshear}
\end{figure}

Fig.~ \ref{volumeandshear}, shows the evolution of the volume of the sample with shear deformation at four values of confining pressure $ p $ and a range of preconsolidation pressures $ p_{\text{max}} \geq p$. The striking feature in these plots is that, in all cases, the volume of the sample attains a {\sl limiting volume}, or {\sl critical state}, at sufficiently large shear deformation. The critical state is attained both under compressive (positive) and tensile (negative) confining pressures. The limiting volume depends on the confining pressure but is independent of the preconsolidation pressure, Fig.~\ref{pvolumeandshear}. The calculations also show that, at the critical state, the sample deforms at a constant shear stress that depends on the confining pressure but is independent of the preconsolidation pressure. Remarkably, the volume initially decreases in under-consolidated samples, $p_{\text{max}} \lesssim 2 p $, and increases in over-consolidated samples, $p_{\text{max}} \gtrsim 2 p $. Similar trends are observed in the evolution of the volumetric strain, Fig.~\ref{pervolumeandshear}.

\section{Mesoscopic Critical-State Model} \label{cont}

The preceding MD data provides a basis for the formulation of a mesoscopic continuum model of the inelasticity of fused silica glass. In particular, the attainment of a critical state in the evolution of volume under pressure-shear loading, Section~\ref{viUri7}, strongly suggests a representation based on critical-state theory of plasticity \cite{Roscoe:1958, Schofield:1968}. A central tenet of critical-state theory is that a solid confined at fixed pressure attains a critical state after sufficient shear deformation beyond which subsequent plastic deformation occurs at constant volume and without further consolidation. In this section, we investigate the ability of critical-state theory to describe the behavior of glass gleaned from molecular dynamics.

\subsection{Finite kinematics}

In view of the large deformations that occur over the pressure range of interest, we formulate the theory in finite kinematics. We assume a standard multiplicative decomposition of the deformation gradient $\bF$ of the form \cite{Lee:1969}
\begin{equation}\label{FeFp}
    \bF = \bF^e \bF^p
\end{equation}
into an elastic part $\bF^e$ and a plastic part $\bF^p$. We denote by $J=\det(\bF)$, $J^e=\det(\bF^e)$ and $J^p=\det(\bF^p)$ the corresponding Jacobians.

\subsection{Equilibrium relations}

We further adopt a thermodynamic formalism \cite{LUBLINER1972237, lubliner1973structure} to describe the local inelastic processes and postulate the existence of a Helmholtz free energy density per unit undeformed volume of the general form
\begin{equation} \label{PhiAdditive}
    {A} = W^e(\bC^e,T) + W^p(J^p,T) ,
\end{equation}
where
\begin{equation}
    \bC^e = \bF^{eT} \bF^e
\end{equation}
is the elastic right Cauchy-Green deformation tensor, $W^e$ is the thermoelastic strain energy density per unit undeformed volume and $W^p$ is the stored energy density per unit undeformed volume. The corresponding equilibrium relations are
\begin{subequations}
\begin{align}
    &
    \bP
    =
    \frac{\partial W}{\partial\bF}
    =
    2 \bF^e \frac{\partial W^e}{\partial \bC^e} \bF^{p-T} ,
    \\ &
    \bY
    =
    -
    \frac{\partial W}{\partial\bF^p}
    =
    \left(
        \bC^e \frac{\partial W^e}{\partial \bC^e}
        +
        \frac{\partial W^e}{\partial \bC^e} \bC^e\right)  \bF^{p-T}
        -
        \frac{\partial W^p}{\partial J^p}
        J^p \bF^{p-T} ,
\end{align}
\end{subequations}
where $\bP$ is the first Piola-Kirchhoff stress tensor and $\bY$ is the thermodynamic driving force conjugate to $\bF^p$. We additionally assume that the elastic behavior of glass is isotropic. In particular,
\begin{equation}
    \bC^e \frac{\partial W^e}{\partial \bC^e}
    =
    \frac{\partial W^e}{\partial \bC^e} \bC^e .
\end{equation}
Using this identity, the rate of dissipation evaluates to
\begin{equation}\label{ch5Udi}
    \bY \cdot \dot{\bF}^p
    =
    J \, \by \cdot \bd^p ,
\end{equation}
where
\begin{equation}
    \by = \mbs{\sigma} - p_c \bI
\end{equation}
is a spatial driving force,
\begin{equation}\label{sPi2cL}
    \bd^p
    =
    \frac{1}{2} (\bl^p + \bl^{pT})
    =
    \frac{1}{2} \big(\dot{\bF}^p \bF^{p-1} + (\dot{\bF}^p \bF^{p-1})^T\big)
\end{equation}
is the plastic rate of deformation tensor,
\begin{equation}
    J
    \mbs{\sigma}
    =
    2
    \bF^e
    \frac{\partial W^e}{\partial\bC^e}(\bC^e,T)
    \bF^{eT}
\end{equation}
is the Cauchy stress and
\begin{equation}
    J p_c = J^p \frac{\partial W^p}{\partial J^p}(J^p,T)
\end{equation}
is a critical pressure.

\subsection{Flow rule}

In view of the structure of the rate-of-dissipation identity (\ref{ch5Udi}), and following the classical kinetic theory of Onsager, we assume the existence of a dual kinetic potential $\psi^*(\by,J^p)$ such that
\begin{equation}\label{gl7wRO}
    \bd^p = \frac{\partial\psi^*}{\partial\by}(\by,J^p) .
\end{equation}
We allow for a dependence of $\psi^*$ on $J^p$ in order to account for the effect of densification of the glass on its flow characteristics.  We also note that objectivity, or invariance under rotations superposed on the spatial configuration, follows from the assumed isotropy of $\psi^*(\cdot,J^p)$. If, in addition, we idealize the kinetics of plastic deformation as rate-independent, then $\psi^*(\by,J^p)$ is the indicator function of an elastic domain ${E}(J^p) \subset \mathbb{R}^{3\times 3}_{\rm sym}$, i.~e.,
\begin{equation}
    \psi^*(\by,J^p)
    =
    I_{{E}(J^p)}(\by)
    =
    \left\{
    \begin{array}{ll}
        0, & \text{if } \by \in {E}(J^p) ,\\
        +\infty, & \text{otherwise}.
    \end{array}
    \right.
\end{equation}
Because of the extended character and lack of differentiability of $I_{{E}(J^p)}(\by)$, the potential relation (\ref{gl7wRO}) needs to be understood in the sense of some appropriate notion of generalized derivative, or flow rule. If ${E}(J^p)$ is convex, the appropriate generalized derivative is supplied by the set-valued subdifferential \cite{rockafellar:1970}
\begin{equation}\label{wRLeb9}
    \bd^p
    \in
    \{
        \br \in \mathbb{R}^{3\times 3}_{\rm sym}\
        \text{ s.~t. }
        (\by - \by^*) \cdot \br \geq 0 ,\
        \forall\text{ }\by^* \in {E}(J^p)
    \} ,
\end{equation}
which embodies Drucker's principle of maximum dissipation, which underlies the classical theory plasticity \cite{Lubliner:1990}.

\subsection{Calibration from MD data}

We proceed to use the data mined from MD, Section~\ref{vlu4To}, to specialize the general framework just outlined to fused silica glass and calibrate the resulting model.

\subsubsection{Elasticity}

For definiteness, we consider elastic strain-energy densities of the neo-Hookean form
\begin{equation}\label{th5uYL}
    W^{e}(\bC^{e})
    =
    \dfrac{\mu(J^{e})}{2} \big( J^{e-2/3} {\rm tr}(\bC^{e}) - 3\big) + f(J^{e}),
\end{equation}
where $\mu(J^{e})$ is a volume-dependent shear modulus and $f(J^{e})$ defines the volumetric equation of state. The Cauchy stress follows from (\ref{th5uYL}) as
\begin{equation}\label{asdf}
\begin{split}
    J \mbs{\sigma}
    =
    2 \bF^{e} \dfrac{\partial W^{e}}{\partial \bC^{e}} \bF^{eT}
    & =
    \Big(
        \dfrac{1}{2} \mu'(J^{e}) (J^{e-2/3} {\rm tr}(\bB^{e})-3)
        +
        f'(J^{e})
    \Big)
    \, J^{e} \bI
    \\ & +
    \mu(J^{e})
    \Big(
        J^{e-2/3} \bB^{e}
        -
        \dfrac{1}{3}J^{e-2/3} {\rm tr}(\bB^{e}) \bI
    \Big) ,
\end{split}
\end{equation}
where
\begin{equation}
    \bB^e
    =
    \bF^e \bF^{eT}
\end{equation}
is the elastic left Cauchy-Green deformation tensor.

\begin{figure}[h]
	\centering
	\begin{subfigure}{0.49\textwidth}\caption{\small }
		\includegraphics[width=0.99\linewidth]{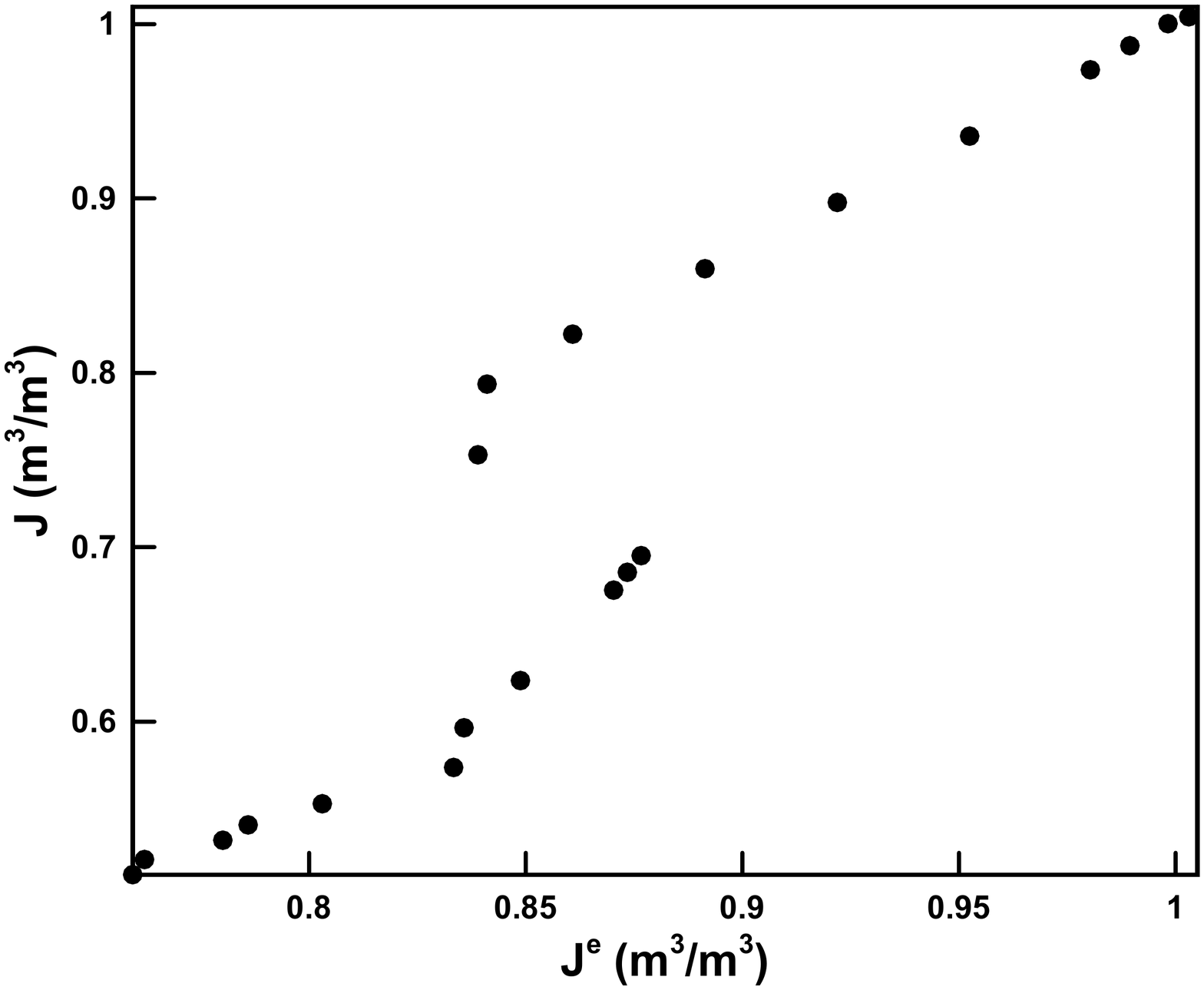}
	\end{subfigure}
	\begin{subfigure}{0.49\textwidth}\caption{\small }
		\includegraphics[width=0.99\linewidth]{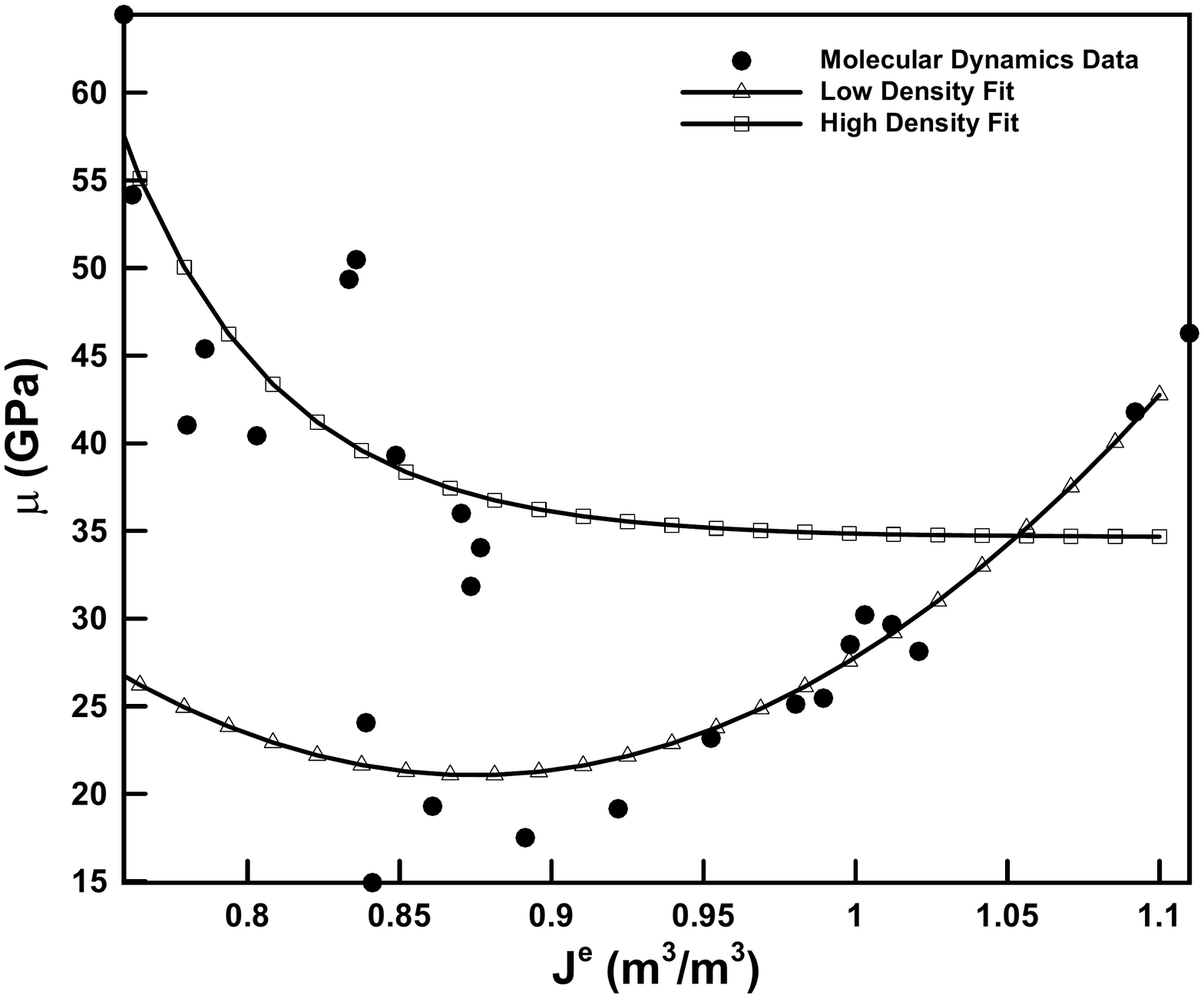}
	\end{subfigure}
    \caption{\small Volumetric MD data during monotonic compressive loading. a) Total volumetric Jacobian $J$ {\sl vs}. elastic Jacobian $ J^{e} $ as deduced from unloading, showing two phases (dense and loose) separated by a densification phase transition. b) Shear modulus $ \mu $ {\sl vs}. $J^{e}$ and fit of each of the phases.} \label{fIu3iu}
\end{figure}

The molecular dynamics data suggests a densification phase transition when the plastic volumetric deformation attains a critical value of $ J^{p} = J^{p}_c \approx 0.9 $, Fig.~\ref{fIu3iu}a. We therefore regard glass as a two-phase material and describe the elasticity of each phase by means of an elastic strain-energy density of the form (\ref{th5uYL}). Specializing \eqref{asdf} to simple elastic shear following a volumetric plastic deformation gives
\begin{equation}
    J\sigma_{12} = \mu(J^{e}) \gamma ,
\end{equation}
in axes aligned with the shearing directions and with $\gamma$ denoting the shear strain. Using this relation in combination with the MD data in Fig.~\ref{shearModulusVersusPressureKondo}a gives the  $\mu$ {\sl vs.} $J^e$ data shown in Fig.~\ref{fIu3iu}b. For definiteness, we fit these data by functions of the form
\begin{equation}
    \mu(J^{e})
    =
    \begin{cases}
        a_{0} + a_{1}J^{e} + a_{2}J^{e^{2}}, & J^{e}\geq J^{p}_c, \\
        b_{1}\exp(b_{2}(J^{e} - 1))+ b_{3}, & \text{otherwise} ,
    \end{cases}
\end{equation}
and obtain the coefficients tabulated in Table~\ref{blu9Ie}. The goodness of the fit is shown in Fig.~\ref{fIu3iu}b. The two-phase structure of the equation of state is also clear from the figure.

\begin{table}[h]
	\centering
	\caption{\small Pressure-dependent shear-modulus parameters}
	\label{blu9Ie}
	\begin{tabular}{lcccccc}
		\hline\noalign{\smallskip}
		&$ a_{0} $ & $ a_{1} $ & $ a_{2} $&$ b_{1} $ & $ b_{2} $ & $ b_{3} $\\
		\noalign{\smallskip}\hline\noalign{\smallskip}
		& 347.15 GPa & -745.82 GPa & 426.46 GPa & 0.20773 GPa &
		-19.498 &34.6439 GPa\\
		\noalign{\smallskip}\hline
	\end{tabular}
\end{table}

\begin{figure}[h]
	\centering
	\begin{subfigure}{0.49\textwidth}\caption{\small }
		\includegraphics[width=0.99\linewidth]{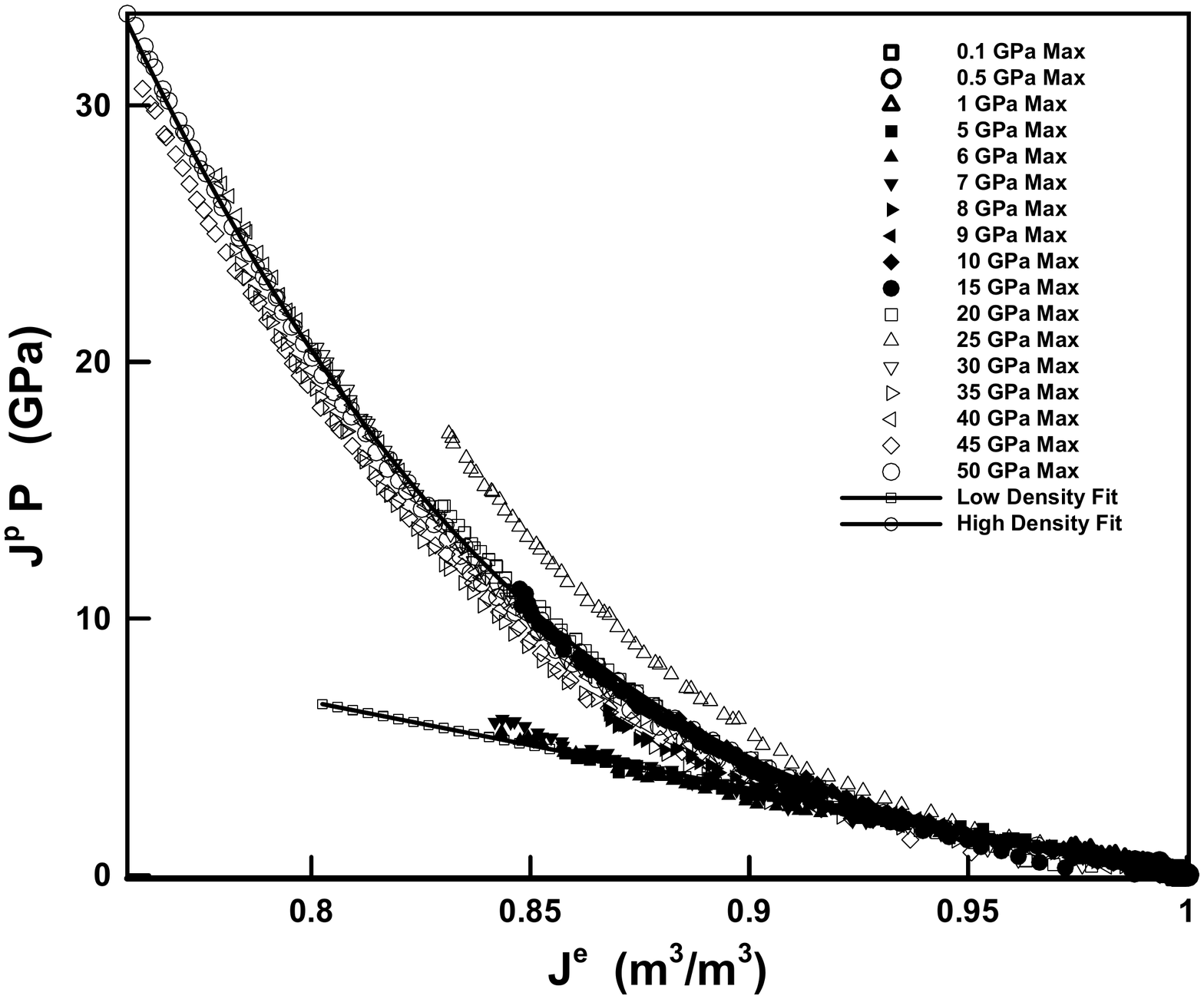}
	\end{subfigure}
	\begin{subfigure}{0.49\textwidth}\caption{\small }
		\includegraphics[width=0.99\linewidth]{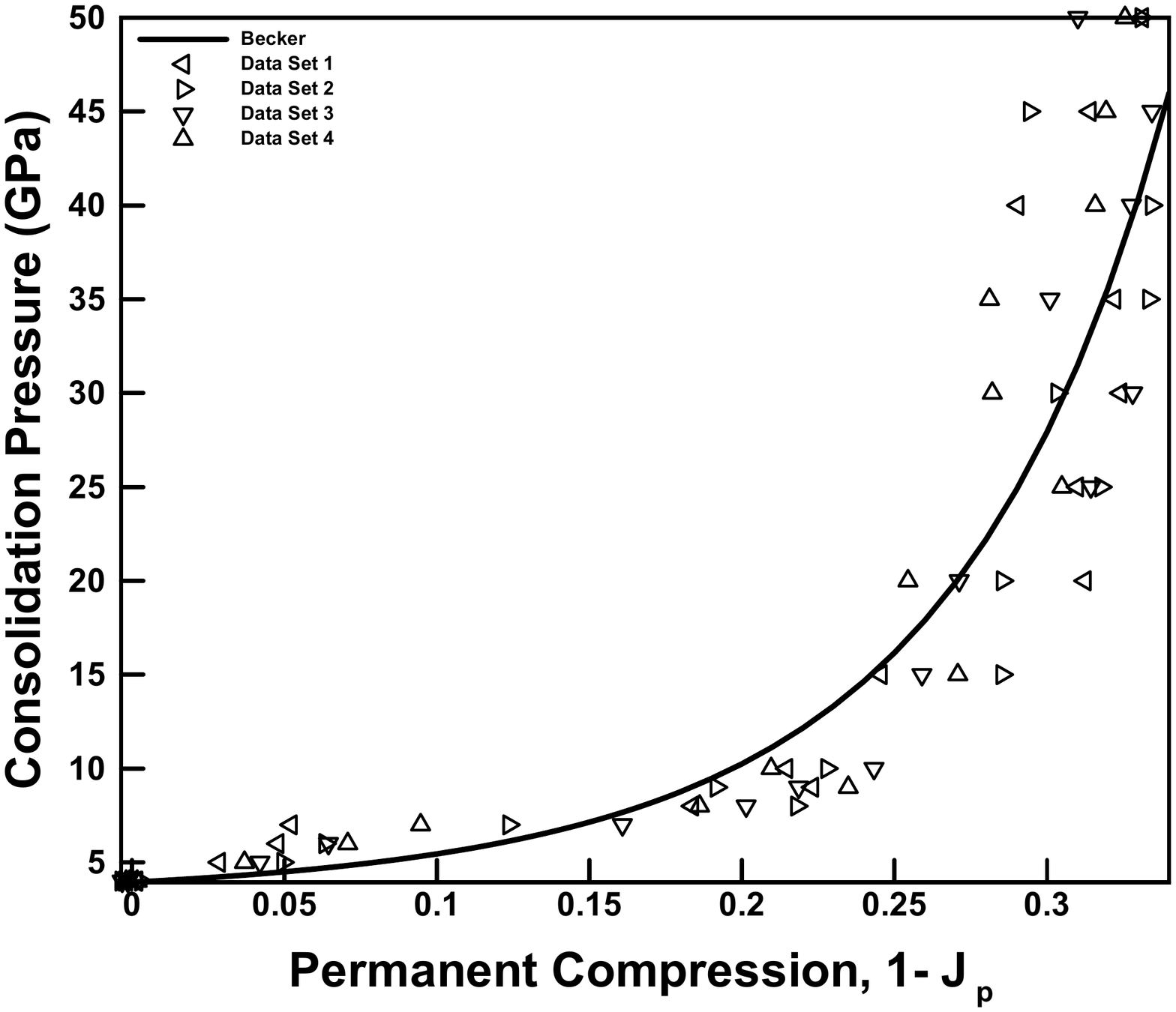}
	\end{subfigure}
    \caption{\small Consolidation MD data during monotonic compressive loading. a) Pressure $p$ {\sl vs}. elastic Jacobian $ J^{e} $ and fits for dense and loose phases. b) Preconsolidation pressure  $p_c$ on permanent densification $1-J^{p} $ and fit.} \label{Diak0e}
\end{figure}

Next, we determine the equation-of-state function $ f(J^e) $ in eq.~(\ref{th5uYL}) by examining the case of pure elastic compression. Specializing \eqref{asdf} to this case, we obtain the relation
\begin{equation}
    -J p = f'(J^{e})J^{e}.
\end{equation}
In this particular case, the MD data of Fig.~\ref{pressurecompression1} reduces to Fig.~\ref{Diak0e}a. We fit these data by functions of the form
\begin{equation}
    f(J^{e})
    =
    \begin{cases}
    \dfrac{c}{2}(J^{e}-1)^{2} , & J^{p}\geq J^{p}_c , \\
    \dfrac{d_1}{2}(J^{e}-1)^{2} + \dfrac{d_2}{4}(J^{e}-1)^4 , & \text{otherwise} ,\\
    \end{cases}
\end{equation}
and obtain the coefficients tabulated in Table~\ref{tab:materialdata2}. The goodness of the fit is also shown in Fig.~\ref{fIu3iu}b.

\begin{table}[h]
	\centering
	\caption{\small Volumetric elastic-energy dependence}
	\label{tab:materialdata2}
	\begin{tabular}{lccc}
		\hline\noalign{\smallskip}
		&$ c $ & $ d_1 $ & $ d_2 $\\
		\noalign{\smallskip}\hline\noalign{\smallskip}
		& -33.75 GPa & -25.167 GPa & -1879.69 GPa \\
		\noalign{\smallskip}\hline
	\end{tabular}
\end{table}

\subsubsection{Elastic domain and yield surface}

\begin{figure}[h]
	\begin{subfigure}{0.40\textwidth}\caption{\small } \includegraphics[width=0.99\linewidth]{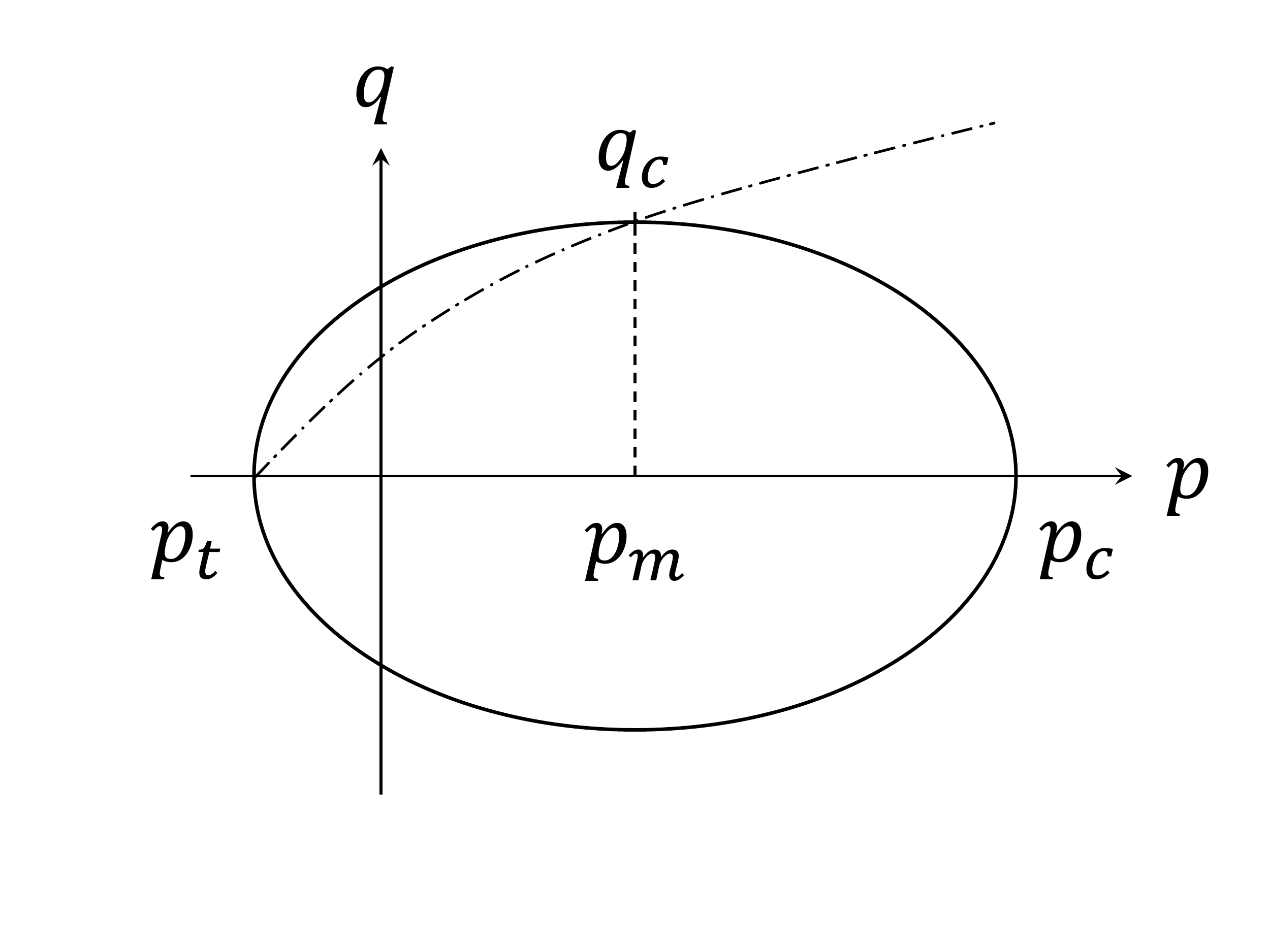}
	\end{subfigure}
	\begin{subfigure}{0.4\textwidth}\caption{\small } \includegraphics[width=0.99\linewidth]{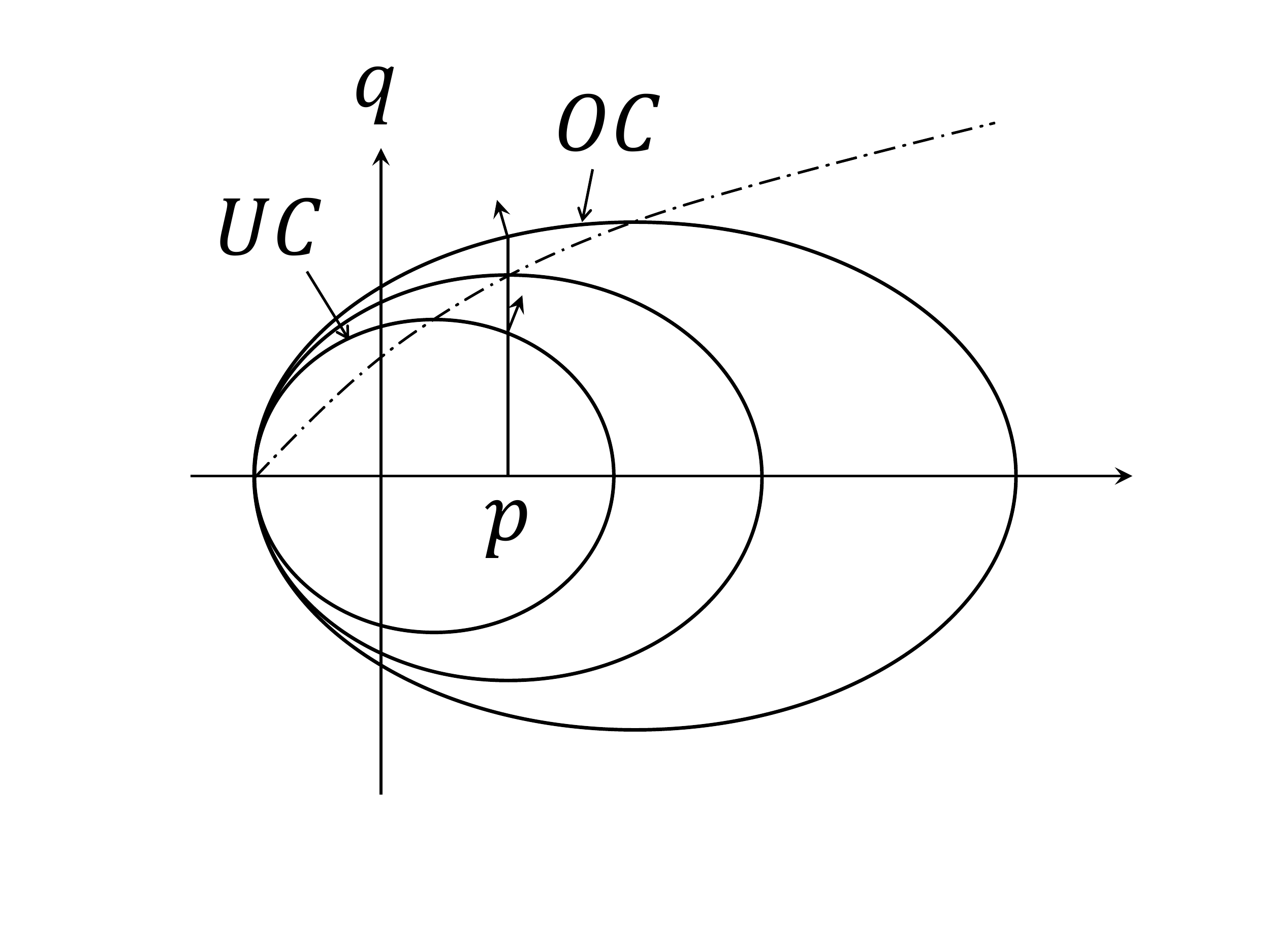}
	\end{subfigure}
	\caption{\small a) Schematic of elastic domain in the $(p,q)$-plane, where $p$ denotes the pressure, $q$ the Mises effective shear stress, $p_t$ the tensile failure pressure, $p_c$ the compressive yield pressure and $q_c$ the shear yield strength. The dash-dot line represents the critical-state line. b) Stress path for pressure-shear test (vertical line at $p$) and directions of plastic deformation rate (arrows) in the over-consolidated case, labeled OC, and under-consolidated case, labeled UC.} \label{0riAsw}
\end{figure}

Under the assumption of rate independence, we model the yield-behavior of glass by means of the elliptic elastic domain
\begin{equation}\label{Zoac3i}
    {E}(J^p)
    =
    \left\{
        \by \in \mathbb{R}^{3\times 3}_{\rm sym},\
        \left(\frac{q}{q_c(J^p)}\right)^2
        +
        \left(\frac{p-(p_c(J^p)+p_t)/2}{(p_c(J^p)+p_t)/2}\right)^2
        \leq
        1
    \right\} ,
\end{equation}
where
\begin{equation}
    q = \sqrt{\frac{1}{2} \bs\cdot\bs}
\end{equation}
is the Mises effective shear stress,
\begin{equation}
    \bs
    =
    \mbs{\sigma} - \frac{1}{3} {\rm tr}(\mbs{\sigma}) \, \bI
    =
    \by - \frac{1}{3} {\rm tr}(\by) \, \bI
\end{equation}
is the stress deviator, $p_t$ is the tensile failure pressure, $p_c$ is the compressive yield pressure, $q_c$ is the shear yield strength and $J^p$ plays the role of an internal variable, cf.~Fig.~\ref{0riAsw}a. Elastic domains of the type (\ref{Zoac3i}) have been used in connection to Cam-Clay models of granular media (cf., e.~g., \cite{Ortiz:2004}) and glasses \cite{Kermouche20083222, Gazonas:2011, Mantisi2012}. The function $p_c(J^p)$ defines the {\sl consolidation relation}. The curve in the $(p,q)$-plane
\begin{equation}\label{b2iuCi}
    q_c = g(p_m) ,
\end{equation}
with
\begin{equation}
    p_m = \frac{p_t + p_c}{2}
\end{equation}
may be obtained by eliminating $J^p$ between $q_c(J^p)$ and $p_c(J^p)$. Evidently, $p_m$
is the pressure at which $q$ attains its maximum value $q_c$ on the yield surface $\partial {E}(J^p)$, cf.~eq.~(\ref{Zoac3i}), and at which, by the flow rule (\ref{wRLeb9}), the plastic strain rate is volume preserving. Thus, the relation (\ref{b2iuCi}) represents the {\sl critical state line} in the $(p,q)$-plane.

\subsubsection{Consolidation curve}

We proceed to identify the consolidation curve $p_c(J^p)$ for fused silica from the MD data shown in Fig.~ \ref{pressurecompression1}. To this end, we identify $J^p$ as the volumetric deformation upon unloading and the corresponding $p_c(J^p)$ as the maximum pressure attained during loading. The resulting data are shown in Fig.~ \ref{fIu3iu}b. We fit these data by means of a power-law relation of the form
\begin{equation}\label{hardening}
    p_{c}
    =
    p_0
    +
    \dfrac{A}{\alpha} (1 - J^{p-\alpha}) ,
\end{equation}
previously used by Becker \cite{Becker:2012} as a volumetric equation of state. In addition, we identify the tensile failure stress $p_t$ from MD calculations as the maximum tensile pressure at which the glass sample is stable. The resulting values of the constants are tabulated in Table \ref{tab:materialdata}. The goodness of the fit is shown in Fig.~\ref{fIu3iu}b.

\begin{table}[h]
	\centering
	\caption{\small Hardening parameters}
	\label{tab:materialdata}
	\begin{tabular}{lcccc}
		\hline\noalign{\smallskip}
		&$ A $ & $ \alpha $ & $ p_0 $ & $ p_t $ \\
		\noalign{\smallskip}\hline\noalign{\smallskip}
		& 8.48613 GPa & 9.2689 &3.02934 GPa & $-10$ GPa\\
		\noalign{\smallskip}\hline
	\end{tabular}
\end{table}

\subsubsection{Evolution towards the critical state}

We verify that a simple elastic domain of the form (\ref{Zoac3i}) and the consolidation curve (\ref{hardening}) are indeed capable of representing the complex yield and flow behavior revealed by the pressure-shear MD data collected in Section~\ref{FRoe1o}. Thus, consider a pressure-shear test at confining pressure $p$ and effective shear stress $q$ increasing monotonically from zero. The corresponding loading path is shown as a vertical line at $p$ in Fig.~\ref{0riAsw}b. The intermediate ellipse in the figure corresponds to the critical state that is eventually attained along the loading path. The figure also depicts two cases, labeled 'under-consolidated' (UC) and 'over-consolidated' (OC). In the under-consolidated case, $p$ lies to right of the initial value of $p_m$, resulting in a plastic strain rate $\mbs{d}^p$ (shown as an arrow in the figure) with a negative, or compressive, volumetric component, ${\rm tr}(\mbs{d}^p) < 0$.\footnote{We recall that, under the pressure sign convention $p = - {\rm tr}(\mbs{\sigma})$, a positive (negative) component of the normal to the yield surface in the $(p,q)$-plane corresponds to a negative (positive), or compressive (tensile), volumetric plastic strain, ${\rm tr}(\mbs{d}^p) < 0$ (${\rm tr}(\mbs{d}^p) > 0$).} By contrast, in the over-consolidated case, $p$ lies to left of the initial value of $p_m$, resulting in a plastic strain rate $\mbs{d}^p$ (also shown as an arrow in the figure) with a positive, or tensile, volumetric component, ${\rm tr}(\mbs{d}^p) > 0$. It thus follows that under-consolidated samples are predicted to decrease their volume, whereas over-consolidated samples are predicted to increase their volume, in accord with the MD data in Fig.~\ref{pervolumeandshear}. From relation (\ref{sPi2cL}), it follows that
\begin{equation}
    \dot{J}^p = J^p \, {\rm tr}(\mbs{d}^p) ,
\end{equation}
and from the monotonicity of the consolidation curve, Fig.~\ref{fIu3iu}b, it follows that $p_c$ increases in the under-consolidated case and decreases in the over-consolidated case. Thus, in both cases the yield surface converges towards the critical-state yield surface, as required. We also note that, following the attainment of the critical state, represented by the intermediate ellipse in Fig.~\ref{phLuy9}b, both the sample volume and the shear stress remain constant, in agreement with the MD data collected in Fig.~\ref{pervolumeandshear} and Fig.~\ref{pvolumeandshear}. We therefore conclude that the MD data for fused silica presented in Section~\ref{vlu4To} is indicative of---and well-represented by---critical state theory of plasticity.

\subsubsection{The anomalous critical-state line of fused silica}

\begin{figure}[h]
	\begin{subfigure}{0.49\textwidth}\caption{\small } \includegraphics[width=0.99\linewidth]{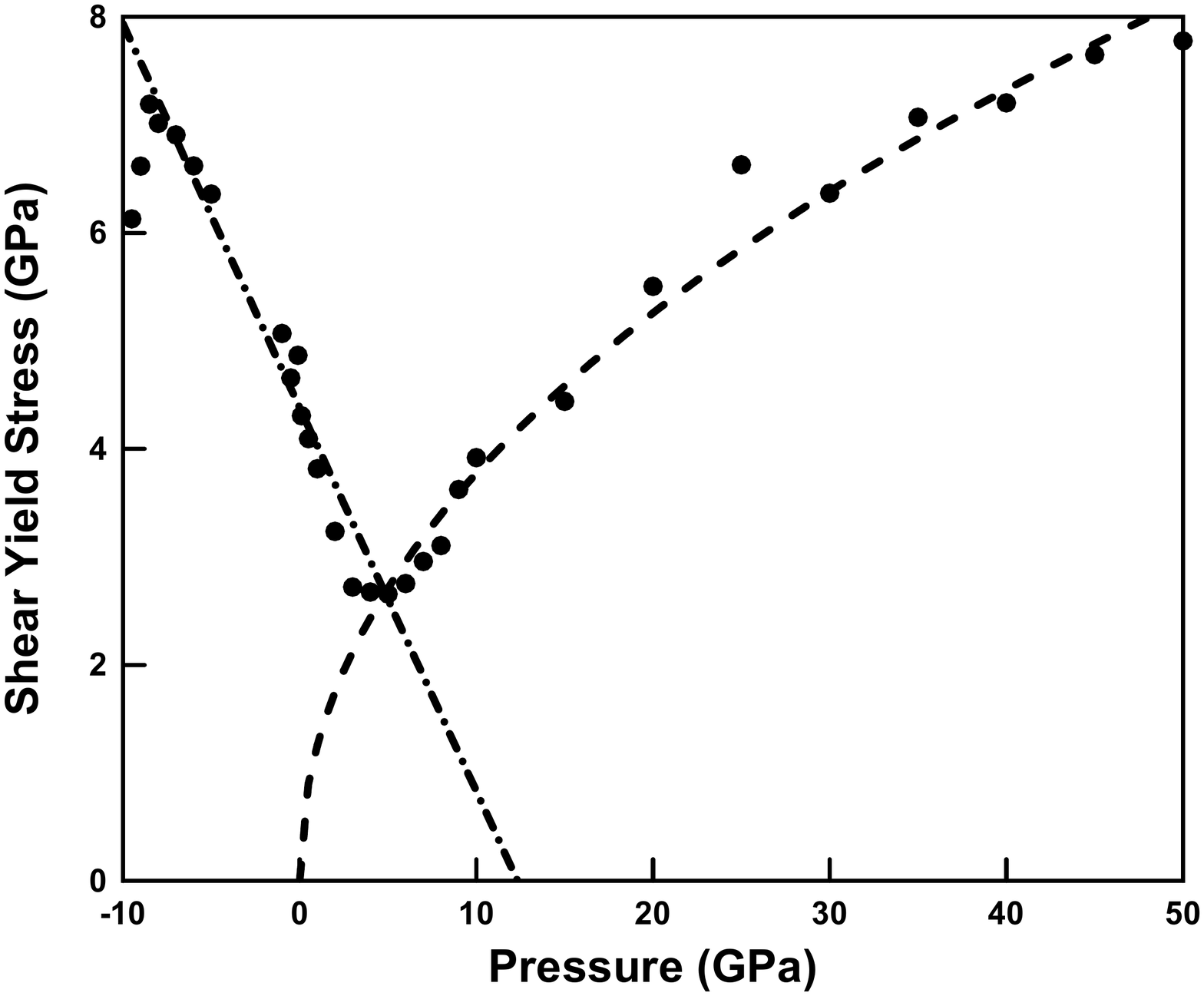}
	\end{subfigure}
	\begin{subfigure}{0.49\textwidth}\caption{\small } \includegraphics[width=0.99\linewidth]{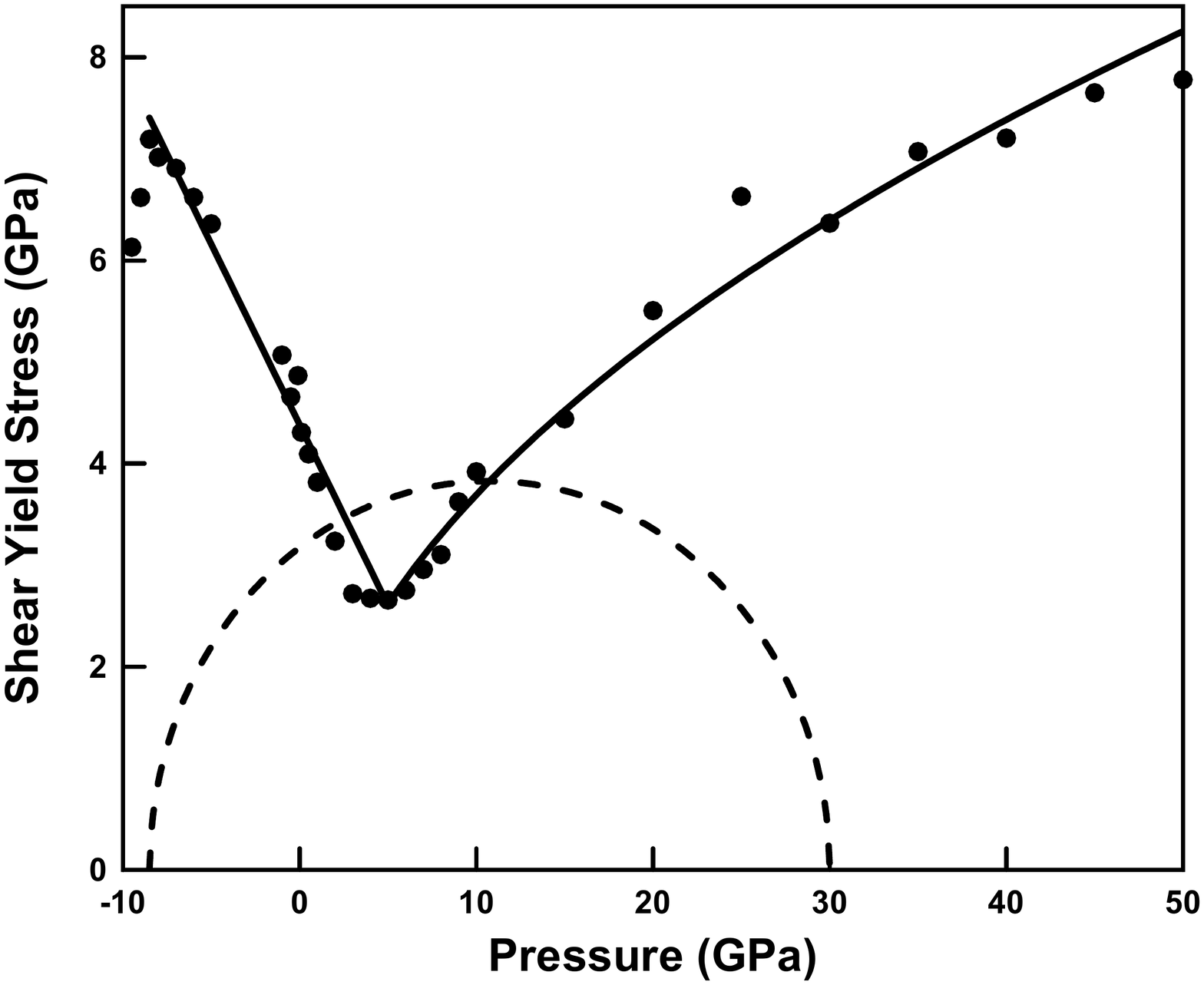}
	\end{subfigure}
    \caption{\small a) Critical state line MD data (dots) and fits. The dash line is the fit in the compressive regime and the dash-dot line is the fit in the tensile regime. b) Critical state line (solid curve) obtained by intersecting the compressive and tensile critical state lines. The dash line represents a typical elastic domain.} \label{phLuy9}
\end{figure}

In order to close the model, the critical state line (\ref{b2iuCi}) remains to be identified. We determine the critical state line, eq.~(\ref{b2iuCi}), from the MD simulations described in Section~\ref{FRoe1o}, by identifying $p_m$ with the confining pressure applied to the sample and the corresponding $q_c$ with the shear stress upon the attainment of the critical state of constant volume.

The data thus obtained is shown in Fig.~\ref{phLuy9}a. The critical-state line thus determined exhibits two clear regimes, one under predominantly compressive pressures and another under predominantly tensile pressures. Remarkably, {\sl in the tensile regime the critical-state line increases with increasing tensile pressure}, which represents anomalous behavior. By contrast, in the compressive regime the critical-state line increases with increasing compressive pressure, or confinement, as expected.

The tensile regime of the critical-state line is well-represented by a linear relation of the form
\begin{equation}\label{TroU6o}
    q = \frac{p_1 - p}{p_1 - p_t} \, q_t ,
\end{equation}
capped vertically at $p = p_t$. The compressive regime of the critical-state line is in turn well-presented by a power law of the form
\begin{equation}\label{c4iUth}
    q = B p^\beta .
\end{equation}
The resulting values of the constants are tabulated in Table \ref{tab:criticalline}. The goodness of the fit is shown in Fig.~\ref{phLuy9}a.

\begin{table}[h]
	\centering
	\caption{\small Critical state line constants}
	\label{tab:criticalline}
	\begin{tabular}{lccccc}
		\hline\noalign{\smallskip}
		&$ p_1 $ & $ q_t $ & $ p_t $ & $ B $ & $ \beta $ \\
		\noalign{\smallskip}\hline\noalign{\smallskip}
		& $12.337$ GPa & $7.402$ GPa & $-8.5$ GPa & $1.168$ $\sqrt{\text{GPa}}$ & $0.5$\\
		\noalign{\smallskip}\hline
	\end{tabular}
\end{table}

The anomalous yield behavior of fused silica under predominantly tensile pressures uncovered by the MD data is indeed consistent with the experimental data of Meade and Jeanloz \cite{MEADE1072} noted in the introduction, Fig.~\ref{8poaSL}b, who attributed the anomaly to changes in coordination at the atomic level. Interestingly, Meade and Jeanloz \cite{MEADE1072} observe an additional region of anomalous shear yield strength behavior at pressures above $30$ GPa, not captured by the present MD calculations. Likely causes of this discrepancy are the large disparity in strain rates between the work of Meade and Jeanloz \cite{MEADE1072}, which was performed at quasi-static loading rates, and the present calculations, which entail large rates of deformation, and possible inadequacies of the interatomic potentials at extremely large pressures and volume reductions.

The intersection of the tensile and compressive critical state lines, eq.~(\ref{TroU6o}) and (\ref{c4iUth}), respectively, results in a {\sl non-convex} combined critical-state line, Fig.~\ref{phLuy9}b. The figure reveals that fused silica is doubly anomalous, on account of the anomalous dependence of the its shear modulus of volumetric deformation, and of the strong non-convexity of its critical-state line.

\section{Microstructure, relaxation and div-quasiconvexification}
\label{Relax}

We now proceed to show that the strongly non-convex critical-state line in Fig.~\ref{phLuy9}b is, in fact, unstable with respect to microstructure formation and that consideration of microstructure results in a stable, or {\sl relaxed}, critical-state line that captures the fine structure of the MD data at the tensile-to-compressive transition. We recall that, as noted in the introduction, several authors \cite{0953-8984-20-24-244128, PhysRevLett.103.065501} have performed molecular dynamics calculations on amorphous solids deforming under shear and found that the resulting deformation field develops fine microstructure in order to accommodate permanent macroscopic deformations, Fig.~\ref{Cho9pr}. In this section, we appeal to notions from the Direct Methods in the Calculus of Variations in order establish a connection between the strong non-convexity of the critical-state line and the development of fine microstructure, and to characterize explicitly and exactly the effective or relaxed behavior at the macroscale. For completeness, a summary of the main mathematical concepts and arguments is consigned to the Appendix. A full mathematical account may be found in the article of Conti {\sl et al.} \cite{Conti:2017}.

We carry out the analysis within the framework of limit analysis \cite{Lubliner:1990}. Thus, we assume that the solid is at {\sl collapse}, i.~e., it deforms plastically at constant applied load. Under these conditions, the instantaneous behavior of the solid is rigid and ideally plastic, i.~e., no instantaneous hardening takes place (ideal plasticity) and (rigid-plastic behavior)
\begin{equation}\label{pLuT9o}
    \mbs{d}^p
    =
    \frac{1}{2} ( \nabla \mbs{v} + \nabla \mbs{v}^T )
    \equiv
    \mbs{e}(\mbs{v}),
\end{equation}
where $\mbs{v} : \Omega \to \mathbb{R}^3$ is the velocity field at collapse, or collapse mode, and $\Omega$ is the domain of the solid at collapse. The corresponding kinematic and static problems of limit analysis \cite{Lubliner:1990} can then be jointly expressed as the saddle-point problem
\begin{equation}\label{wluVL0}
    \inf_{\mbs{v}}
    \sup_{\mbs{\sigma}}
    \Big\{
        \int_\Omega \mbs{\sigma} \cdot \nabla \mbs{v} \, dx
        \, : \,
        \mbs{\sigma}(x) \in E(J^p(x)),
        \ \mbs{v} = \mbs{g}
        \ \text{on } \partial \Omega
    \Big\} ,
\end{equation}
where the minimization and maximization take place over suitable spaces of velocities and stresses, respectively, $J^p$ accounts for the state of consolidation of the solid, $\mbs{g}$ is a prescribed velocity field over the boundary and we assume that the solid is free of body forces. We recall that the inner maximum problem in (\ref{wluVL0}) embodies Drucker's principle of maximum dissipation and the static principle of classical plasticity, whereas the outer minimum problem embodies the kinematic principle of classical plasticity.

We further note that, for a solid obeying critical-state theory of plasticity, instantaneous rigid-ideally plastic behavior implies, in particular, instantaneous constancy of volume, which in turn requires that the solid be either locally rigid or at critical state. This condition sets the requirement that $\sigma(x) \in K$ a.~e.~in $\Omega$, where $K$ is the domain bounded by the critical-state line. Since the critical-state line is the locus of points in stress space at which material behavior is ideally plastic, $K$ may be regarded as a {\sl limit domain} in the sense of hardening plasticity (cf., e.~g., \cite{Martin:1975} for a lucid introduction to limit surfaces in hardening plasticity). Thus, at collapse (\ref{wluVL0}) specializes to
\begin{equation}\label{rierL4}
    \inf_{\mbs{v}}
    \sup_{\mbs{\sigma}}
    \Big\{
        \int_\Omega \mbs{\sigma} \cdot \nabla \mbs{v} \, dx
        \, : \,
        \mbs{\sigma}(x) \in K,
        \ \mbs{v} = \mbs{g}
        \ \text{on } \partial \Omega
    \Big\} ,
\end{equation}
The maximization with respect $\mbs{\sigma}$ may be effected pointwise, whereupon the problem (\ref{rierL4}) reduces to the kinematic problem
\begin{equation}
    \inf_{\mbs{v}}
    \Big\{
        \int_\Omega \phi(\mbs{e}(\mbs{v})) \, dx
        \, : \,
        \ \mbs{v} = \mbs{g}
        \ \text{on } \partial \Omega
    \Big\} ,
\end{equation}
where
\begin{equation}
    \phi(\mbs{d}^p)
    =
    \sup_{\mbs{\sigma} \in K}
    \mbs{\sigma} \cdot \mbs{d}^p
\end{equation}
is the limit plastic dissipation potential.

This classical theory of limit analysis is mathematically well-developed provided that the limit domain $K$ is {\sl convex}, in which case no microstructure occurs. In order extend the theory to non-convex domains and microstructure formation, we reformulate the saddle-point problem (\ref{rierL4}) as
\begin{equation}\label{Q2ukle}
    \sup_{\mbs{\sigma}}
    \inf_{\mbs{v}}
    \Big\{
        \int_\Omega \mbs{\sigma} \cdot \nabla \mbs{v} \, dx
        \, : \,
        \mbs{\sigma} \in K,
        \ \mbs{v} = \mbs{g}
        \ \text{on } \partial \Omega
    \Big\} ,
\end{equation}
where we have simply inverted the order of the maximum and minimum problems. We recall that, in the convex case, problems (\ref{Q2ukle}) and (\ref{rierL4}) are equivalent by the inf-sup theorem \cite{Ekeland:1999}, but not so in the non-convex case. An integration by parts gives (\ref{Q2ukle}) in the equivalent form
\begin{equation}\label{ji8blU}
    \sup_{\mbs{\sigma}}
    \inf_{\mbs{v}}
    \Big\{
        \int_{\partial \Omega}
            \mbs{\sigma}\mbs{\nu} \cdot \mbs{g}
        \, d\mathcal{H}^2
        -
        \int_\Omega {\rm div} \mbs{\sigma} \cdot \mbs{v} \, dx
        \, : \,
        \mbs{\sigma} \in K,
         \ \mbs{v} = \mbs{g}
        \ \text{on } \partial \Omega
    \Big\} ,
\end{equation}
where $d\mathcal{H}^2$ denotes the element of area on the boundary $\partial\Omega$. Evidently, for the supremum to be non-trivial we must have ${\rm div} \, \mbs{\sigma} = {\bf 0}$, i.~e., the stress field must be in equilibrium, whereupon (\ref{ji8blU}) reduces to the static problem
\begin{equation}\label{drL4hi}
    \sup_{\mbs{\sigma}}
    \Big\{
        \int_{\partial \Omega}
            \mbs{\sigma}\mbs{\nu} \cdot \mbs{g}
        \, d\mathcal{H}^2
        \, : \,
        \mbs{\sigma} \in K,
        \ {\rm div} \mbs{\sigma} = {\bf 0}
    \Big\} .
\end{equation}
The question of existence of solutions of problem (\ref{drL4hi}) may be ascertained by recourse to the direct method of the Calculus of Variations \cite{Dacorogna:1989:DMC:63481}. Existence of solutions is indicative of {\sl stability} of the material with respect to microstructure. Stability in turn necessitates some appropriate notion of convexity to be satisfied by the limit domain $K$. In the present setting, the appropriate notion is {\sl symmetric ${\rm div}$-quasiconvexity} \cite{FonsecaMueller:1999, Conti:2017}, cf.~Appendix A, a notion of convexity in symmetric stress space that accounts for the equilibrium constraint ${\rm div} \mbs{\sigma} = {\bf 0}$.

Equally as important as establishing existence is the treatment of cases that depart from the preceding program, specifically, solids for which $K$ fails to be symmetric ${\rm div}$-quasiconvex. In such cases, the supremum in (\ref{drL4hi}) may be attained arbitrarily closely by weakly-convergent sequences of stress fields, but the supremum itself may  not be attained by any one stress field. The weakly-convergent maximizing sequences are typically characterized by increasingly fine microstructure, a situation reminiscent of the fine patterns computed by \cite{0953-8984-20-24-244128}. The weak limits of the maximizing sequences can then be identified as the macroscopically observable, or average, stress fields. The problem is, then, to characterize all macroscopic stress fields that are attainable as weak limits of sequences of maximizing microscopic stress-field sequences. This characterization determines the effective yield behavior of the solid at the macroscale.

Based on standard theory \cite{Dacorogna:1989:DMC:63481} we expect that the macroscopic states thus defined satisfy the {\sl relaxed problem}
\begin{equation}\label{7Hiaxi}
    \sup_{\mbs{\sigma}}
    \Big\{
        \int_{\partial \Omega}
            \mbs{\sigma}\mbs{\nu} \cdot \mbs{g}
        \, d\mathcal{H}^2
        \, : \,
        \mbs{\sigma} \in \bar{K},
        \ {\rm div} \mbs{\sigma} = 0
    \Big\} ,
\end{equation}
for some effective limit domain $\bar{K}$. Evidently, $\bar{K}$ must contain $K$ and be symmetric ${\rm div}$-quasiconvex in order for the supremum of the effective problem (\ref{7Hiaxi}) to be attained. In addition, $\bar{K}$ must be as small as possible in order for the solutions of the effective problem (\ref{7Hiaxi}) to be weak limits of maximizing sequences of the unrelaxed problem (\ref{drL4hi}). These constraints identify $\bar{K}$ as the symmetric ${\rm div}$-quasiconvex envelope of $K$, and can be visualized  as the smallest symmetric ${\rm div}$-quasiconvex set containing $K$.

The remaining problem of interest is to determine the symmetric ${\rm div}$-quasiconvex envelope $\bar{K}$ of the limit surface of fused silica, eqs.~(\ref{TroU6o}) and (\ref{c4iUth}), Fig.~\ref{phLuy9}. An explicit and exact construction of $\bar{K}$ has been derived by Conti {\sl et al.} \cite{Conti:2017}. They show that
the curves
\begin{equation}\label{b1lAyo}
    q = \Big({s} + \frac{3}{4}(p-{r})^2\Big)^{1/2}
\end{equation}
in $(p,q)$-plane represent rank-$2$ connections between states of constant stress in traction equilibrium, and that the curves bound symmetric ${\rm div}$-quasiconvex sets in the $(p,q)$-plane. Evidently, the smallest such set containing $K$, or {\sl rank-$2$ envelope} of $K$, contains $\bar{K}$. The mathematical challenge is to show that the rank-$2$ envelope of $K$ is in fact $\bar{K}$. This equivalence has been proven by Conti {\sl et al.} \cite{Conti:2017}.

Specifically, the rank-$2$ envelope of the limit domain $K$ for fused silica is obtained by fitting a curve of the form (\ref{b1lAyo}) so as to smooth out the transition between the tensile and compressive regimes of the critical-state line. The conditions that determine the extreme rank-$2$ connection are
\begin{subequations}
\begin{align}
    &
    q_t^2 = {s} + \frac{3}{4}(p_t-{r})^2 ,
    \\ &
    q^2 = {s} + \frac{3}{4}(p-{r})^2 ,
    \\ &
    q = B p^\beta,
    \\ &
    \beta B p^{\beta-1}
    =
    \frac{1}{q} \frac{3}{4}(p-{r}) ,
\end{align}
\end{subequations}
to be solved for ${r}$, ${s}$, $p$ and $q$. The values of these variables computed from Tables~\ref{tab:materialdata} and \ref{tab:criticalline} are shown in Table~\ref{qleNi5}.

\begin{table}[h]
	\centering
	\caption{\small The rank-$2$ envelope of fused silica glass.}
	\label{qleNi5}
	\begin{tabular}{lcccc}
		\hline\noalign{\smallskip}
		&$ {r} $ & $ {s} $ & $ p_{\rm min} $ & $ p_{\rm max} $ \\
		\noalign{\smallskip}\hline\noalign{\smallskip}
		& $5.176$ GPa & $7.674$ GPa${}^2$ & $4.141$ GPa & $6.084$ GPa\\
		\noalign{\smallskip}\hline
	\end{tabular}
\end{table}

\begin{figure}[h]
	\begin{subfigure}{0.49\textwidth}\caption{\small } \includegraphics[width=0.99\linewidth]{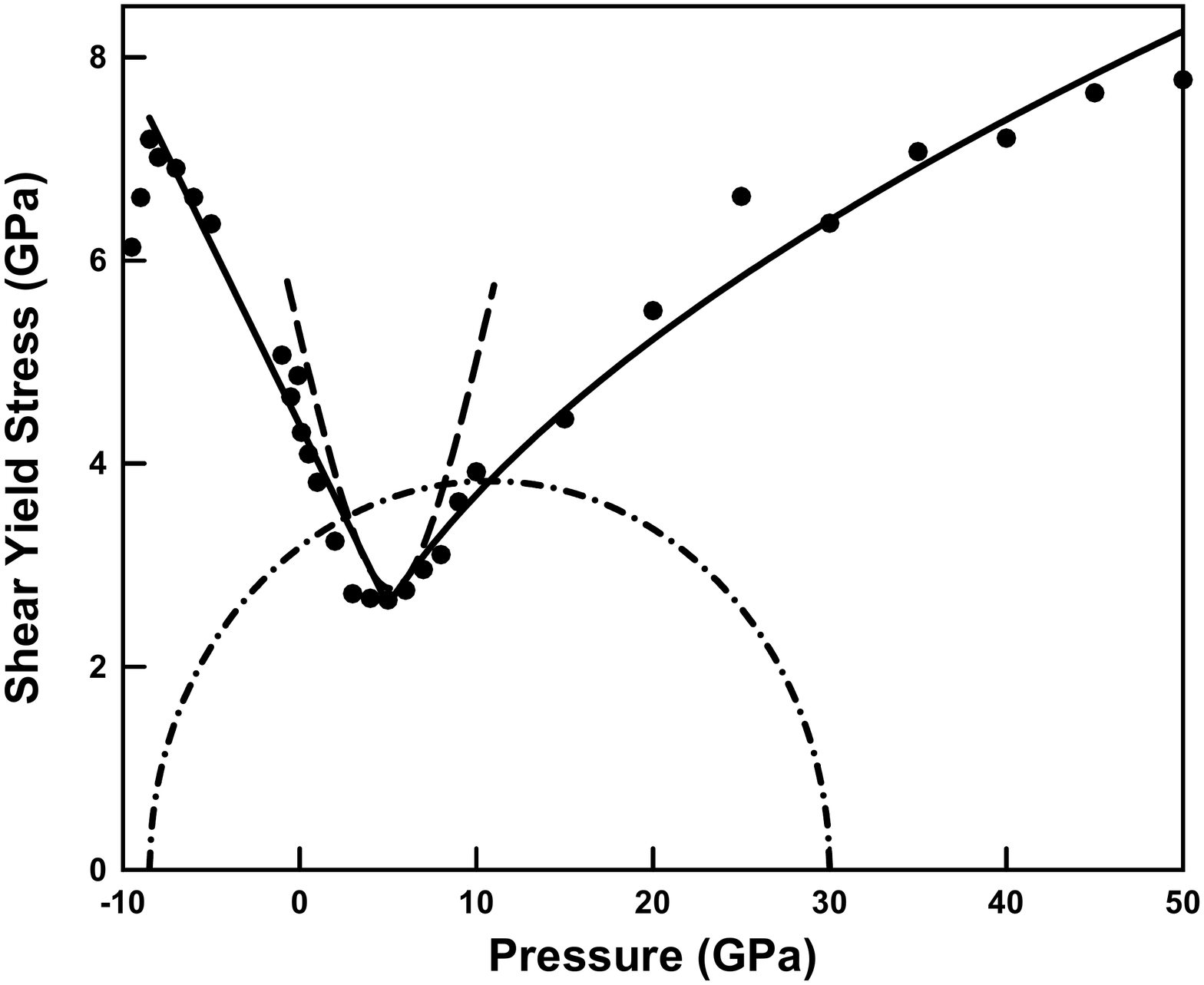}
	\end{subfigure}
	\begin{subfigure}{0.49\textwidth}\caption{\small } \includegraphics[width=0.99\linewidth]{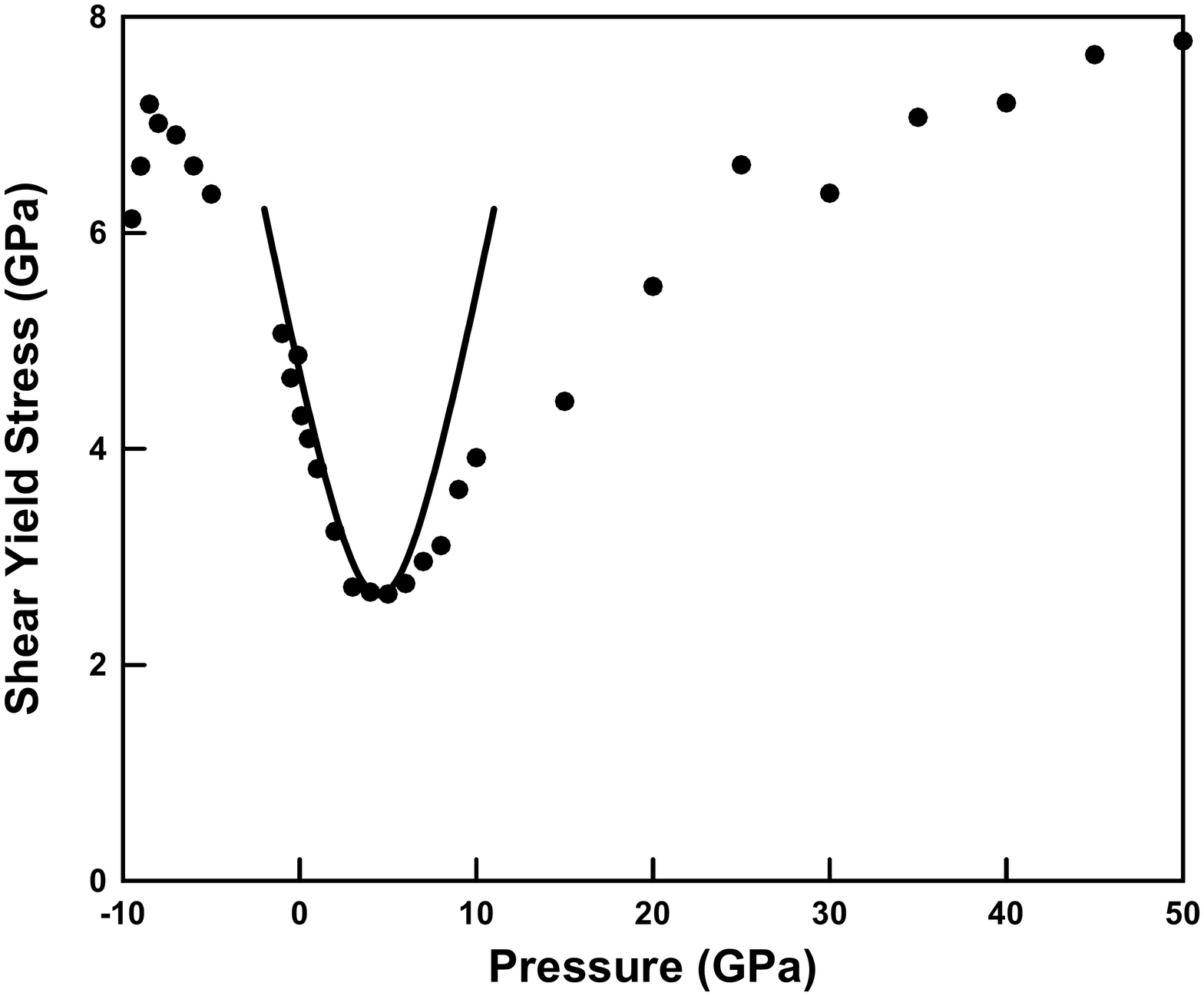}
	\end{subfigure}
    \caption{\small a) Relaxed critical-state line showing rank-2 connection envelope (dash line). b) Rank-2 connection captures the fine structure of the MD data at the tension-to-compression transition point.} \label{Gl9pri}
\end{figure}

The resulting envelope is shown in Fig.~\ref{Gl9pri}a. It bears emphasis that the relaxed limit domain $\bar{K}$ is not convex, which illustrates the fact that symmetric ${\rm div}$-quasiconvex sets are a strictly larger class than convex sets. We also note that $\bar{K} \neq K$, which shows that, indeed, $K$ is not symmetric ${\rm div}$-quasiconvex, or stable against microstructure, as surmised. Fig.~\ref{Gl9pri}b shows the rank-$2$ connection curve in isolation together with the MD data. The comparison suggests that the rank-$2$ envelope construction indeed captures the fine structure of the MD data at the tension-to-compression transition point, which, in hindsight, the unrelaxed model in Fig.~\ref{phLuy9} fails to do. Conversely, we conclude that the fine structure of the MD data at the tension-to-compression transition point is the result of accommodation at the microstructural level.

\section{Summary and concluding remarks}

We have developed a critical-state model of fused silica plasticity on the basis of data mined from Molecular Dynamics (MD) calculations. The MD data is suggestive of an irreversible densification transition in volumetric compression resulting in permanent, or plastic, densification upon unloading. The MD data also reveals an evolution towards a critical state of constant volume under pressure-shear deformation. The trend towards constant volume is from above, when the glass is overconsolidated, or from below, when it is underconsolidated. We have shown that these characteristic behaviors are well-captured by a critical-state model of plasticity, where the densification law for glass takes the place of the classical consolidation law of granular media and the locus of constant-volume states defines the critical-state line.

A salient feature of the critical-state line of fused silica, as identified from MD data, that renders its yield behavior anomalous---and raises it from the commonplace---is that it is strongly non-convex, owing to the existence of two well-differentiated phases, at low and high pressures. This anomalous yield strength of fused silica is indeed consistent with---and born out by---the measurements of \cite{MEADE1072}. The strong non-convexity of yield in turn explains the patterning observed by \cite{0953-8984-20-24-244128} in molecular dynamics calculations of amorphous solids deforming in shear.

The proclivity of fused silica for patterning at the microscale raises the question of its effective behavior at the macroscale, i.~e., the average stress and deformation conditions that are attainable when microstructure is accounted for. Remarkably, this question can be rigorously and exactly ascertained for fused silica within the framework of limit analysis and the calculus of variations \cite{Conti:2017}. We recall that stress solutions of the static problem of limit analysis are subject to an equilibrium, or divergence, constraint. The problem is, therefore, to determine all macroscopic states of stress attainable as averages of microscopic stress fields that are within yield and at equilibrium. Conti {\sl et al.} \cite{Conti:2017} have shown that the effective or macroscopic critical-state line thus defined can be computed explicitly and exactly through a rank-$2$ envelope construction in the $(p,q)$-plane. This remarkable result effectively {\sl upscales} the microscopic critical state model delineated by the MD data to the macroscale. The rank-$2$ envelope indeed captures the fine structure of the critical-state line, as gleaned from MD data, at the tension-to-compression transition, which further underscores the importance of microstructure in shaping the macroscopic, or effective, behavior of fused silica. The effective or macroscopic model of fused silica is stable with respect to microstructure, defines well-posed boundary-value problems and is, therefore, suitable for use in large-scale continuum calculations.

\section*{Acknowledgements}

WS, SH and MO gratefully acknowledge support from the US Office of Naval Research through grant N000141512453. SC is grateful for the support of the Deutsche Forschungsgemeinschaft through the Sonderforschungsbereich 1060 {\sl ``The mathematics of emergent effects''}.

\begin{appendix}

\section{Relaxation of the limit-analysis problem}

For completeness, we summarize the main concepts and arguments leading to the computation of the relaxed critical-state line and limit domain $\bar{K}$. Further mathematical details may be found in the article of \cite{Conti:2017}.

We begin by introducing the dissipation functional $F : L^\infty(\Omega, \mathbb{R}^{3\times 3}_{\rm sym}, {\rm div})$ $\to$ $\overline{\mathbb{R}}$ defined as
\begin{equation}\label{StoeF7}
    F(\mbs{\sigma})
    =
    \left\{
    \begin{array}{ll}
        \int_{\partial \Omega}
            \mbs{\sigma}\mbs{\nu} \cdot \mbs{g}
        \, d\mathcal{H}^2 , &
        \text{if } \mbs{\sigma} \in K \text{ almost everywhere in $\Omega$} , \\
        -\infty , & \text{otherwise} ,
    \end{array}
    \right.
\end{equation}
where $L^\infty(\Omega, \mathbb{R}^{3\times 3}_{\rm sym}, {\rm div})$ is the space of essentially bounded stress fields over $\Omega$ with zero distributional divergence endowed with its weak${}^*$ topology and we assume $\Omega$ to be Lipschitz and bounded. Then, problem (\ref{drL4hi}) is equivalent to
\begin{equation}\label{souv2E}
    \sup_{\mbs{\sigma} \in L^\infty(\Omega, \mathbb{R}^{3\times 3}_{\rm sym}, {\rm div})}
    F(\mbs{\sigma}) .
\end{equation}

The question of existence of solutions of problem (\ref{souv2E}) may be ascertained by recourse to the direct method of the Calculus of Variations \cite{Dacorogna:1989:DMC:63481}. Thus, if $K$ is bounded the functional $F$ is clearly weakly coercive in $L^\infty(\Omega, \mathbb{R}^{3\times 3}_{\rm sym}, {\rm div})$. In addition, if $\mbs{g} \in L^1(\partial\Omega, \mathbb{R}^3)$, the space of integrable velocity fields over $\partial\Omega$, then the dissipation function
\begin{equation}
    D(\mbs{\sigma})
    =
    \int_{\partial \Omega}
        \mbs{\sigma}\mbs{\nu} \cdot \mbs{g}
    \, d\mathcal{H}^2
\end{equation}
is weakly continuous in $L^\infty(\Omega, \mathbb{R}^{3\times 3}_{\rm sym}, {\rm div})$ by the trace theorem for $W^{1,1}(\Omega, \mathbb{R}^3)$ (cf., e.~g., \cite{Ambrosio:2000}, p.~168).

In order to apply Tonelli's theorem \cite{Tonelli:1921}, there remains to identify conditions under which $F$ is upper-semicontinuous on $L^\infty(\Omega, \mathbb{R}^{3\times 3}_{\rm sym}, {\rm div})$. We recall that $F$ is upper-semincontinuous if $\limsup_{h\to\infty} F(\mbs{\sigma}_h) \leq F(\mbs{\sigma})$ for every $\mbs{\sigma} \in L^\infty(\Omega, \mathbb{R}^{3\times 3}_{\rm sym}, {\rm div})$ and every sequence $(\mbs{\sigma}_h)$ converging weak${}^*$ to $\mbs{\sigma}$ in $L^\infty(\Omega, \mathbb{R}^{3\times 3}_{\rm sym}, {\rm div})$. We expect upper-semicontinuity to necessitate some appropriate notion of convexity of $K$. The appropriate notion is {\sl symmetric ${\rm div}$-quasiconvexity}, which is a special case of ${\mathcal A}$-quasiconvexity, see \cite{FonsecaMueller:1999} and \cite{Conti:2017} for the mathematical treatment.

\begin{defn}[Symmetric ${\rm div}$-quasiconvex function]
A function $f : \mathbb{R}^{3\times 3}_{\rm sym} \to \overline{\mathbb{R}}$ is symmetric ${\rm div}$-quasiconvex if
\begin{equation}
    f(\mbs{\sigma}) \leq \int_{(0,1)^3} f(\mbs{\sigma}+\mbs{\xi}) \, dx ,
\end{equation}
for all $\mbs{\sigma} \in \mathbb{R}^{3\times 3}_{\rm sym}$ and all $\mbs{\xi} \in C^\infty_{\rm per}([0,1]^3, \mathbb{R}^{3\times 3}_{\rm sym})$ such that ${\rm div} \, \mbs{\xi} = {\bf 0}$ and $\int_{(0,1)^3} \mbs{\xi}\, dx=\bf 0$.
\end{defn}

This notion of convexity may be transferred to sets.

\begin{defn}[Symmetric ${\rm div}$-quasiconvex set]
A compact set $K \subset \mathbb{R}^{3\times 3}_{\rm sym}$ is symmetric ${\rm div}$-quasiconvex if there is a symmetric ${\rm div}$-quasiconvex function $g\in C^0(\mathbb{R}^{3\times 3}_{\rm sym};[0,\infty))$ such that $K=\{\mbs\sigma: g(\mbs\sigma)=0\}$.
\end{defn}

Evidently, every convex function, respectively convex set, is a symmetric ${\rm div}$-quasiconvex function, respectively symmetric ${\rm div}$-quasiconvex set, but the converse, as we shall see, is not true. The relevance of symmetric ${\rm div}$-quasi-\-convexity to problem (\ref{souv2E}) stems from the following connection.

\begin{thm}[${\rm div}$-quasiconvexity and upper-semicontinuity]\label{riA8ro}
Suppose that the compact set $K \subset \mathbb{R}^{3\times 3}_{\rm sym}$ is symmetric ${\rm div}$-quasiconvex. Then, the functional (\ref{StoeF7}) is weak${}^*$ upper semicontinuous in $L^\infty(\Omega, \mathbb{R}^{3\times 3}_{\rm sym}, {\rm div})$.
\end{thm}

This theorem is in the spirit of the classical theorems of Morrey \cite{Morrey1:952}, which put forth a equivalence between quasiconvexity and lower-semicontinuity of energy functionals. The proof of the theorem is based on the results of Fonseca and M\"uller \cite{FonsecaMueller:1999}  and may be found in \cite{Conti:2017}. Existence then follows from an application of Tonelli's theorem \cite{Tonelli:1921}.

\begin{thm}[Existence]\label{sWoaS1}
Let $\Omega \subset \mathbb{R}^3$ be bounded and Lipschitz. Suppose that $K \subset \mathbb{R}^{3\times 3}_{\rm sym}$ is a nonempty compact symmetric ${\rm div}$-quasiconvex set. Let $\mbs{g} \in L^1(\partial\Omega, \mathbb{R}^3)$. Then, the static problem (\ref{souv2E}) of limit analysis has solutions.
\end{thm}

Suppose now that $K$ fails to be symmetric ${\rm div}$-quasiconvex. Based on standard theory \cite{Dacorogna:1989:DMC:63481} we expect that the weak limits of maximizing sequences, representing the macroscopic states of  solids with increasingly fine microstructure, satisfy the relaxed problem
\begin{equation}\label{g0ewLE}
    \sup_{\mbs{\sigma} \in L^\infty(\Omega, \mathbb{R}^{3\times 3}_{\rm sym}, {\rm div})}
    \bar{F}(\mbs{\sigma}) ,
\end{equation}
where the relaxed functional $\bar{F} : L^\infty(\Omega, \mathbb{R}^{3\times 3}_{\rm sym}, {\rm div})$ $\to$ $\overline{\mathbb{R}}$ has the form
\begin{equation}\label{v1eThl}
    \bar{F}(\mbs{\sigma})
    =
    \left\{
    \begin{array}{ll}
        \int_{\partial \Omega}
            \mbs{\sigma}\mbs{\nu} \cdot \mbs{g}
        \, d\mathcal{H}^2 , &
        \text{if } \mbs{\sigma} \in \bar{K}
        \text{ almost everywhere in $\Omega$} , \\
        - \infty , & \text{otherwise} ,
    \end{array}
    \right.
\end{equation}
for some effective limit domain $\bar{K}$. Evidently, $\bar{K}$ must contain $K$ and be symmetric ${\rm div}$-quasiconvex in order for $\bar{F}$ to be upper-semicontinuous and the supremum in the effective problem (\ref{g0ewLE}) to be attained. In addition, $\bar{K}$ must be as small as possible in order for the solutions of the effective problem (\ref{g0ewLE}) to be weak limits of maximizing sequences of the unrelaxed problem (\ref{souv2E}). These constraints lead to the following notion of envelope.

\begin{defn}[Symmetric ${\rm div}$-quasiconvex envelope]
The symmetric ${\rm div}$-quasiconvex envelope of a compact set $K \subset \mathbb{R}^{3\times 3}_{\rm sym}$ is the set
\begin{equation}
\begin{split}
\bar{K}
    =
    \{&
        \mbs{\sigma} \in \mathbb{R}^{3\times 3}_{\rm sym} \, : \,
       g(\mbs\sigma) \le \max g(K) \\
       &\text{ for all  symmetric ${\rm div}$-quasiconvex  $g\in C^0(\mathbb{R}^{3\times 3}_{\rm sym};[0,\infty))$}
    \} .
    \end{split}
\end{equation}
\end{defn}

The remaining problem of interest is to determine the symmetric ${\rm div}$-quasiconvex envelope $\bar{K}$ of sets $K$ in the $(p,q)$-plane. For sets of a specific form, a  construction of $\bar{K}$ has been put forth by \cite{Conti:2017}. Here we limit ourselves to summarizing the main arguments and refer the interested reader to \cite{Conti:2017} for mathematical details.

A main building block of the explicit construction of $\bar{K}$ is the following classical result of \cite{Tartar:1985}.

\begin{thm} [Tartar'85]\label{siUjL4}
The function $f(\mbs{\sigma}) = 2 |\mbs{\sigma}|^2 - {\rm tr}(\mbs{\sigma})^2$ is symmetric ${\rm div}$-quasiconvex.
\end{thm}

We recall that the critical-state surface of fused silica is isotropic and is defined by its trace, or critical-state line, on the $(p,q)$-plane. From Tartar's theorem~\ref{siUjL4}, \cite{Conti:2017} show the following.

\begin{thm}\label{hiaT0l}
The set $\{ \mbs{\sigma} \in \mathbb{R}^{3\times 3}_{\rm sym} \, : \, q^2 \leq {s} + \frac{3}{4}(p-{r})^2\}$, with ${r}$, ${s} \in \mathbb{R}$, is symmetric ${\rm div}$-quasiconvex.
\end{thm}

The curves $q = ({s} + \frac{3}{4}(p-{r})^2)^{1/2}$ in $(p,q)$-plane represent rank-$2$ connections, or connections between stress states in equilibrium. By theorem~\ref{hiaT0l}, the curves bound symmetric ${\rm div}$-quasiconvex sets in the $(p,q)$-plane. Therefore, the smallest such set containing $K$, or {\sl rank-$2$ envelope} of $K$, contains $\bar{K}$. \cite{Conti:2017} show that the rank-$2$ envelope of $K$ and $\bar{K}$ in fact coincide, which effectively replaces the computation of $\bar{K}$ by the much easier task of constructing the rank-$2$ envelope of $K$.

For fused silica with $K$ determined from MD data, the rank-$2$ envelope construction of $\bar{K}$ is given in Section~\ref{Relax}, Table~\ref{qleNi5}.

\end{appendix}

\bibliography{research}
\bibliographystyle{unsrt}

\end{document}